\documentclass[a4paper,twocolumn,11pt,accepted=2025-05-05]{quantumarticle}
\pdfoutput=1

\usepackage{resizegather}
\usepackage{amssymb}
\usepackage{amsmath}

\ExplSyntaxOn

\NewDocumentCommand{\pFq}{O{}mmmmm}
 {
  \group_begin:
  \keys_set:nn { hypergeometric } { #1 }
  \hypergeometric_print:nnnnn { #2 } { #3 } { #4 } { #5 } { #6 }
  \group_end:
 }
\NewDocumentCommand{\hypergeometricsetup}{m}
 {
  \keys_set:nn { hypergeometric } { #1 }
 }

\tl_new:N \l_hypergeometric_divider_tl
\tl_new:N \l_hypergeometric_left_tl
\tl_new:N \l_hypergeometric_right_tl

\keys_define:nn { hypergeometric }
 {
  symbol .tl_set:N = \l_hypergeometric_symbol_tl,
  symbol .initial:n = \phi,
  separator .tl_set:N = \l_hypergeometric_separator_tl,
  separator .initial:n = {},
  skip .tl_set:N = \l_hypergeometric_skip_tl,
  skip .initial:n = 8,
  divider .choice:,
  divider/semicolon .code:n = \tl_set:Nn \l_hypergeometric_divider_tl { \;; },
  divider/bar .code:n = \tl_set:Nn \l_hypergeometric_divider_tl { \;\middle|\; },
  divider .initial:n = semicolon,
  fences .choice:,
  fences/brack .code:n = 
   \tl_set:Nn \l_hypergeometric_left_tl {[}
   \tl_set:Nn \l_hypergeometric_right_tl {]},
  fences/parens .code:n = 
   \tl_set:Nn \l_hypergeometric_left_tl {(}
   \tl_set:Nn \l_hypergeometric_right_tl {)},
  fences .initial:n = brack,
 }

\cs_new_protected:Nn \hypergeometric_print:nnnnn
 {
  {} \sb {#1} \l_hypergeometric_symbol_tl \sb { #2 }
  \left\l_hypergeometric_left_tl
  \genfrac .. 
           {0pt} 
           {} 
           { \__hypergeometric_process:n { #3 } } 
           { \__hypergeometric_process:n { #4 } } 
  \l_hypergeometric_divider_tl
  #5
  \right\l_hypergeometric_right_tl
 }

\cs_new_protected:Nn \__hypergeometric_process:n
 {
  \clist_use:nn { #1 }
   {
    {\l_hypergeometric_separator_tl}
    \mspace { \l_hypergeometric_skip_tl mu }
   }
 }

\ExplSyntaxOff

\usepackage{graphicx}
\usepackage{placeins}
\usepackage{mathrsfs}
\usepackage{bbm}
\usepackage{dsfont}
\usepackage{marvosym}

\usepackage[permil]{overpic}

\usepackage[dvipsnames]{xcolor}

\usepackage{todonotes}
\usepackage[normalem]{ulem}
\usepackage{dcolumn}
\usepackage{hyperref}
\usepackage{rotating}
\usepackage{mathtools}
\usepackage[T1]{fontenc}
\usepackage{caption}
\usepackage{accents}

\usetikzlibrary{decorations.pathreplacing}
\usetikzlibrary{arrows}
\usetikzlibrary{shapes,snakes}
\usetikzlibrary{calc}
\usetikzlibrary{shapes.geometric}
\usetikzlibrary{decorations}
\usetikzlibrary{arrows.meta}
\usetikzlibrary{fadings}
\tikzfading[name=fade out, 
    inner color=transparent!0,
    outer color=transparent!100]
\pgfdeclaredecoration{lightning bolt}{draw}{
\state{draw}[width=\pgfdecoratedpathlength]{
  \pgfpathmoveto{\pgfpointorigin}%
  \pgfpathlineto{\pgfpoint{\pgfdecoratedpathlength*0.6}%
    {-\pgfdecoratedpathlength*.1}}%
  \pgfpathlineto{\pgfpoint{\pgfdecoratedpathlength*0.55}{0pt}}%
  \pgfpathlineto{\pgfpoint{\pgfdecoratedpathlength}{0pt}}%
  \pgfpathlineto{\pgfpoint{\pgfdecoratedpathlength*0.4}%
    {\pgfdecoratedpathlength*.1}}%
  \pgfpathlineto{\pgfpoint{\pgfdecoratedpathlength*0.45}{0pt}}%
  \pgfpathclose%
}%
}

\makeatletter
\newcommand*{\centerfloat}{%
  \parindent \z@
  \leftskip \z@ \@plus 1fil \@minus \textwidth
  \rightskip\leftskip
  \parfillskip \z@skip}
\makeatother

\usepackage[braket]{qcircuit}

\usepackage{amsthm}

    \theoremstyle{definition}

\usepackage{physics}
\usepackage{qcircuit}
\usepackage{subcaption}
\usepackage{enumitem}
\newcommand{\bigslant}[2]{{\raisebox{.2em}{$#1$}\left/\raisebox{-.2em}{$#2$}\right.}}

\newcommand{\transp}{\mathsf{T}}

\newcommand{\G}[2]{G^{#2}_{#1}}
\newcommand{\Phit}[2]{\Phi^{#2}_{#1}}
\newcommand{\tend}{\textrm{end}}

\newcommand{\kg}[1]{{\color{black} #1}}

\hypergeometricsetup{
  separator={,},
  divider=bar,
}

\begin{document}

\title{On noise in swap ASAP repeater chains: exact analytics, distributions and tight approximations}

\author{K. Goodenough}
\affiliation{College of Information and Computer Science, University of Massachusetts Amherst, 140 Governors Dr, Amherst, Massachusetts 01002, USA}
\author{T. Coopmans}
\affiliation{Leiden Institute of Advanced Computer Science, Leiden University, Leiden, The Netherlands}
\author{D. Towsley}
\affiliation{College of Information and Computer Science, University of Massachusetts Amherst, 140 Governors Dr, Amherst, Massachusetts 01002, USA}

\maketitle

\begin{abstract}
	Losses are one of the main bottlenecks for the distribution of entanglement in quantum networks, which can be overcome by the implementation of quantum repeaters. The most basic form of a quantum repeater chain is the swap ASAP repeater chain. In such a repeater chain, elementary links are probabilistically generated and deterministically swapped as soon as two adjacent links have been generated. As each entangled state is waiting to be swapped, decoherence is experienced, turning the fidelity of the entangled state between the end nodes of the chain into a random variable. Fully characterizing the (average) fidelity as the repeater chain grows is still an open problem. Here, we analytically investigate the case of equally-spaced repeaters, where we find exact analytic formulae for all moments of the fidelity up to $25$ segments. We obtain these formulae by providing a general solution in terms of a \emph{generating function}; a function whose $n$'th term in its Maclaurin series yields the moments of the fidelity for $n$ segments. We generalize this approach as well to a \emph{global cut-off} policy --- a method for increasing fidelity at the cost of longer entanglement delivery times --- allowing for fast optimization of the cut-off parameter by eliminating the need for Monte Carlo simulation. We furthermore find simple approximations of the average fidelity that are exponentially tight, and, for up to 10 segments, the full distribution of the delivered fidelity. We use this to analytically calculate the secret-key rate, both with and without binning methods.
\end{abstract}

\section{Introduction}
Quantum communication allows for benefits inaccessible with classical communication alone, key examples including long-baseline telescopes~\cite{gottesman2012longer, sajjad2024quantum}, secret sharing~\cite{hillery1999quantum, markham2008graph} and distributed quantum computation~\cite{cirac1999distributed, serafini2006distributed}. These protocols all rely on end-users sharing entanglement. A quantum internet~\cite{wehner2018quantum} would allow for the distribution of such entanglement over large distances. One way to achieve this distribution is to transmit states over optical fibers; this has the benefit of being able to use pre-existing classical infrastructure~\cite{rabbie2022designing}.

One bottleneck of quantum communication is losses, where the success probability of a single photon arriving end-to-end over a distance $L$ is given by $p=~\exp(-\frac{L}{L_\textrm{att}})$, $L_\textrm{att}\approx 22$~kilometers. A naive amplification of the signal to overcome this (as one does in the classical case) is prohibited by the no-cloning theorem. Fortunately, \emph{quantum repeaters}~\cite{briegel1998quantum} provide a solution. That is, by splitting the total length into $n$ segments and generating entanglement over the segments, end-to-end entanglement can be created by performing so-called entanglement swaps~\cite{azuma2022quantum}. Assuming that entanglement is swapped as soon as possible (\emph{swap ASAP}) one can show that the state delivery rate now scales as $\mathcal{O}\left(^n\hspace{-2mm}\sqrt{p}\right)$ (where we assume here and in the rest of the paper that the swaps are deterministic). Thus, while the rate still decays exponentially with the distance, the delivery rate increases significantly compared to the case without repeaters.

Although repeaters have the ability to increase the delivery rate, they also introduce additional \emph{noise}, lowering the quality of the delivered entanglement. There are two contributions to this additional noise~\cite{rozpkedek2018parameter, rozpkedek2019near}. First, noise arising from imperfections in the devices (such as state preparation); second, noise arising from memory decoherence. Note that the first type of noise is deterministic\footnote{\kg{Or, at the very least, approximately independent of when the different elementary links succeed.}}. In contrast, the second type of noise is random. This is due to the fact that a node needs to wait for a random amount of time until both its incident links are present before a swap can be performed. This random noise causes the fidelity of the end-to-end state to be a random variable. 

The trade-off between increasing the rate and average fidelity complicates understanding what can be done with near-term devices, especially in combination with so-called \emph{cut-off policies}. A cut-off policy is one that sets a threshold for when to discard the present entanglement. After the discard, the entanglement generation is started anew. Discarding generated entanglement reduces the rate at which entanglement is delivered, but, by a clever choice of a cut-off policy, will mitigate the time spent on distributing end-to-end entanglement with too low a fidelity. A cut-off policy thus allows for a trade-off between the rate and quality of the delivered end-to-end entanglement. Cut-off policies and their trade-offs have been studied, for example, in~\cite{rozpkedek2018parameter, rozpkedek2019near, collins2007multiplexed, avis2022requirements, li2020efficient, khatri2019practical, santra2018quantum, davies2023tools, azuma2021tools, praxmeyer2013reposition, kamin2023exact, shchukin2019waiting, reiss2023deep, zang2023entanglement}.

However, analytic expressions for the expected fidelity in the presence or absence of cut-off policies have been derived only up to $8$ segments~\cite{kamin2023exact}. Here, we analytically calculate the average fidelity of the delivered state for homogeneous swap ASAP repeater chains consisting of up $25$ segments. Furthermore, we derive a closed-form expressions for the probability distribution of the fidelity for up to $10$ segments. While the analytical expressions grow unwieldy very fast, we find a closed-form expression of a so-called \emph{generating function} --- a function whose Maclaurin series coefficients correspond to the average noise parameters. This, in a sense, fully characterizes the average noise for arbitrary length swap ASAP repeater chains. Using tools from analytic combinatorics, we use this generating function to find simple approximations that are provably asymptotically tight, and which in practice are indistinguishable from exact results already for three and more segments. As a caveat, we assume for the exact analytics that the end nodes experience no decoherence, which holds when performing quantum key distribution (QKD)~\cite{bennett2020quantum} (when the states at the end nodes can be measured directly) or can be justified in the case where the end nodes have significantly more resources than the repeater chain, see also~\cite{kamin2023exact} for a similar setup.

We note that a specific type of generating function, the probability generating function, has been used before in the noise analysis for more general quantum repeater protocols, ranging up to $n=8$ segments for a full analysis of all moments of the fidelity~\cite{kamin2023exact}. Our generating function approach is more powerful, since it captures all probability generating functions at once; its $n$'th term is exactly the probability generating function associated with $n$ segments. This thus provides an explanation for the expressions from~\cite{kamin2023exact}, where the expressions were found through brute-force calculation.
By taking appropriate derivatives/expansions of our generating function, we compute an analytical expression of the average fidelity for up to 25 segments within seconds, and find an analytical expression for the distribution of the fidelity for up to 10 repeater segments.

We furthermore show how most of our tools can be applied to a \emph{global} cut-off policy. This cut-off policy does not require any communication between the nodes, making it especially relevant for near-term quantum networks.

We show how our tools can be used to 1) understand the effect of adding or removing repeaters for a given distance, 2) derive the distribution of the noise and how it relates to the cut-off threshold, 3) calculate the secret-key rate exactly for up to 10 segments, with and without so-called \emph{binning strategies}, 4) understand the interaction between decoherence and losses in a manner that is independent of the number of repeaters.

All of our analytical results can be found in the provided Mathematica files~\cite{mathematicafiles}.

We leave most of the technical details to the Appendix, and focus on the general concepts in the main text. We start by making explicit our model in Section \ref{sec:model}, where we also discuss the global cut-off policy. In Section \ref{sec:recursion} we give a general overview of how to calculate exact analytics (with and without a global cut-off). Section \ref{sec:gen_function} presents an overview of the derivation of the approximation for the average fidelity, which is based on techniques from analytic combinatorics. For our final result we focus in Section~\ref{sec:distribution}, not on the average fidelity, but on the fidelity distribution. We close each of the previous three sections by applying our tools to numerical exploration of repeater chains. We conclude with a discussion on possible extensions and touch upon follow-up work that extends some of our results numerically to both the inhomogeneous and multipartite setting in Section~\ref{sec:discussion}.

\section{Swap ASAP repeater model}\label{sec:model}
In this section we first present the basic setup of a swap ASAP protocol in the case of qubit depolarizing noise. We then discuss how our results extend to a large set of other noise models. Afterwards, we discuss the cut-off policy relevant for this work. We note that a similar setup was also investigated in~\cite{kamin2023exact}.

\subsection{Model details}
The aim of this section is to lay out our model of a swap ASAP repeater chain. A repeater chain consists of $n$ segments, separated by $n-1$ repeater nodes. We order the $n$ segments from left to right by $i=1,\ldots, n$. In a swap ASAP protocol, elementary link generation attempts are made in parallel for each segment in a given \textit{round}. We assume that rounds are discretized (with duration equal to a single timestep), and label the round in which segment $i$ succeeds by $t_i$, where $t_i\geq 1$. \kg{This is a common model in quantum networking~\cite{pant2019routing, fittipaldi2022linear, van2023entanglement, van2024utilizing}}\footnote{We note that a continuous-time model is also a possible model, which is especially accurate for small $p$ (see e.g.~\cite{vardoyan2019stochastic}). While this model is out-of-scope for this manuscript, the techniques used in this manuscript (such as the recursion relation and the generating function approach) could be applied to the continuous-time setting as well.}. \kg{We note that this requires synchronization of all of the nodes in the repeater chain. This synchronization can be challenging when scaling up to larger repeater chains. Furthermore, quantum networks would require the nodes along all flows to synchronize their entanglement generation, which could become impractical for larger networks.}

Link generation succeeds with a uniform probability $p$ across all segments. We assume that link generation is attempted again only \emph{after} successful end-to-end entanglement generation or after reaching a \emph{cut-off condition} (see the next subsection). This is not only done to aid in the analysis, but it also prevents congestion\footnote{In the setting where there is only a single memory for each repeater node per connection, newly created entanglement may need to idle in memory as older states are kept around until end-to-end entanglement has been established.} and is closer to the capabilities of near-term devices. \kg{For future quantum networks however, the idling of segments can lead to an under-utilization of resources, and thus to a lower entanglement throughput.}

\begin{figure}[h!]
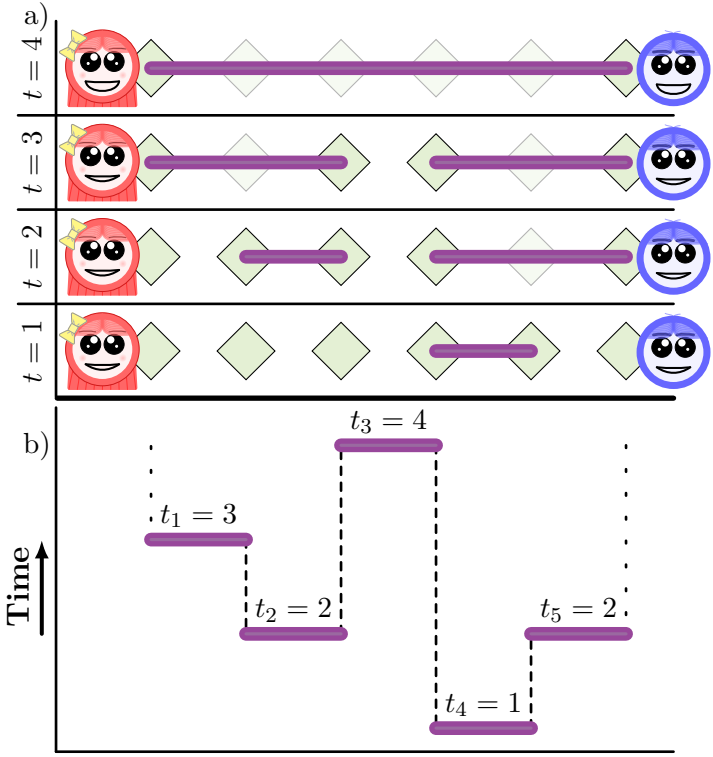

\centerfloat
\begin{tikzpicture}[scale=1.25, line cap=round, entanglement/.style={shade, shading=ring, line width = 0.7mm, Purple,  double=RedViolet!40!Periwinkle!90,double distance=1.2pt, repeater/.style={diamond, NavyBlue}}]

\draw[line width=1pt] (-0.0,-0.25) -- (-0.0,3.4);
\draw[line width=1pt] (-0.0,3.515) -- (-0.0,7.5);

\draw[line width=1pt] (0.0,-0.25) -- (6.5,-0.25);
\node[rotate=90] at (-0.4,1.5) {\large\textbf{Time}};
  \draw [-{Latex[length=2.5mm]}, line width = 1.5pt] (-0.15,1)   -- (-0.15,2);

\node[] at (-0.2,3.0) {b)};
\node[] at (-0.2,7.55) {a)};

\node[rotate=90] at (-0.25,4.0) {$t = 1$};
\node[rotate=90] at (-0.25,5.0) {$t = 2$};
\node[rotate=90] at (-0.25,6.0) {$t = 3$};
\node[rotate=90] at (-0.25,7.0) {$t = 4$};

\draw[line width=1pt, dashed] (1, 6) -- (1.9,6) node[right] {};
\draw[line width=1pt, dashed] (2.1, 6) -- (2,6) node[right] {};

\node[diamond, draw, scale=1.5, fill = LimeGreen!20] (d) at (1,7) {};
\node[diamond, draw, scale=1.5, opacity=0.3, fill = LimeGreen!20] (d) at (2,7) {};
\node[diamond, draw, scale=1.5, opacity=0.3, fill = LimeGreen!20] (d) at (3,7) {};
\node[diamond, draw, scale=1.5, opacity=0.3, fill = LimeGreen!20] (d) at (4,7) {};
\node[diamond, draw, scale=1.5, opacity=0.3, fill = LimeGreen!20] (d) at (5,7) {};
\node[diamond, draw, scale=1.5, fill = LimeGreen!20] (d) at (6,7) {};

\node[diamond, draw, scale=1.5, fill = LimeGreen!20] (d) at (1,6) {};
\node[diamond, draw, scale=1.5, opacity=0.3, fill = LimeGreen!20] (d) at (2,6) {};
\node[diamond, draw, scale=1.5, fill = LimeGreen!20] (d) at (3,6) {};
\node[diamond, draw, scale=1.5, fill = LimeGreen!20] (d) at (4,6) {};
\node[diamond, draw, scale=1.5, opacity=0.3, fill = LimeGreen!20] (d) at (5,6) {};
\node[diamond, draw, scale=1.5, fill = LimeGreen!20] (d) at (6,6) {};

\node[diamond, draw, scale=1.5, fill = LimeGreen!20] (d) at (1,5) {};
\node[diamond, draw, scale=1.5, fill = LimeGreen!20] (d) at (2,5) {};
\node[diamond, draw, scale=1.5, fill = LimeGreen!20] (d) at (3,5) {};
\node[diamond, draw, scale=1.5, fill = LimeGreen!20] (d) at (4,5) {};
\node[diamond, draw, scale=1.5, opacity=0.3, fill = LimeGreen!20] (d) at (5,5) {};
\node[diamond, draw, scale=1.5, fill = LimeGreen!20] (d) at (6,5) {};

\node[diamond, draw, scale=1.5, fill = LimeGreen!20] (d) at (1,4) {};
\node[diamond, draw, scale=1.5, fill = LimeGreen!20] (d) at (2,4) {};
\node[diamond, draw, scale=1.5, fill = LimeGreen!20] (d) at (3,4) {};
\node[diamond, draw, scale=1.5, fill = LimeGreen!20] (d) at (4,4) {};
\node[diamond, draw, scale=1.5, fill = LimeGreen!20] (d) at (5,4) {};
\node[diamond, draw, scale=1.5, fill = LimeGreen!20] (d) at (6,4) {};


\draw[line width=5pt, entanglement] (1, 7) -- (6,7) node[right] {};

\draw[line width=1pt] (-0.4,6.5) -- (6.5,6.5);
\draw[line width=5pt, entanglement] (1, 6) -- (3,6) node[right] {};
\draw[line width=5pt, entanglement] (4, 6) -- (6,6) node[right] {};
\draw[line width=1pt] (-0.4,5.5) -- (6.5,5.5);
\draw[line width=5pt, entanglement] (2, 5) -- (3,5) node[right] {};
\draw[line width=5pt, entanglement] (4, 5) -- (6,5) node[right] {};
\draw[line width=1pt] (-0.4,4.5) -- (6.5,4.5);
\draw[line width=5pt, entanglement] (4, 4) -- (5,4) node[right] {};

\node[scale=2.2] at (0.45, 4.0) {\input{Alice_normal.tex}};
\node[scale=2.2] at (0.45, 5.0) {\input{Alice_normal.tex}};
\node[scale=2.2] at (0.45, 6.0) {\input{Alice_normal.tex}};
\node[scale=2.2] at (0.45, 7.0) {\input{Alice.tex}};

\node[scale=2.2] at (6.5, 4.0) {\input{Bob_normal.tex}};
\node[scale=2.2] at (6.5, 5.0) {\input{Bob_normal.tex}};
\node[scale=2.2] at (6.5, 6.0) {\input{Bob_normal.tex}};
\node[scale=2.2] at (6.5, 7.0) {\input{Bob.tex}};

\draw[line width=1pt, dashed] (2, 2) -- (2,1) node[right] {};
\draw[line width=1pt, dashed] (3, 1) -- (3,3) node[right] {};
\draw[line width=1pt, dashed] (4, 3) -- (4,0) node[right] {};
\draw[line width=1pt, dashed] (5, 0) -- (5,1) node[right] {};

\draw[line width=1pt,  dash pattern=on \pgflinewidth off 8pt] (1, 1.95) -- (1,3) node[right] {};
\draw[line width=1pt,  dash pattern=on \pgflinewidth off 8pt] (6, 3) -- (6,1) node[right] {};


\draw[line width=2pt] (0.013,3.5) -- (6.5,3.5);
\draw[line width=5pt, entanglement] (1, 2) -- (2,2) node[above] {};
\draw[line width=5pt, entanglement] (2, 1) -- (3,1) node[right] {};
\draw[line width=5pt, entanglement] (3, 3) -- (4,3) node[right] {};
\draw[line width=5pt, entanglement] (4, 0) -- (5,0) node[right] {};
\draw[line width=5pt, entanglement] (5, 1) -- (6,1) node[right] {};

\node[] at (1.5,2.25) {$t_1 = 3$};
\node[] at (2.5,1.25) {$t_2 = 2$};
\node[] at (3.5,3.25) {$t_3 = 4$};
\node[] at (4.5,0.25) {$t_4 = 1$};
\node[] at (5.5,1.25) {$t_5 = 2$};

\end{tikzpicture}
\caption{The used model for a swap ASAP repeater chain. In the top figure a) we show the evolution of the present entanglement in the network. At time $t=1$ the fourth link gets created, at time $t=2$ both link 2 and 5 are generated. Since we are considering the swap ASAP model, link 4 and 5 are swapped at this time. At time $t=3$ link 1 is created and swapped with link 2. Finally, at time $t=4$ the third link is created, which is immediately swapped on both sides, leading to end-to-end entanglement. In b) we summarize the total evolution of a single run. The decoherence experienced corresponds to the waiting time between the different links, indicated by dashed lines. The dotted lines correspond to having to wait until all links have been created, which are not relevant when performing tasks such as QKD.}
\label{fig:model}
\end{figure}

As soon as a repeater holds two entangled pairs a swap is performed, which we assume to be deterministic. Fig.~\ref{fig:model} illustrates a swap ASAP repeater protocol, where $t_1 = 3$, $t_2=2$, $t_3=4$, $t_4=1$ and $t_5=2$.

The decoherence experienced in a given run will be determined by the number of rounds each node has to wait for until it can swap, along with the quality of the memory. 
We model noise in the system as depolarizing noise. That is, our states are of the form $\Lambda\ket{\psi}\hspace{-1mm}\bra{\psi} + (1-\Lambda)\frac{\mathbb{I}}{4}$, where $\sqrt{2}\ket{\psi} = \ket{00}+\ket{11}$ and $\Lambda \in [0, 1]$. From $\Lambda$, we calculate the fidelity to be $\Lambda + \frac{1-\Lambda}{4} = \frac{3}{4}\Lambda+\frac{1}{4}$. The usage of the parameter $\Lambda$ is convenient --- swapping two states with parameters $\Lambda'$ and $\Lambda''$ gives a new state with parameter $\Lambda'\cdot\Lambda''$.
Furthermore, decoherence over a single time step corresponds to the map $\Lambda \mapsto \lambda \Lambda $ for some $\lambda\in \left[0, 1\right]$. Here and in what follows, we focus on $\Lambda$, since it completely characterizes the quantum state, such that other quality indicators (such as fidelity) can be extracted from it.

Let us consider a concrete example. In Fig.~\ref{fig:model} the decoherence experienced by the memory in the repeater holding links $2$ and $3$ equals $\lambda^{\left|2-4\right|}=\lambda^2$, where $\lambda$ is the noise parameter associated with a single round of decoherence.

Importantly, we ignore the noise that the end nodes incur if they had to keep their states stored until all other links have succeeded. This can be justified exactly when QKD is performed (in which case the states at the end nodes can be measured directly), or when the end nodes are assumed to have significantly better equipment or error correction. See~\cite{kamin2023exact} for a treatment that includes this noise.

With knowledge of which swaps are performed and how long certain memories decohere, it is possible to calculate the noise parameter $\Lambda$ describing the final state. We show in the next section how to calculate the expected value of the final noise parameter. Finally, from linearity of expectation, the average fidelity is $\mathds{E}\left[F\right] = \frac{3}{4}\mathds{E}\left[\Lambda\right]+ \frac{1}{4}$. 

We note that this model also includes noisy state generation \kg{and noisy swaps/measurements}. Namely, if $\Lambda$ is the end-to-end noise parameter for the case of perfect state generation, then the average noise parameter is given by $\Lambda\cdot \left(\prod_{i=1}^{n}\lambda_{\textrm{gen}, i}\right)\left(\prod_{i=1}^{n-1}\lambda_{\textrm{swap},i}\right)$, where $\lbrace{\lambda_{\textrm{gen}, i}\rbrace}$ are the noise parameters corresponding to the $n$ noisy links\kg{, and  $\lbrace{\lambda_{\textrm{swap}, i}\rbrace}$ the noise parameters for the swaps/measurements}. This results in a constant multiplicative factor to the $\Lambda$ parameter. In what follows however, we will assume for simplicity that \kg{the only source of noise is decoherence, unless explicitly mentioned}.

\subsection{Extension to general noise models}
Above, we chose the single-qubit depolarizing channel as the noise model. The properties that make this noise model amenable to analysis are the facts that decoherence corresponds to multiplication of the $\lambda$ parameter, and that swapping two states corresponds to multiplication of the $\lambda$ parameters.
We derive in Appendix \ref{sec:general_noise} that if the  state is a two-qubit maximally-entangled state, then \emph{any} qubit Pauli channel $\mathcal{N}$ has these properties.
Consequently, the analysis in this work straightforwardly extends to any qubit Pauli noise model.
The parameter update is achieved by a change of variables of $\mathcal{N}$: we start out with the usual parameterization of a Pauli channel,
\[
\mathcal{N}(\rho) = p_I  \cdot \rho + p_X  \cdot X \rho X^{\dagger} + p_Y  \cdot Y \rho Y^{\dagger} + p_Z  \cdot Z\rho Z^{\dagger}\ ,
\]
where $\{p_I, p_X, p_Y, p_Z\}$ 
is a probability distribution over the Pauli matrices $I, X, Y, Z$ and $\rho$ a single-qubit density matrix.
We now re-parameterize $\mathcal{N} = \mathcal{N}_{\vec{\lambda}}$ using four new variables $\vec{\lambda} = (\lambda_1, \lambda_2, \lambda_3, \lambda_4)$, where
\begin{align*}
\lambda_1 &= p_I + p_X + p_Y + p_Z = 1,\\
\lambda_2 &= p_I + p_X - p_Y - p_Z\ ,\\
\lambda_3 &= p_I - p_X + p_Y - p_Z\ ,\\
\lambda_4 &= p_I - p_X - p_Y + p_Z\ .
\end{align*}
To verify that this change of variables satisfies the above criteria, it is straightforward to calculate that first applying $\mathcal{N}_{\vec{\lambda}}$, followed by applying $\mathcal{N}_{\vec{\lambda}'}$, is the same as applying $\mathcal{N}_{\left(\lambda_1 \cdot \lambda_1', \lambda_2 \cdot \lambda_2', \lambda_3 \cdot \lambda_3', \lambda_4 \cdot \lambda_4'\right)}$. As a simple example, consider dephasing noise (which was also considered in~\cite{kamin2023exact}), where

\begin{align}
\mathcal{N}\left(\cdot \right)  &=p\left(\cdot \right)+ \left(1-p\right) Z\left(\cdot \right)Z^\dagger\\
&=\left(\frac{1+\lambda_{2}}{2}\right)\left(\cdot \right) +  \left(\frac{1-\lambda_{2}}{2}\right) Z\left(\cdot \right)Z^\dagger\ \\
&= \mathcal{N}_{\left(1, 1-2p, 1-2p, 1\right)} \left(\cdot \right)\ .
\end{align}

Multiplicativity under swapping is also satisfied for the general qubit Pauli channel. Denote by $\rho(\vec{\lambda})$ the resultant state after $\mathcal{N}_{\vec{\lambda}}$ has acted on one of the qubits of a maximally entangled state. An entanglement swap on the states $\rho(\vec{\lambda})$ and $\rho(\vec{\lambda}')$ then yields the state $\rho(\lambda_1 \cdot \lambda_1', \lambda_2 \cdot \lambda_2', \lambda_3 \cdot \lambda_3', \lambda_4 \cdot \lambda_4')$ after correction, independently of the required correction. Since all the $\lambda_i$ terms are multiplied point-wise, all of the analyses in this paper for the $\lambda$ parameter for depolarizing noise can be directly translated to the individual $\lambda_i$ parameters.
In Appendix \ref{sec:general_noise}, we also examine the case where the link between nodes is given by a maximally-entangled state on two \emph{qudits}, i.e.~the more general case where the local Hilbert space has dimension $\geq 2$.
Specifically, we show that the above multiplicativity properties hold for $X$-symmetric channels.
These are channels corresponding to a probabilistic application of $d$-dimensional qudit Pauli operators $X^a Z^b$ (with $a, b\in \{0, 1, \dots, d-1\}$), where the probability for $X^a Z^b$ is equal to the probability for $X^{-a} Z^b$.
The qubit case follows since any qubit Pauli channel is $X$-symmetric. Our analysis thus is not only applicable to dephasing (also considered in~\cite{kamin2023exact}), but also to (biased) depolarizing noise and relevant noise models for GKP qudits, see~\cite{noh2020fault, albert2018performance, shaw2024logical, noh2018quantum}. Our derivation is based on interpreting the composition of qudit Pauli channels as a type of convolution, which allows us to use tools from Fourier theory on finite Abelian groups.

\subsection{The global cut-off policy}
We now describe the cut-off policy considered in this paper, which we refer to as the \emph{global} cut-off policy. This policy does not require any of the nodes to communicate information to \kg{non-neighboring nodes}, making it particularly appealing for near-term quantum networks. We depict the global cut-off policy visually in Fig.~\ref{fig:global_cutoff}. For completeness, we discuss in Appendix~\ref{sec:policies} a number of other policies, but which are outside the scope of this paper.

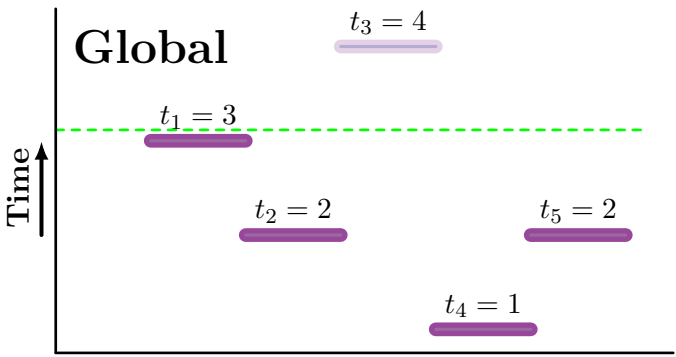
\begin{figure}[h!]
    \centerfloat
\begin{tikzpicture}[scale=1.25, line cap=round, entanglement/.style={shade, shading=ring, line width = 0.7mm, Purple,  double=RedViolet!40!Periwinkle!90,double distance=1.2pt}, repeater/.style={diamond, NavyBlue}, noentanglement/.style={shade, shading=ring, line width = 0.7mm, Purple!20,  double=RedViolet!10!Periwinkle!50,double distance=1.2pt, repeater/.style={diamond, NavyBlue}}]
\draw[line width=1pt, dashed, green] (0.0,2.115) -- (6.2,2.115);

\draw[line width=1pt] (-0.0,-0.25) -- (-0.0,3.4);

\node[scale=1.5] at (1,3) {\large\textbf{Global}};

\draw[line width=1pt] (0.0,-0.25) -- (6.5,-0.25);
\node[rotate=90] at (-0.4,1.5) {\large\textbf{Time}};
  \draw [-{Latex[length=2.5mm]}, line width = 1.5pt] (-0.15,1)   -- (-0.15,2);






\draw[line width=5pt, entanglement] (1, 2) -- (2,2) node[above] {};
\draw[line width=5pt, entanglement] (2, 1) -- (3,1) node[right] {};
\draw[line width=5pt, noentanglement] (3, 3) -- (4,3) node[right] {};
\draw[line width=5pt, entanglement] (4, 0) -- (5,0) node[right] {};
\draw[line width=5pt, entanglement] (5, 1) -- (6,1) node[right] {};

\node[] at (1.5,2.25) {$t_1 = 3$};
\node[] at (2.5,1.25) {$t_2 = 2$};
\node[] at (3.5,3.25) {$t_3 = 4$};
\node[] at (4.5,0.25) {$t_4 = 1$};
\node[] at (5.5,1.25) {$t_5 = 2$};

\node[diamond, scale=6] at (5.2,1.25) {};

\end{tikzpicture}
    \caption{Graphical description of the global cut-off policy. Each of the nodes has agreed on a parameter $T_c$. For the global cut-off, a reset is performed when after $T_c$ rounds end-to-end entanglement has not been established.}
    \label{fig:global_cutoff}
\end{figure}
The global cut-off policy imposes a constraint on how long it takes to establish end-to-end entanglement. That is, starting from a chain with no entanglement present, a timer is initialized to zero and increases by one after every elementary link generation attempt. Then the ordinary swap ASAP protocol is followed. Two events can occur. If end-to-end entanglement is established before the timer reaches $T_c$, the timer is reset and the swap ASAP protocol is performed again. If, however, the timer reaches $T_c$ (i.e.~the cut-off is reached), the present entanglement is thrown away and the swap ASAP protocol starts anew.

\kg{We assume for simplicity that classical communication is instantaneous, such that nodes know when to start entanglement generation again. We note that in general the effect of classical communication delays cannot be disregarded, especially in more complex distribution schemes~\cite{haldar2024reducing, li2024optimising}. In Appendix~\ref{sec:exp_delivery_time} we additionally calculate the expected generation time in the case where all links only start generating again when the cut-off timer has been reached (but will not further consider this case in our results). This is especially appealing since it minimizes the classical communication between the nodes.}

Fig.~\ref{fig:global_cutoff} shows an example of the global cut-off policy. In this case, the third link did not succeed before the timer $T_c = 3$, and the entanglement is discarded.

\section{Recursion relation for the swap ASAP repeater model}\label{sec:recursion}
With the considered model of the swap ASAP repeater protocol in hand, we will analyze the average end-to-end fidelity. \kg{We show that the average end-to-end fidelity can be interpreted as a \emph{probability generating function}, which was shown before in~\cite{kamin2023exact}. We find a recursive formulation of the probability generating function, which allows us to go beyond the expressions for swap ASAP chains of length $n=8$ from~\cite{kamin2023exact}. Furthermore, this recursive relation is key for the derivation of the \emph{generating function} of the next section, which is strictly more powerful than the \emph{probability} generating function discussed in this section and in~\cite{kamin2023exact}}.

Subsection~\ref{sec:analytics_no_cutoff} considers the no cut-off policy. Subsection~\ref{sec:analytics_global_cutoff} describes how this approach can be extended to a global cut-off. We further discuss why a similar approach fails for the local cut-off policy (see Appendix~\ref{sec:policies}), and how to calculate the higher-order moments of $\Lambda$ (which was also done before in~\cite{kamin2023exact}). In Subsection~\ref{sec:first_results} we use our results to explore the performance of swap ASAP repeater chains.

\subsection{Analytics without a cut-off policy}\label{sec:analytics_no_cutoff}
First, we need to calculate the probability $P\left(T=\overline{t}\right)$ of observing a specific \emph{instance} $\overline{t}$. Here an instance corresponds to a sequence $\overline{t} = \left(t_1, t_2, \ldots, t_n\right)$,
where $t_i\in \lbrace{1, 2,\ldots,\rbrace}$ indicate the round in which link $i$ succeeds, see Section \ref{sec:model}. A single link $i$ succeeds in round $t_i$ with probability $p\left(1-p\right)^{t_i-1}= \left(1-q\right)q^{t_i-1}$, where $q\equiv 1- p$. An instance $\overline{t} = \left(t_1, t_2, \ldots, t_n\right)$ for $n$ links occurs with probability

\begin{align}
  P\left(T=\overline{t}\right)=&\prod_{i=1}^{n}\left(1-q\right)q^{t_i-1} \\
  = & \left(\frac{1-q}{q}\right)^n q^{\sum_{i=1}^n t_i} \ ,
\end{align}

for $\overline{t}\in\lbrace{1, 2, \ldots, \rbrace}^{n}$.
Let us now consider the decoherence for a given instance $\overline{t}$. Each node that is not an end node has two neighboring links, which succeed at times $t_i$ and $t_{i+1}$. With the swap ASAP protocol, the swap occurs when the latest of the two links got created, i.e.~at time $\max(t_i, t_{i+1})$. The link created first was created at time $\min(t_i, t_{i+1})$. Thus, the corresponding node waits $\max(t_i, t_{i+1}) - \min(t_i, t_{i+1}) = \left|t_i-t_{i+1}\right|$ rounds, during which the associated memory experiences decoherence.

Since a memory experiences a decay of $\lambda$ for each time step, every node that is not an end node adds a multiplicative factor of $\lambda^{\left|t_{i}-t_{i+1}\right|}$, leading to a total decay parameter of $\Lambda = \lambda^{\sum_{i}^{n-1}\left|t_i-t_{i+1}\right|} = \lambda^{K(\overline{t})}$, where $K(\overline{t}) \equiv \sum_{i}^{n-1}\left|t_i-t_{i+1}\right|$ is the aggregate waiting time. We will also refer to $K$ as the \emph{roughness} parameter of $\overline{t}$.

The end nodes have to keep their state stored until end-to-end entanglement has been established. By definition, end-to-end entanglement is established at $\max(\overline{t})$, so that the first and last node store entanglement for $\max(\overline{t})-t_1$ and $\max(\overline{t})-t_n$ rounds, respectively. This leads to an additional total decoherence parameter of $\lambda^{2\max(\overline{t}) - t_1-t_n}$. The average noise parameter $\mathbb{E}\left[\lambda_\textrm{end}\right]$ can then be written as

\begin{gather}
\left(\frac{1-q}{q}\right)^n \sum_{\overline{t}}\left(q^{\sum_{i}^nt_i}\right)\cdot\left(\lambda^{\left(2\max(\overline{t})-t_1-t_n\right)+\sum_{i=1}^{n-1}\left|t_i-t_{i+1}\right|}\right) \nonumber\ .
\end{gather}

As mentioned in the previous section, we now drop the $\lambda^{2\max(\overline{t}) - t_1-t_n}$ factor. 

We are interested in calculating the average noise parameter for $n$ segments, i.e.~

\begin{align}
\mathbb{E}\left[\Lambda_n\right]=&\left(\frac{1-q}{q}\right)^n \sum_{\overline{t}}\left(q^{\sum_{i=1}^nt_i}\right)\cdot\left(\lambda^{\sum_{i=1}^{n-1}\left|t_i-t_{i+1}\right|}\right)\nonumber \\
=&~\sum_{k=0}^{\infty}P\left(K_n=k\right)\lambda^{k}\nonumber\\
=&~H_{K_n}\left(\lambda\right)\nonumber\\
\equiv&~H_{n}\ \label{eq:probgen},
\end{align}
where $H_{X}\left(z\right)$ is shorthand for the \emph{probability generating function} of the random variable $X$ and the subscript indicates the associated
number of segments $n$.
It follows that $\mathds{E}\left[\Lambda_n\right]$ is the \emph{probability generating function} of the roughness parameter $K_n=\sum_{i=1}^{n-1}\left|t_i-t_{i+1}\right|$, first observed in~\cite{kamin2023exact}. This interpretation is useful for two reasons. First, any method to calculate $\mathds{E}\left[\Lambda_n\right]$ (or $\mathds{E}\left[\Lambda_n\left(T_c\right)\right]$), allows for the determination of any of the higher-order moments $\mathds{E}\left[\left(\Lambda_n\right)^m\right]$.
This is done by repeating such a method but with $\lambda$ replaced by $\lambda^m$, since

\begin{align}
\mathbb{E}\left[\left(\Lambda_n\right)^m\right]
=&~\sum_{k=0}^{\infty}P\left(K_n=k\right)\left(\lambda^{k}\right)^m\nonumber\\
=&~\sum_{k=0}^{\infty}P\left(K_n=k\right)\left(\lambda^{m}\right)^k\nonumber\nonumber \ ,
\end{align}
noted first in~\cite{kamin2023exact}.
Second, it is well known that a probability mass function can be reconstructed from the derivatives of its associated probability generating function; we exploit this idea in Section~\ref{sec:distribution}.
Note that the interpretation of the average noise as the probability generating function of the aggregate waiting time is not specific to swap ASAP repeater chains. This was first noted in~\cite{kamin2023exact}, where they investigated several different methods of entanglement distributions for repeater chains. It would be of interest to see similar tools exploited for further variants of entanglement distribution.

To derive an analytical expression of $\mathds{E}\left[\Lambda_n\right] = H_n$, it is convenient to define

\begin{gather}
H_n^{t} = \left(\frac{1-q}{q}\right)^n\sum_{\mathclap{\substack{\overline{t}~\textrm{s.t.}\\t_n =t}}}\left(q^{\sum_{i=1}^nt_i}\right)\cdot\left( \lambda^{\sum_{i=1}^{n-1}\left|t_i-t_{i+1}\right|}\right)\ .
\end{gather}
This is convenient since

\begin{gather}
H_n^{t} = \frac{1-q}{q}\sum_{t'=1}^{\infty}q^t\lambda^{\left|t-t'\right|}H_{n-1}^{t'}\label{eq:recursion}\ .
\end{gather}

That is, we can write $H_n^{t}$ in terms of $H_{n-1}^{t}$, which in turn can be written in terms of $H_{n-2}^{t}$, etc. The approach is as follows: start with $H_1^t=\left(\frac{1-q}{q}\right)q^t$ (since there is only one link, there is only the $q$ term), and then calculate $H_n^{t}$ recursively using \eqref{eq:recursion}. Finally, we use that

\begin{gather}
H_n = \sum_{t=1}^{\infty}H_n^{t}\label{eq:recstep1}\ ,
\end{gather}
to calculate the expectation value of the noise parameter $\Lambda_n$ for $n$ segments.

The recursion takes a particular simple form; we show in Appendix \ref{sec:app_recursion} that for $n\geq 1$, $H_n^{t}$ is always a linear combination of terms of the form $q^{at }\lambda^{b t}$, where $a, b \in\mathbb{Z}$. We can thus think of $H_n^{t}$ as an element of the vector space $V$ consisting of all formal (real) linear combinations of elements of the form $q^{a t }\lambda^{b t}$. Furthermore, we show that the recursion taking $H_{n-1}^t$ to $H_{n}^t$ is a linear map $M$ from $V$ to itself. This allows for a fast determination of $H_n^t$. Finally, the map \eqref{eq:recstep1} taking $H_n^t$ to $H_n$ is a linear form, which can also be conveniently calculated.

\kg{We note that the above procedure also works for inhomogeneous repeater chains, i.e.~chains with varying entanglement generation success probabilities (due to for example non-equidistant nodes), and varying coherence parameters of the memories. This can be done by allowing the map $M_i$ to vary at each step of the recursion, which we will discuss in more detail in a follow-up work.}

While we have explicit analytical expressions for $\mathds{E}\left[\Lambda_n\right]$ up to $n=25$ segments, the expressions themselves are rather long. Instead of reporting them here, we provide all of our analytical expressions in Mathematica files~\cite{mathematicafiles}.

\subsubsection{Analytics for the global cut-off policy}\label{sec:analytics_global_cutoff}
Let $\mathds{E}\left[\Lambda_n\left(T_c\right)\right]$ denote the average of $\Lambda_n$ with a global cut-off parameter of $T_c$. To calculate $\mathds{E}\left[\Lambda_n\left(T_c\right)\right]$, we sum over all instances that satisfy the global cut-off constraint, divided by the probability that entanglement was generated before the cut-off $T_c$ was reached. That is, we calculate 

\begin{align}&\mathds{E}\left[\Lambda_n\left(T_c\right)\right]\nonumber \\
=&\left(\frac{1-q}{q\left(1-q^{T_c}\right)}\right)^n\sum_{\mathclap{\substack{\overline{t}~\textrm{s.t.}\\\max(\overline{t}) \leq T_c }}}\left(q^{\sum_{i}^nt_i}\right)\cdot\left( \lambda^{\sum_{i}^{n-1}\left|t_i-t_{i+1}\right|}\right) \nonumber\\
\equiv&~\bar{H}_n\ \nonumber.
\end{align}
We define $\bar{H}_{n}^{t}$ similar to the no cut-off case, i.e.~
\begin{gather}
\bar{H}_n^{t} = \left(\frac{1-q}{q\left(1-q^{T_c}\right)}\right)^n\sum_{\mathclap{\substack{\overline{t}~\textrm{s.t.}\\t_n =t \\t_i\leq T_c}}}\left(q^{\sum_{i}^nt_i}\right)\cdot\left( \lambda^{\sum_{i}^{n-1}\left|t_i-t_{i+1}\right|}\right)\ \nonumber .
\end{gather}

Now, \eqref{eq:recursion} and~\eqref{eq:recstep1} generalize straightforwardly,

\begin{gather}
\bar{H}_n^{t} = \left(\frac{1-q}{q\left(1-q^{T_c}\right)}\right)\sum_{t'=1}^{T_c}q^t\lambda^{\left|t-t'\right|}\bar{H}_{n-1}^{t'}\ \label{eq:recglobal},\\
\bar{H}_n = \sum_{t=1}^{T_c}\bar{H}_n^{t}\ . 
\end{gather}

In Appendix~\ref{sec:app_recursion} we show that the desired properties from the case without a cut-off are preserved. That is, the maps $\bar{H}_n^t\mapsto \bar{H}_{n+1}^{t}$ and $\bar{H}_n^t\mapsto \bar{H}_n$ are best thought of as a linear map and as a linear form, allowing for the analytical determination of $\mathds{E}\left[\Lambda_{n}\left(T_c\right)\right]$.

For the derivations in the appendices we drop the $\left(\frac{1-q}{q}\right)^n$ term, i.e.~we define the quantity
\begin{align}
Z_n = \sum_{\mathclap{\substack{\overline{t}}}}\left(q^{\sum_{i}^nt_i}\right)\cdot\left( \lambda^{\sum_{i}^{n-1}\left|t_i-t_{i+1}\right|}\right)\ \label{eq:partition_function}.
\end{align}
Not only does this simplify some of the calculations, but the $Z_n$ can also be interpreted as the \emph{partition function} of the so-called solid-on-solid model~\cite{owczarek1993exact, rozycki2003rsos, owczarek2009exact, owczarek2010exact}. In follow-up work we exploit this connection, bringing to bear powerful techniques from statistical physics to numerically calculate noise during (multipartite) entanglement distribution.

There are also other cut-off policies one could consider besides the global cut-off policy; we list several of them in Appendix~\ref{sec:policies}. A natural policy is the \emph{local} cut-off policy, where one limits the time a single qubit is allowed to decohere. One is tempted to think that a similar recursive formulation is possible for the local cut-off policy as well. However, such a formulation requires that the sum in~\eqref{eq:recglobal} starts at $\max(1, t-T_c)$, greatly complicating its calculation.

\subsection{Parameter exploration}\label{sec:first_results}
The secret-key rate is the rate at which one can generate secret bits between two parties. It is a convenient metric with a clear operational interpretation, which relies on both the (average) quality of the state and the (average) delivery time. In this section, we explore the effect that the cut-off has on the secret-key rate for several different parameter regimes. Appendix~\ref{sec:exp_delivery_time} contains details for the calculations of the secret-key rate.
We note that for up to 8 repeater segments, the work from~\cite{kamin2023exact} includes an extensive numerical analysis of the secret-key rate not only for swap ASAP protocols, but several additional policies as well.

\begin{figure}[h]
    \centering
    \includegraphics[width=0.48\textwidth, trim={1.75cm 0 0 0},clip]{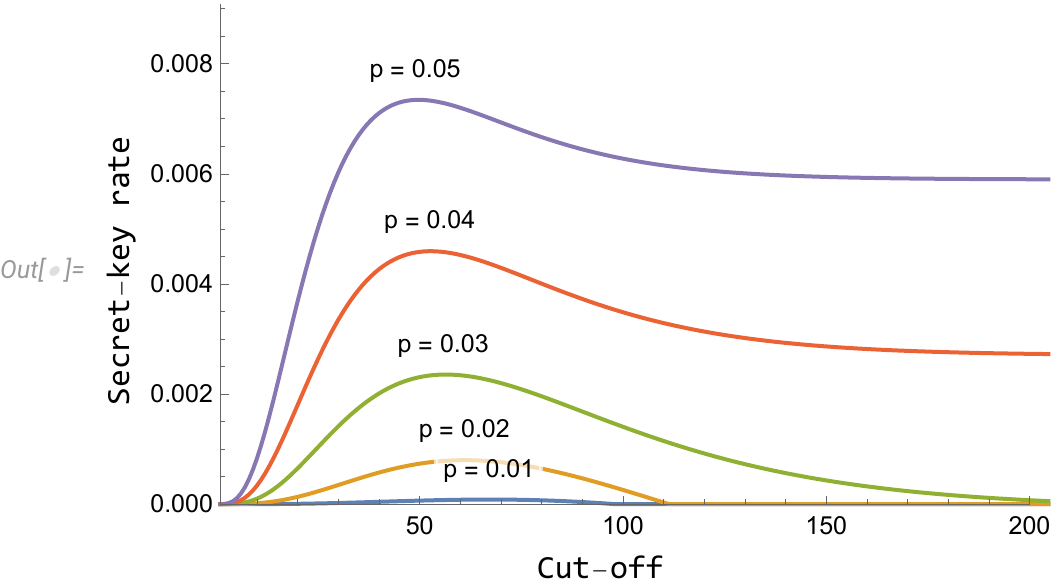}
    \caption{Secret-key rate with a global cut-off policy as a function of the cut-off parameter. Parameters used are $n = 5$ and $\lambda= 0.998$.}
    \label{fig:skrvscutoff}
\end{figure}

We start by fixing $n=5$ segments and a decoherence parameter of $\lambda=0.998$. In Fig.~\ref{fig:skrvscutoff} we vary the success probability $p$ and the cut-off. We find that for success probabilities of $p=0.04-0.05$, the cut-off brings a modest increase (where the case of no cut-off corresponds to the cut-off going to infinity). For success probabilities smaller than $p=0.03$ we find that a cut-off is necessary to be able to generate any amount of secret key~\cite{rozpkedek2018parameter, rozpkedek2019near}. Observe that optimizing the cut-off is vital for maximizing secret-key rate --- something that our methods now enable.

\begin{figure}[h!]
    \centering
    \includegraphics[width=0.48\textwidth, trim={1.75cm 0 0 0},clip]{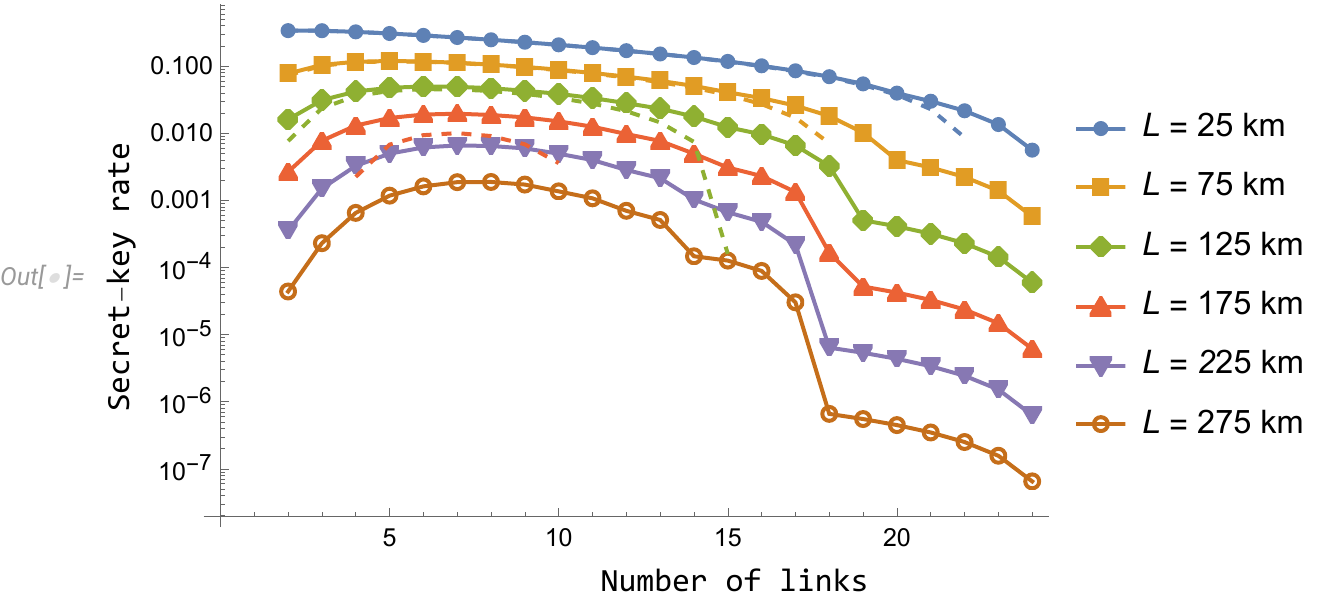}
    \caption{Secret-key rate optimized over the global cut-off parameter for several distances $L$, for $\lambda = 0.995$ and $\lambda_\textrm{gen} = 0.999$. The dashed lines correspond to the secret-key rate without any cut-off. The abrupt behavior corresponds to the optimal cut-off changing.}
    \label{fig:bigplot}
\end{figure}

Fig.~\ref{fig:bigplot} illustrates how such an optimization can be used for parameter exploration. There, we show the secret-key rate optimized over the global cut-off parameter as a function of the number of segments for several total distances. The parameters used are $\lambda = 0.995$ and $\lambda_\textrm{gen} = 0.999$, where the last parameter corresponds to imperfect state generation.

\section{Generating function approach}\label{sec:gen_function}
The methods from the previous section provide a recipe to calculate an analytical form of the average noise. However, the expressions grow unwieldy very fast, seemingly without any structure to them~\cite{kamin2023exact}. Thus, while the recursive calculation allows for an exact, quantitative understanding of the noise in swap ASAP repeater chains, it does not directly yield a qualitative understanding.

In this section we present a method for overcoming this, based on generating functions (not to be confused with the probability generating function \kg{from the previous section}). This approach also yields a faster way to compute the analytical formulae found with previous methods. We first introduce generating functions, and how they the understanding of noise in repeater chains. Afterwards we use this approach to develop tight approximations and investigate the performance of repeater chains.

\subsection{Outline of the approach}
Generating functions are the main object of study in analytic combinatorics~\cite{flajolet2009analytic,  pemantle2013analytic}, \kg{and can be used to extrapolate important information about sequences $\left(a_n\right)$. This is done by associating a complex function $G(x)$ to the sequence $\left(a_n\right)$ in the following manner,}

\begin{align}
G(x)= \sum_{n=0}^\infty a_n x^n \label{eq:general_gen_function}\ ,
\end{align}
\kg{where $a_n = \mathds{E}\left[\Lambda_n\right]$ in our case of interest.} For general sequences $\left(a_n\right)$, one first find a closed-form expression for the associated generating function $G$. Given such a closed-form expression, it is possible to extract \textit{complex analytic} behavior about $G$. In turn, this complex analytic behavior of $G$ dictates the asymptotic behavior of $ \mathds{E}\left[\Lambda_n\right]$.

\kg{Let us make this more concrete. Let $G$ be the generating function associated with a sequence $\left(a_n\right)$, where $a_n>0$. Let $\rho$ be the singularity that is both real and closest to the origin (see Pringsheim's theorem~\cite{flajolet2009analytic, remmert1991theory}). Furthermore, assume that $\rho$ is a simple pole of $G$, which is always the case in this manuscript. Then }

\begin{align}
\lim_{n\rightarrow\infty}\frac{-\textrm{Res}_{\rho}\left(G\right)\left(\frac{1}{\rho}\right)^{n+1}}{a_n} = 1 \label{eq:asymptotic_comparison}\ .
\end{align}
\kg{In other words, $a_n$ is asymptotically equivalent to $-\textrm{Res}_{\rho}\left(G\right)\left(\frac{1}{\rho}\right)^{n+1}$, where $\textrm{Res}_{\rho}\left(G\right) =\lim_{x\rightarrow \rho}\left(x-\rho\right)G(x)$ is the residue of $G$ at $\rho$. This last equality holds since $\rho$ is a simple pole~\cite{lang2013complex}. These claims follow from standard techniques in analytic combinatorics, in particular Theorem IV.10 from~\cite{flajolet2009analytic}. More specifically, this follows from expanding $f(z)$ in Theorem IV.10 around $z=\rho$ for the case of a simple pole, see Example IV.7.~in~\cite{flajolet2009analytic} for an explicit demonstration.}
\kg{In Appendix~\ref{sec:fib_gen} we give an example of this statement when the sequence $\left(a_n\right)$ is given by the Fibonacci numbers. We show there how the generating function approach straightforwardly yields both asymptotic approximations and even exact expressions for the Fibonacci numbers.}

\kg{Let us now return to the generating function associated to swap ASAP repeater chains, i.e.~$G=\sum_{n=1}^{\infty}\mathds{E}\left[\Lambda_n\right]x^n$.} At this point, it is not clear whether a nice closed-form expression for $G$ should exist, especially since we have argued that the individual coefficients $\mathds{E}\left[\Lambda_n\right]$ become increasingly unwieldy as $n$ grows. This is, however, a powerful feature of generating functions --- they often behave much better than their coefficients considered individually~\cite{flajolet2009analytic, pemantle2013analytic}. This good behavior is not only what allows us to extract information about the individual coefficients, but also allows us to find $G$ in the first place.

Using the recursive formulation from the previous section, we show in Appendix~\ref{sec:appendix_gen_function} that 
\begin{align}
G =  x\cdot \Phi_1 +x^2\left(\frac{\left(1-q\right)^2\lambda}{\left(\lambda-q\right)\left(1-\lambda q\right)}\right) \frac{\Phi_2\cdot \Phi_4}{1-x\cdot \Phi_3}\ \nonumber,
\end{align}
where $\Phi_i = \Phi_i(x) = \Phi_i(x, \lambda, q),~i=1, 2, 3,4$ is a particular instance of a so-called $q$-hypergeometric series~\cite{gasper2004basic}, and is a function of $x$, $\lambda$ and $q$ (see \eqref{eq:final_gen_func} to \eqref{eq:final_gen_func4} for a full description). Since $\Phi_i$ is an entire function in $x$ for fixed $\lambda, q$ (with $\lambda \neq q$), the only singularity arises when $1-x\cdot\Phi_3 = 0$. Although characterizing analytical solutions to equations involving ($q$-)hypergeometric series is generally a difficult endeavor~\cite{driver2008zeros, dominici2013real}, it is straightforward to find solutions numerically. Let $x=\rho$ be the smallest real solution to $1-x\cdot \Phi_3 = 0$. \kg{Using the statement from Eq.~\eqref{eq:asymptotic_comparison}, we find that }
\begin{gather}
\mathds{E}\left[\Lambda_n\right] \sim  \left[-\textrm{Res}_\rho\left(G\right)\right]\cdot \left(\frac{1}{\rho}\right)^{n+1}\nonumber\\
= \left[\left(\frac{\left(1-q\right)^2\lambda \rho}{\left(\lambda-q\right)\left(1-\lambda q\right)}\right)\cdot \left( \frac{\Phi_2(\rho)\cdot \Phi_4(\rho)}{\Phi_3(\rho)+\rho \cdot \Phi_3'(\rho)}\right)\right]\cdot  \left(\frac{1}{\rho}\right)^n\nonumber\\
\equiv \left[A(\lambda, q)\right]\cdot B(\lambda, q)^n\ \label{eq:approximation} ,
\end{gather}
where $\sim$ means that the ratio between the two sides converges to $1$ as $n\rightarrow \infty$. \kg{The first equality follows from $\textrm{Res}_\rho\left(f/g\right) = \lim_{x\rightarrow \rho }\left(x-\rho\right)\frac{f(x)}{g(x)} = \frac{f(\rho)}{g'(\rho)}$, which holds because of L'H\^{o}pital's rule and the fact that $\rho$ is a simple zero of $g$, i.e.~$g(\rho) = 0$, while $g'(\rho) \neq 0$.}

The approximation in \eqref{eq:approximation} has a clear conceptual interpretation --- the quality of the state decays exponentially in the number of segments $n$, where the decay rate is given by $B(\lambda, q)$ for each added segment. The $A(\lambda, q)$ term, on the other hand, can be thought of as a correction term. Note that both terms are independent of $n$. In the following subsection we numerically show that this approximation is not only tight in the asymptotic limit; it is already accurate enough for practical purposes for small $n$.

It is possible to approximate $\mathds{E}\left[\Lambda_n\right]$ by including more than just the singularity closest to the origin, leading to a sum of exponential terms~\cite{flajolet2009analytic}. In practice, we find that taking the first singularity suffices.

All of the above generalizes straightforwardly to the case of a global cut-off, where there exist analogous functions $\overline{A}(\lambda, q, T_c)$ and $\overline{B}(\lambda, q, T_c)$ as well --- see Appendices~\ref{sec:appendix_gen_function} and~\ref{sec:appendix_cutoff_approx} for the derivation of the associated generating function.

We conclude this subsection with two remarks. First, with the generating function $G$ in hand, it is easier to find analytical expressions for $\mathds{E}\left[\Lambda_n\right]$ than with the recursive relation from Section~\ref{sec:recursion}. This allows us to analytically calculate $\mathds{E}\left[\Lambda_{25}\right]$ in a matter of seconds with Mathematica --- something that was impossible with the recursive approach~\cite{mathematicafiles}. Second, we note that the function $G$ is a \emph{multivariate} generating function~\cite{pemantle2013analytic}, where the associated combinatorial object can be interpreted as a type of lattice path $L$. The associated parameters $q$, $\lambda$ and $x\left(\frac{1-q}{q}\right)$ count the area $A(L)$ under the path, the roughness $K(L)$ and the width $n(L)$ of the path, respectively. This result may be of independent interest to combinatorialists.

\subsection{Parameter exploration}\label{sec:gen_func_exploration}
In this subsection we first explore the approximation found in \eqref{eq:approximation}, before using it to understand how $\lambda$ and $q$ affect $\mathds{E}\left[\Lambda_n\right]$ qualitatively.

\begin{figure}
    \centering
    \includegraphics[width=0.45\textwidth, trim={0cm 0 0 0},clip]{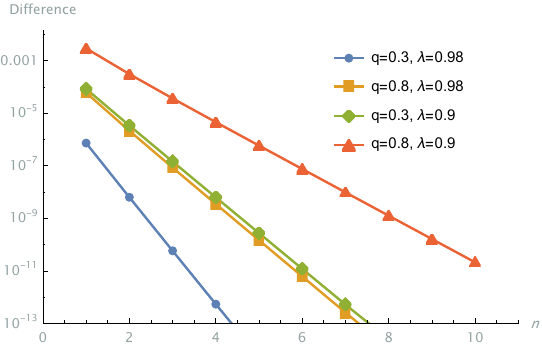}
    \caption{Difference between the approximation of Eq.~\eqref{eq:approximation} and the exact average noise parameter $\mathds{E}\left[\Lambda_n\right]$ as a function of $n$, for four different sets of parameters.}
    \label{fig:convergence}
\end{figure}

Fig.~\ref{fig:convergence} shows the difference between the exact average noise and \eqref{eq:approximation} for four sets of values of $\lambda$ and $q$. We find that \eqref{eq:approximation} provides an approximation that converges exponentially fast, independent of the parameters used.

From \eqref{eq:approximation} and Fig.~\ref{fig:convergence} we find that $\mathds{E}\left[\Lambda_n\right]$ decays exponentially in the number of segments. In particular, $B\left(\lambda, q\right)$ can be thought of as a decay rate --- every additional segment added reduces the average $\Lambda$ parameter by a factor of $B\left(\lambda, q\right)$. The decay rate $B\left(\lambda, q\right)$ thus allows us to assign a metric to a given repeater chain for a fixed $\lambda$ and $q$, \emph{independent of the number of segments}.

\begin{figure}
    \centering \includegraphics[width=0.45\textwidth, trim={2.95cm 0 0.2cm 2.4cm},clip]{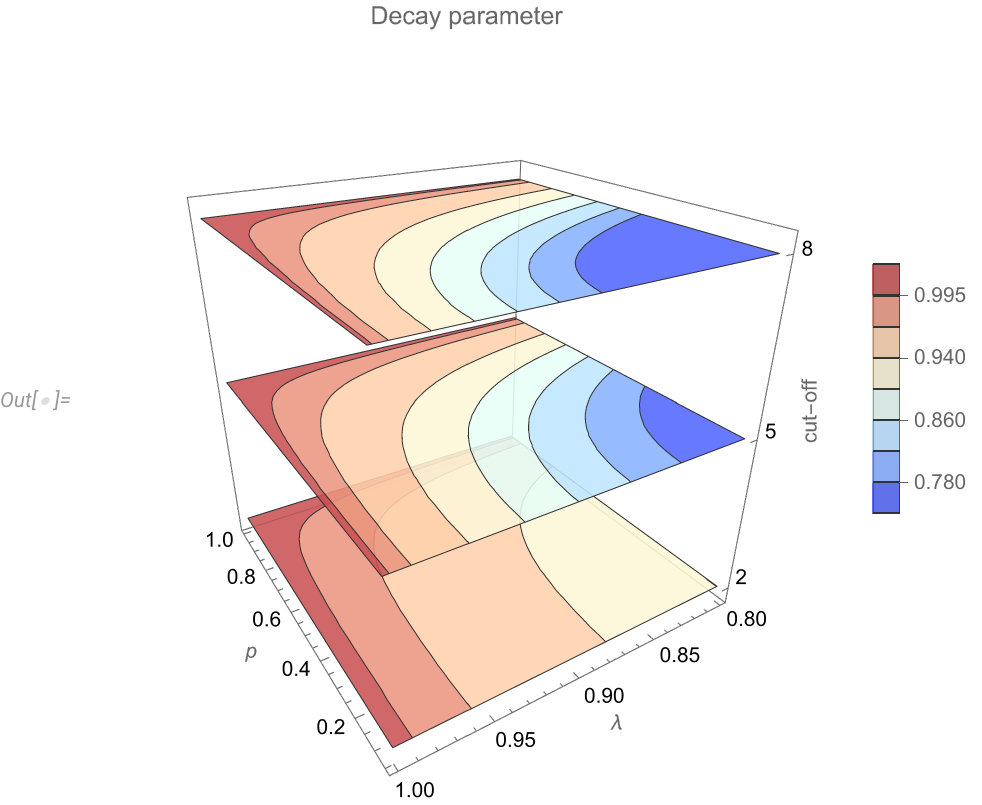}
	\caption{Several slices of the $\overline{B}(\lambda, q, T_c)$ function for $T_c = 2, 6, 10$. The  $\overline{B}(\lambda, q, T_c)$ is a proxy for the decay rate of the quality of the state as additional segments get added.}
    \label{fig:sliceplot}
\end{figure}

As we show in Appendix~\ref{sec:appendix_cutoff_approx}, a similar approximation works for the global cut-off as well. That is, we find a similar function for the decay parameter rate $\overline{B}\left(\lambda, q, T_c\right)$, and analyze how $\overline{B}(\lambda, q, T_c)$ varies as a function of $\lambda$ and $q$ in Fig.~\ref{fig:sliceplot}, for three values of the cut-off, $T_c = 2$, $6$ and $10$. We find that this tight approximation allows for a fast qualitative understanding of noise in swap ASAP repeater chains, which would have been impossible through a Monte Carlo simulation.

\section{Distribution of the noise}\label{sec:distribution}
In this section we show that knowledge of $\mathds{E}\left[\Lambda_n\right]$ in principle suffices to reconstruct the underlying probability distribution of the $\Lambda_n$ parameter. We then apply our results to a so-called `binning' approach to quantum key distribution.

\subsection{Derivation}
We will use here the found analytical forms of $\mathds{E}\left[\Lambda_n\right]$ to find the actual distribution of the noise. As noted before in the paper and in~\cite{kamin2023exact}, the key insight is that $\mathbb{E}\left[\Lambda_n\right]$ is the probability generating function (not to be confused with the generating function from the previous section) of the random variable corresponding to the roughness $K$. That is,

\begin{align}
\mathbb{E}\left[\Lambda_n\right]=&~\left(\frac{1-q}{q}\right)^n \sum_{\overline{t}}\left(q^{\sum_{i}^nt_i}\right)\cdot\left(\lambda^{\sum_{i}^{n-1}\left|t_i-t_{i+1}\right|}\right)\nonumber \\
=&~\sum_{k=0}^{\infty}P\left(K=k\right)\lambda^{k}\nonumber\\
=&~\sum_{k=0}P\left(\Lambda_n =\lambda^k\right)\lambda^{k} \nonumber \ .
\end{align}
From this it follows that

\begin{align}
k!\cdot P\left(\Lambda_n =\lambda^k\right) = \left.\left(\mathds{E}\left[\lambda_n\right]^{(k)}\right)\right|_{\lambda=0}\ .
\end{align}
Here $\mathds{E}\left[\Lambda_n\right]^{(k)}$ is the $k$'th derivative of $\mathds{E}\left[\Lambda_n\right]$ with respect to $\lambda$. Using Mathematica, $P\left(\Lambda_n =\lambda^k\right)$ can be calculated analytically for all $k$ for up to $n=10$ segments~\cite{mathematicafiles}.

\begin{figure}[h]
    \centering \includegraphics[width=0.49\textwidth, trim={1.8cm 0 1.0cm -1cm},clip]{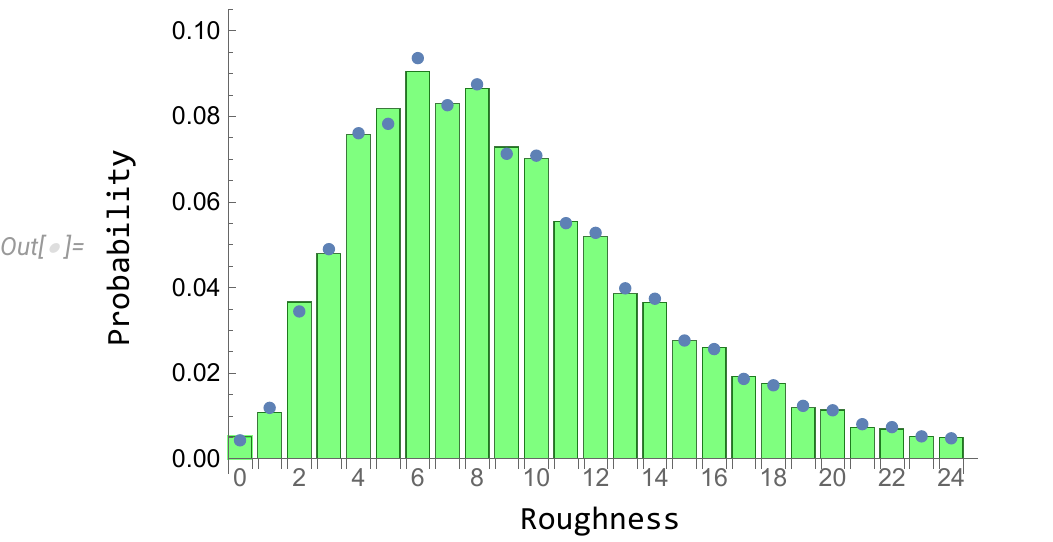}
    \caption{Verification of the analytical distribution of the roughness $K=\sum_{i=1}^{n-1}\left|t_i-t_{i+1}\right|$ for $q=0.6$ and $n=6$ segments (blue markers), along with an estimation found through a Monte Carlo simulation over $10^5$ runs (green bars). The Monte Carlo simulation converges to the exact distribution as the number of runs increases.}
    \label{fig:monte_carlo}
\end{figure}

In Fig.~\ref{fig:monte_carlo}, we validate our analytical results through Monte Carlo simulation. That is, we compare the exact distribution $P(K=k)$ of the roughness for $6$ segments and $q=0.6$ to an estimate of the distribution from a Monte Carlo simulation over $10^5$ timesteps. We observe excellent agreement, and find convergence to the analytical distribution as we increase the number of runs. One interesting feature is the `non-smooth' behavior of $P(K = k)$ (e.g.~$P(K = 6)\geq P(K = 7)$, but $P(K = 7) \leq P(K = 8)$) which becomes more pronounced for larger $p$.

\begin{figure}[h]
    \centering \includegraphics[width=0.49\textwidth, trim={1.8cm 0 0.0cm -1cm},clip]{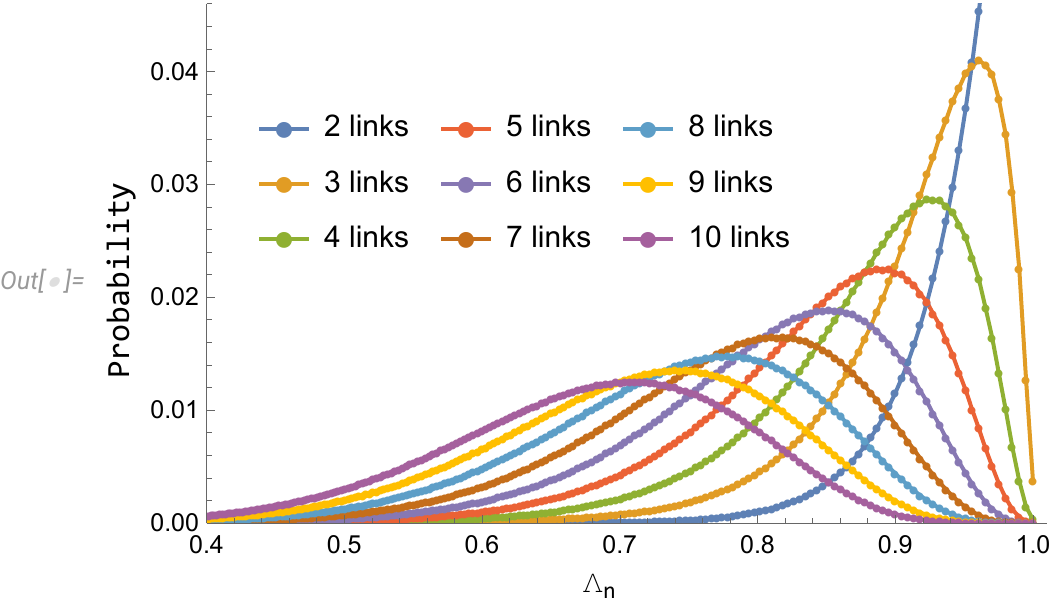}
    \caption{Distribution of the final $\Lambda_n$ parameter for $q=0.9$, $\lambda=0.995$ for several number of segments.}
    \label{fig:noise_distr}
\end{figure}

Noting that $P(K=k)=P(\Lambda_n = \lambda^k)$, we show in Fig.~\ref{fig:noise_distr} the distribution of $\Lambda_n$ for $p=0.1$ and $\lambda = 0.995$ for $n$ ranging from $2$ to $10$. Note that the `non-smooth' behavior has disappeared since $p$ is relatively small.

In principle a similar approach will work for the case with a global cut-off; however, due to the associated expressions becoming more complex, it was not possible to find analytical expressions of the $k$'th derivative.

\subsection{Parameter exploration}
As an application of the analytical expressions of the distribution on $\Lambda_n$, we consider secret key generation using a so-called binning method; see, for example,~\cite{jing2020quantum, goodenough2023near}. That is, the collected classical data are separated into `bins' according to the available information about the associated generated state. Here, this available information will contain the roughness parameter $K(\overline{t})$, which describes the quality of the state. Note that the data need not be received at the time of measurement, which otherwise would come at the cost of increased decoherence.

How does this binning help? Let $\textrm{SKF}\left(\Lambda\right)$ be the secret-key fraction, i.e.~the fraction of bits that can be extracted from an asymptotic number of copies of a state with parameter $\Lambda$ (see Appendix~\ref{sec:exp_delivery_time} for more details). Without binning, the secret-key fraction is given by 
$\textrm{SKF}\left(\mathds{E}\left[\Lambda\right]\right)$. With binning, the secret-key fraction becomes the weighted average. That is,

\begin{align}
&\sum_{k=0}^\infty P(\Lambda = \lambda^k)\cdot \textrm{SKF}\left(\lambda^k\right)\\
=&\sum_{k=0}^{k_\textrm{max}} P(\Lambda=\lambda^k)\cdot \textrm{SKF}\left(\lambda^k\right)\nonumber\\
\geq&\, \textrm{SKF}\left(\mathds{E}\left[\Lambda\right]\right)\nonumber \ ,
\end{align}
where in the first step we use the fact that the secret-key fraction is zero for $\Lambda \leq \Lambda_\textrm{min}\approx 0.778$, such that $k_\textrm{max}=\left\lceil \frac{\log(\Lambda_\textrm{min})}{\log(\lambda)}\right\rceil$. The second step, relies on the fact that the secret-key fraction is convex as a function of $\Lambda$.

\begin{figure}[h]
    \centering \includegraphics[width=0.48\textwidth, trim={1.6cm 0 -0.5cm -0.5cm},clip]{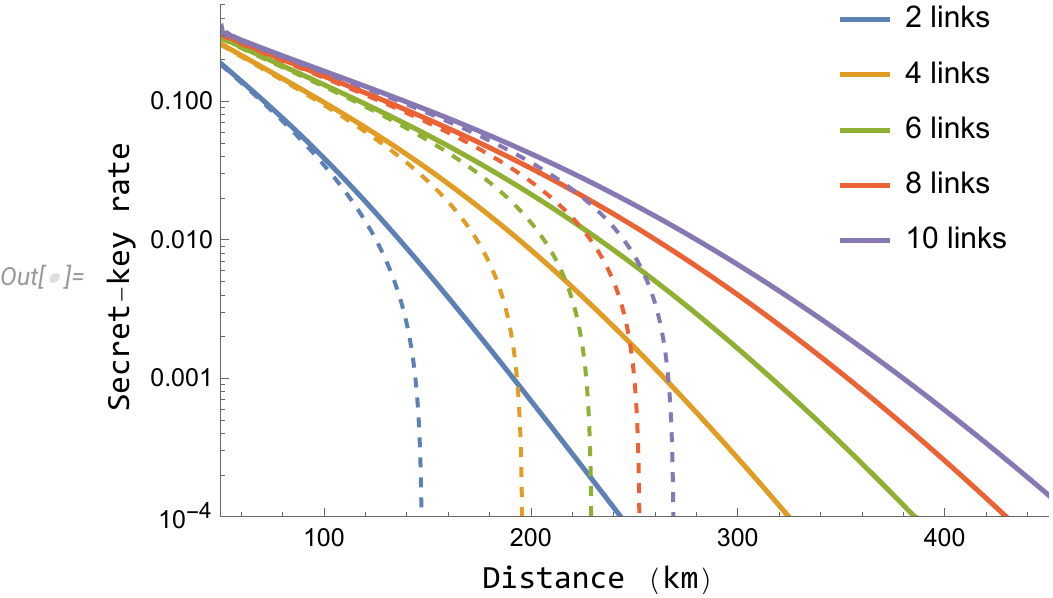}
    \caption{Secret-key rate with (solid) and without (dashed) a binning strategy (see main text) as a function of the distance, for several number of segments $n$ and $\lambda = 0.99$.}
    \label{fig:binning_comparison}
\end{figure}

Fig.~\ref{fig:binning_comparison} shows the secret-key rate with and without the binning method as a function of distance. The data shown correspond to $\lambda=0.99$ and several values of $n$. We observe that the secret-key rate is non-zero for any arbitrary distance. This should not come as a surprise --- the probability $P(K = 0) = \sum_{t=1}^\infty \left(p\left(1-p\right)^{t-1}\right)^n = \frac{p^n}{1-(1-p)^n}\approx p^n$ is small but strictly positive for any $p>0$. 

\section{Discussion}\label{sec:discussion}
In this work we have characterized the noise in homogeneous swap ASAP repeater chains. We have done so by providing analytical formulae and tight approximations of the average fidelity (even when considering a global cut-off) along with the complete distribution of the noise, for a general class of noise models.
The numerical evaluation of the formulae goes beyond existing work and reaches 25 repeater segments if one is interested in the average fidelity of the end-to-end link only, and to 10 repeater segments for a full characterization of the noise distribution.
Since near-term quantum repeater experiments will most likely be based on swapping as soon as possible together with a cut-off policy, these results yield an understanding of what to expect for near-term networks. In particular, the optimization of the global cut-off (which would once be a costly Monte Carlo simulation) can now be done efficiently.

Due to the flexibility of the generating function approach, we anticipate that similar techniques as used here can be applied to other entanglement distribution problems\kg{, such as perhaps the swap ASAP protocol without idling (see~\ref{sec:model}),} or the setup considered in~\cite{avis2023analysis}.

In follow-up work, we investigate both the inhomogeneous bipartite setting and the inhomogeneous multipartite setting. This
follow-up work hinges on interpreting \eqref{eq:partition_function} as the (classical) partition function of a particular physical system --- the \emph{solid-on-solid model}~\cite{owczarek1993exact, rozycki2003rsos, owczarek2009exact, owczarek2010exact}. By then exploiting techniques from statistical physics, we find numerically tight approximations on the average noise.

The generating function provides a powerful tool to understand noise in repeater chains, and it is natural to extend this to more complex scenarios. For example, is there a way to extend the generating function approach to incorporate additional noise on Alice's and Bob's memories? Using Jensen's inequality one can provide a lower bound on $\mathds{E}\left[\Lambda_n\right]$ including the decoherence on Alice and Bob, but an analytical form for the associated generating function seems currently out of reach.
Furthermore, other swapping schemes and cut-off policies besides the global cut-off policy have been analytically studied for small $n$, see~\cite{kamin2023exact}. The expressions found in~\cite{kamin2023exact} were rational functions in $q$ and $\lambda$, similar to the swap ASAP scenario (with a global cut-off policy). It is thus natural to wonder whether these expressions can be obtained using the generating function approach as well.

\section*{Acknowledgements}
We thank Guus Avis, Aliza Brinton, Patrick Emonts, Vicente Lenz, Filip Rozp\k{e}dek, Matthew Skrzypczyk, and Gayane Vardoyan for discussions and feedback on the manuscript. We also thank Mark C.~Wilson for discussions on multivariate generating functions and their associated limit laws. Finally, we thank Philippe Flajolet and Robert Sedgewick for the free dissemination of their book.

This research was supported in part by the NSF grant CNS-1955744, NSF-ERC Center for Quantum Networks grant EEC-1941583, the MURI ARO Grant W911NF2110325. Tim Coopmans is supported by the Dutch National Growth Fund, as part of the Quantum Delta NL program.

\bibliography{references}

\begin{thebibliography}{10}
\providecommand{\url}[1]{#1}
\csname url@samestyle\endcsname
\providecommand{\newblock}{\relax}
\providecommand{\bibinfo}[2]{#2}
\providecommand{\BIBentrySTDinterwordspacing}{\spaceskip=0pt\relax}
\providecommand{\BIBentryALTinterwordstretchfactor}{4}
\providecommand{\BIBentryALTinterwordspacing}{\spaceskip=\fontdimen2\font plus
\BIBentryALTinterwordstretchfactor\fontdimen3\font minus \fontdimen4\font\relax}
\providecommand{\BIBforeignlanguage}[2]{{%
\expandafter\ifx\csname l@#1\endcsname\relax
\typeout{** WARNING: IEEEtran.bst: No hyphenation pattern has been}%
\typeout{** loaded for the language `#1'. Using the pattern for}%
\typeout{** the default language instead.}%
\else
\language=\csname l@#1\endcsname
\fi
#2}}
\providecommand{\BIBdecl}{\relax}
\BIBdecl

\bibitem{gottesman2012longer}
D.~Gottesman, T.~Jennewein, and S.~Croke, ``Longer-baseline telescopes using quantum repeaters,'' \emph{Physical review letters}, vol. 109, no.~7, p. 070503, 2012. doi: \href{https://doi.org/10.1103/PhysRevLett.109.070503}{10.1103/PhysRevLett.109.070503}\,.

\bibitem{sajjad2024quantum}
A.~Sajjad, M.~R. Grace, and S.~Guha, ``Quantum limits of parameter estimation in long-baseline imaging,'' \emph{Physical Review Research}, vol.~6, no.~1, p. 013212, 2024. doi: \href{https://doi.org/10.1103/PhysRevResearch.6.013212}{10.1103/PhysRevResearch.6.013212}\,.

\bibitem{hillery1999quantum}
M.~Hillery, V.~Bu{\v{z}}ek, and A.~Berthiaume, ``Quantum secret sharing,'' \emph{Physical Review A}, vol.~59, no.~3, p. 1829, 1999. doi: \href{https://doi.org/10.1103/PhysRevA.59.1829}{10.1103/PhysRevA.59.1829}\,.

\bibitem{markham2008graph}
D.~Markham and B.~C. Sanders, ``Graph states for quantum secret sharing,'' \emph{Physical Review A}, vol.~78, no.~4, p. 042309, 2008. doi: \href{https://doi.org/10.1103/PhysRevA.78.042309}{10.1103/PhysRevA.78.042309}\,.

\bibitem{cirac1999distributed}
J.~Cirac, A.~Ekert, S.~Huelga, and C.~Macchiavello, ``Distributed quantum computation over noisy channels,'' \emph{Physical Review A}, vol.~59, no.~6, p. 4249, 1999. doi: \href{https://doi.org/10.1103/PhysRevA.59.4249}{10.1103/PhysRevA.59.4249}\,.

\bibitem{serafini2006distributed}
A.~Serafini, S.~Mancini, and S.~Bose, ``Distributed quantum computation via optical fibers,'' \emph{Physical review letters}, vol.~96, no.~1, p. 010503, 2006. doi: \href{https://doi.org/10.1103/PhysRevLett.96.010503}{10.1103/PhysRevLett.96.010503}\,.

\bibitem{wehner2018quantum}
S.~Wehner, D.~Elkouss, and R.~Hanson, ``Quantum internet: A vision for the road ahead,'' \emph{Science}, vol. 362, no. 6412, 2018. doi: \href{https://doi.org/10.1126/science.aam9288}{10.1126/science.aam9288}\,.

\bibitem{rabbie2022designing}
J.~Rabbie, K.~Chakraborty, G.~Avis, and S.~Wehner, ``Designing quantum networks using preexisting infrastructure,'' \emph{npj Quantum Information}, vol.~8, no.~1, p.~5, 2022. doi: \href{https://doi.org/10.1038/s41534-021-00501-3}{10.1038/s41534-021-00501-3}\,.

\bibitem{briegel1998quantum}
H.-J. Briegel, W.~D{\"u}r, J.~I. Cirac, and P.~Zoller, ``Quantum repeaters: the role of imperfect local operations in quantum communication,'' \emph{Physical Review Letters}, vol.~81, no.~26, p. 5932, 1998. doi: \href{https://doi.org/10.1103/PhysRevLett.81.5932}{10.1103/PhysRevLett.81.5932}\,.

\bibitem{azuma2022quantum}
K.~Azuma, S.~E. Economou, D.~Elkouss, P.~Hilaire, L.~Jiang, H.-K. Lo, and I.~Tzitrin, ``Quantum repeaters: From quantum networks to the quantum internet,'' \emph{arXiv preprint arXiv:2212.10820}, 2022. doi: \href{https://doi.org/10.1103/RevModPhys.95.045006}{10.1103/RevModPhys.95.045006}\,.

\bibitem{rozpkedek2018parameter}
F.~Rozp{\k{e}}dek, K.~Goodenough, J.~Ribeiro, N.~Kalb, V.~C. Vivoli, A.~Reiserer, R.~Hanson, S.~Wehner, and D.~Elkouss, ``Parameter regimes for a single sequential quantum repeater,'' \emph{Quantum Science and Technology}, vol.~3, no.~3, p. 034002, 2018.

\bibitem{rozpkedek2019near}
F.~Rozp{\k{e}}dek, R.~Yehia, K.~Goodenough, M.~Ruf, P.~C. Humphreys, R.~Hanson, S.~Wehner, and D.~Elkouss, ``Near-term quantum-repeater experiments with nitrogen-vacancy centers: Overcoming the limitations of direct transmission,'' \emph{Physical Review A}, vol.~99, no.~5, p. 052330, 2019.

\bibitem{collins2007multiplexed}
O.~Collins, S.~Jenkins, A.~Kuzmich, and T.~Kennedy, ``Multiplexed memory-insensitive quantum repeaters,'' \emph{Physical review letters}, vol.~98, no.~6, p. 060502, 2007. doi: \href{https://doi.org/10.1103/PhysRevLett.98.060502}{10.1103/PhysRevLett.98.060502}\,.

\bibitem{avis2022requirements}
G.~Avis, F.~F. da~Silva, T.~Coopmans, A.~Dahlberg, H.~Jirovsk{\'a}, D.~Maier, J.~Rabbie, A.~Torres-Knoop, and S.~Wehner, ``Requirements for a processing-node quantum repeater on a real-world fiber grid,'' \emph{arXiv preprint arXiv:2207.10579}, 2022. doi: \href{https://doi.org/10.1038/s41534-023-00765-x}{10.1038/s41534-023-00765-x}\,.

\bibitem{li2020efficient}
B.~Li, T.~Coopmans, and D.~Elkouss, ``Efficient optimization of cut-offs in quantum repeater chains,'' in \emph{2020 IEEE International Conference on Quantum Computing and Engineering (QCE)}.\hskip 1em plus 0.5em minus 0.4em\relax IEEE, 2020, pp. 158--168. doi: \href{https://doi.org/10.1109/QCE49297.2020.00029}{10.1109/QCE49297.2020.00029}\,.

\bibitem{khatri2019practical}
\BIBentryALTinterwordspacing
S.~Khatri, C.~T. Matyas, A.~U. Siddiqui, and J.~P. Dowling, ``Practical figures of merit and thresholds for entanglement distribution in quantum networks,'' \emph{Phys. Rev. Research}, vol.~1, p. 023032, Sep 2019. doi: \href{https://doi.org/10.1103/PhysRevResearch.1.023032}{10.1103/PhysRevResearch.1.023032}\,. [Online]. Available: \url{https://link.aps.org/doi/10.1103/PhysRevResearch.1.023032}
\BIBentrySTDinterwordspacing

\bibitem{santra2018quantum}
\BIBentryALTinterwordspacing
S.~Santra, L.~Jiang, and V.~S. Malinovsky, ``Quantum repeater architecture with hierarchically optimized memory buffer times,'' \emph{Quantum Science and Technology}, vol.~4, no.~2, p. 025010, mar 2019. doi: \href{https://doi.org/10.1088/2058-9565/ab0bc2}{10.1088/2058-9565/ab0bc2}\,. [Online]. Available: \url{https://doi.org/10.1088%2F2058-9565%2Fab0bc2}
\BIBentrySTDinterwordspacing

\bibitem{davies2023tools}
B.~Davies, T.~Beauchamp, G.~Vardoyan, and S.~Wehner, ``Tools for the analysis of quantum protocols requiring state generation within a time window,'' \emph{arXiv:2304.12673}, 2023. doi: \href{https://doi.org/10.1109/TQE.2024.3358674}{10.1109/TQE.2024.3358674}\,.

\bibitem{azuma2021tools}
\BIBentryALTinterwordspacing
K.~Azuma, S.~B\"auml, T.~Coopmans, D.~Elkouss, and B.~Li, ``Tools for quantum network design,'' \emph{AVS Quantum Science}, vol.~3, no.~1, p. 014101, 2021. doi: \href{https://doi.org/10.1116/5.0024062}{10.1116/5.0024062}\,. [Online]. Available: \url{https://doi.org/10.1116/5.0024062}
\BIBentrySTDinterwordspacing

\bibitem{praxmeyer2013reposition}
L.~Praxmeyer, ``Reposition time in probabilistic imperfect memories,'' \emph{arXiv preprint arXiv:1309.3407}, 2013.

\bibitem{kamin2023exact}
L.~Kamin, E.~Shchukin, F.~Schmidt, and P.~van Loock, ``Exact rate analysis for quantum repeaters with imperfect memories and entanglement swapping as soon as possible,'' \emph{Physical Review Research}, vol.~5, no.~2, p. 023086, 2023. doi: \href{https://doi.org/10.1103/PhysRevResearch.5.023086}{10.1103/PhysRevResearch.5.023086}\,.

\bibitem{shchukin2019waiting}
E.~Shchukin, F.~Schmidt, and P.~van Loock, ``Waiting time in quantum repeaters with probabilistic entanglement swapping,'' \emph{Physical Review A}, vol. 100, no.~3, p. 032322, 2019. doi: \href{https://doi.org/10.1103/PhysRevA.100.032322}{10.1103/PhysRevA.100.032322}\,.

\bibitem{reiss2023deep}
\BIBentryALTinterwordspacing
S.~D. Rei\ss{} and P.~van Loock, ``Deep reinforcement learning for key distribution based on quantum repeaters,'' \emph{Phys. Rev. A}, vol. 108, p. 012406, Jul 2023. doi: \href{https://doi.org/10.1103/PhysRevA.108.012406}{10.1103/PhysRevA.108.012406}\,. [Online]. Available: \url{https://link.aps.org/doi/10.1103/PhysRevA.108.012406}
\BIBentrySTDinterwordspacing

\bibitem{zang2023entanglement}
A.~Zang, X.~Chen, A.~Kolar, J.~Chung, M.~Suchara, T.~Zhong, and R.~Kettimuthu, ``Entanglement distribution in quantum repeater with purification and optimized buffer time,'' in \emph{IEEE INFOCOM 2023 - IEEE Conference on Computer Communications Workshops (INFOCOM WKSHPS)}, 2023, pp. 1--6. doi: \href{https://doi.org/10.1109/INFOCOMWKSHPS57453.2023.10226122}{10.1109/INFOCOMWKSHPS57453.2023.10226122}\,.

\bibitem{bennett2020quantum}
\BIBentryALTinterwordspacing
C.~H. Bennett and G.~Brassard, ``Quantum cryptography: Public key distribution and coin tossing,'' \emph{Theoretical Computer Science}, vol. 560, pp. 7--11, 2014. doi: \href{https://doi.org/https://doi.org/10.1016/j.tcs.2014.05.025}{https://doi.org/10.1016/j.tcs.2014.05.025}\,., theoretical Aspects of Quantum Cryptography – celebrating 30 years of BB84. [Online]. Available: \url{https://www.sciencedirect.com/science/article/pii/S0304397514004241}
\BIBentrySTDinterwordspacing

\bibitem{mathematicafiles}
K.~Goodenough, ``Swap asap analytics,'' \url{https://github.com/KDGoodenough/swapASAPAnalytics}, 2024.

\bibitem{pant2019routing}
M.~Pant, H.~Krovi, D.~Towsley, L.~Tassiulas, L.~Jiang, P.~Basu, D.~Englund, and S.~Guha, ``Routing entanglement in the quantum internet,'' \emph{npj Quantum Information}, vol.~5, no.~1, p.~25, 2019. doi: \href{https://doi.org/10.1038/s41534-019-0139-x}{10.1038/s41534-019-0139-x}\,.

\bibitem{fittipaldi2022linear}
P.~Fittipaldi, A.~Giovanidis, and F.~Grosshans, ``A linear algebraic framework for quantum internet dynamic scheduling,'' in \emph{2022 IEEE International Conference on Quantum Computing and Engineering (QCE)}.\hskip 1em plus 0.5em minus 0.4em\relax IEEE, 2022, pp. 447--453. doi: \href{https://doi.org/10.1109/QCE53715.2022.00066}{10.1109/QCE53715.2022.00066}\,.

\bibitem{van2023entanglement}
E.~A. Van~Milligen, E.~Jacobson, A.~Patil, G.~Vardoyan, D.~Towsley, and S.~Guha, ``Entanglement routing over networks with time multiplexed repeaters,'' \emph{arXiv preprint arXiv:2308.15028}, 2023.

\bibitem{van2024utilizing}
E.~A. Van~Milligen, C.~N. Gagatsos, E.~Kaur, D.~Towsley, and S.~Guha, ``Utilizing probabilistic entanglement between sensors in quantum networks,'' \emph{Physical Review Applied}, vol.~22, no.~6, p. 064085, 2024. doi: \href{https://doi.org/10.1103/PhysRevApplied.22.064085}{10.1103/PhysRevApplied.22.064085}\,.

\bibitem{vardoyan2019stochastic}
G.~Vardoyan, S.~Guha, P.~Nain, and D.~Towsley, ``On the stochastic analysis of a quantum entanglement switch,'' \emph{ACM SIGMETRICS Performance Evaluation Review}, vol.~47, no.~2, pp. 27--29, 2019. doi: \href{https://doi.org/10.1145/3374888.3374899}{10.1145/3374888.3374899}\,.

\bibitem{noh2020fault}
K.~Noh and C.~Chamberland, ``Fault-tolerant bosonic quantum error correction with the surface--gottesman-kitaev-preskill code,'' \emph{Physical Review A}, vol. 101, no.~1, p. 012316, 2020. doi: \href{https://doi.org/10.1103/PhysRevA.101.012316}{10.1103/PhysRevA.101.012316}\,.

\bibitem{albert2018performance}
V.~V. Albert, K.~Noh, K.~Duivenvoorden, D.~J. Young, R.~Brierley, P.~Reinhold, C.~Vuillot, L.~Li, C.~Shen, S.~M. Girvin \emph{et~al.}, ``Performance and structure of single-mode bosonic codes,'' \emph{Physical Review A}, vol.~97, no.~3, p. 032346, 2018. doi: \href{https://doi.org/10.1103/PhysRevA.97.032346}{10.1103/PhysRevA.97.032346}\,.

\bibitem{shaw2024logical}
M.~H. Shaw, A.~C. Doherty, and A.~L. Grimsmo, ``Logical gates and read-out of superconducting gottesman-kitaev-preskill qubits,'' \emph{arXiv preprint arXiv:2403.02396}, 2024.

\bibitem{noh2018quantum}
K.~Noh, V.~V. Albert, and L.~Jiang, ``Quantum capacity bounds of gaussian thermal loss channels and achievable rates with gottesman-kitaev-preskill codes,'' \emph{IEEE Transactions on Information Theory}, vol.~65, no.~4, pp. 2563--2582, 2018. doi: \href{https://doi.org/10.1109/TIT.2018.2873764}{10.1109/TIT.2018.2873764}\,.

\bibitem{haldar2024reducing}
S.~Haldar, P.~Barge, X.~Cheng, K.-C. Chang, B.~Kirby, S.~Khatri, C.~W. Wong, and H.~Lee, ``Reducing classical communication costs in multiplexed quantum repeaters using hardware-aware quasi-local policies,'' 2024. doi: \href{https://doi.org/10.1038/s42005-025-02029-w}{10.1038/s42005-025-02029-w}\,.

\bibitem{li2024optimising}
J.~Li, T.~Coopmans, P.~Emonts, K.~Goodenough, J.~Tura, and E.~van Nieuwenburg, ``Optimising entanglement distribution policies under classical communication constraints assisted by reinforcement learning,'' \emph{arXiv preprint arXiv:2412.06938}, 2024.

\bibitem{owczarek1993exact}
A.~L. Owczarek and T.~Prellberg, ``Exact solution of the discrete (1+ 1)-dimensional sos model with field and surface interactions,'' \emph{journal of Statistical Physics}, vol.~70, pp. 1175--1194, 1993. doi: \href{https://doi.org/10.1007/BF01049427}{10.1007/BF01049427}\,.

\bibitem{rozycki2003rsos}
B.~R{\'o}zycki and M.~Napi{\'o}rkowski, ``The rsos model for a slit with different walls,'' \emph{Journal of Physics A: Mathematical and General}, vol.~36, no.~16, p. 4551, 2003.

\bibitem{owczarek2009exact}
A.~Owczarek and T.~Prellberg, ``Exact solution of the discrete (1+ 1)-dimensional rsos model with field and surface interactions,'' \emph{Journal of Physics A: Mathematical and Theoretical}, vol.~42, no.~49, p. 495003, 2009. doi: \href{https://doi.org/10.1088/1751-8113/42/49/495003}{10.1088/1751-8113/42/49/495003}\,.

\bibitem{owczarek2010exact}
A.~L. Owczarek and T.~Prellberg, ``Exact solution of the discrete (1+ 1)-dimensional rsos model in a slit with field and wall interactions,'' \emph{Journal of Physics A: Mathematical and Theoretical}, vol.~43, no.~37, p. 375004, 2010. doi: \href{https://doi.org/10.1088/1751-8113/43/37/375004}{10.1088/1751-8113/43/37/375004}\,.

\bibitem{flajolet2009analytic}
P.~Flajolet and R.~Sedgewick, \emph{Analytic combinatorics}.\hskip 1em plus 0.5em minus 0.4em\relax cambridge University press, 2009. doi: \href{https://doi.org/10.1017/CBO9780511801655}{10.1017/CBO9780511801655}\,.

\bibitem{pemantle2013analytic}
R.~Pemantle and M.~C. Wilson, \emph{Analytic combinatorics in several variables}.\hskip 1em plus 0.5em minus 0.4em\relax Cambridge University Press, 2013, no. 140. doi: \href{https://doi.org/10.1017/9781108874144}{10.1017/9781108874144}\,.

\bibitem{remmert1991theory}
R.~Remmert, \emph{Theory of complex functions}.\hskip 1em plus 0.5em minus 0.4em\relax Springer Science \& Business Media, 1991, vol. 122. doi: \href{https://doi.org/10.1007/978-1-4612-0939-3}{10.1007/978-1-4612-0939-3}\,.

\bibitem{lang2013complex}
S.~Lang, \emph{Complex analysis}.\hskip 1em plus 0.5em minus 0.4em\relax Springer Science \& Business Media, 2013, vol. 103. doi: \href{https://doi.org/10.1007/978-3-642-59273-7}{10.1007/978-3-642-59273-7}\,.

\bibitem{gasper2004basic}
G.~Gasper and M.~Rahman, \emph{Basic hypergeometric series}.\hskip 1em plus 0.5em minus 0.4em\relax Cambridge university press, 2004, vol.~96. doi: \href{https://doi.org/10.1017/CBO9780511526251.004}{10.1017/CBO9780511526251.004}\,.

\bibitem{driver2008zeros}
K.~Driver and K.~Jordaan, ``Zeros of the hypergeometric polynomial f (-n, b; c; z),'' \emph{arXiv preprint arXiv:0812.0708}, 2008.

\bibitem{dominici2013real}
D.~Dominici, S.~J. Johnston, and K.~Jordaan, ``Real zeros of 2f1 hypergeometric polynomials,'' \emph{Journal of Computational and Applied Mathematics}, vol. 247, pp. 152--161, 2013.

\bibitem{jing2020quantum}
Y.~Jing, D.~Alsina, and M.~Razavi, ``Quantum key distribution over quantum repeaters with encoding: Using error detection as an effective postselection tool,'' \emph{Physical Review Applied}, vol.~14, no.~6, p. 064037, 2020. doi: \href{https://doi.org/10.1103/PhysRevApplied.14.064037}{10.1103/PhysRevApplied.14.064037}\,.

\bibitem{goodenough2023near}
K.~Goodenough, S.~de~Bone, V.~L. Addala, S.~Krastanov, S.~Jansen, D.~Gijswijt, and D.~Elkouss, ``Near-term $ n $ to $ k $ distillation protocols using graph codes,'' \emph{arXiv preprint arXiv:2303.11465}, 2023. doi: \href{https://doi.org/10.1109/JSAC.2024.3380094}{10.1109/JSAC.2024.3380094}\,.

\bibitem{avis2023analysis}
G.~Avis, F.~Rozp{\k{e}}dek, and S.~Wehner, ``Analysis of multipartite entanglement distribution using a central quantum-network node,'' \emph{Physical Review A}, vol. 107, no.~1, p. 012609, 2023. doi: \href{https://doi.org/10.1103/PhysRevA.107.012609}{10.1103/PhysRevA.107.012609}\,.

\bibitem{inesta2023optimal}
{\'A}.~G. I{\~n}esta, G.~Vardoyan, L.~Scavuzzo, and S.~Wehner, ``Optimal entanglement distribution policies in homogeneous repeater chains with cutoffs,'' \emph{npj Quantum Information}, vol.~9, no.~1, p.~46, 2023.

\bibitem{kepler1966strena}
J.~Kepler, \emph{Strena seu de Niue sexangula}.\hskip 1em plus 0.5em minus 0.4em\relax Gottfried Tampach, 1966.

\bibitem{schneer1960kepler}
C.~Schneer, ``Kepler's new year's gift of a snowflake,'' \emph{Isis}, vol.~51, no.~4, pp. 531--545, 1960. doi: \href{https://doi.org/10.1086/349411}{10.1086/349411}\,.

\bibitem{stanley2011enumerative}
R.~P. Stanley, ``Enumerative combinatorics volume 1 second edition,'' \emph{Cambridge studies in advanced mathematics}, 2011.

\bibitem{shahbeigi2021log}
F.~Shahbeigi, D.~Amaro-Alcal{\'a}, Z.~Pucha{\l}a, and K.~{\.Z}yczkowski, ``Log-convex set of lindblad semigroups acting on n-level system,'' \emph{Journal of Mathematical Physics}, vol.~62, no.~7, p. 072105, 2021. doi: \href{https://doi.org/10.1063/5.0009745}{10.1063/5.0009745}\,.

\bibitem{weyl1927quantenmechanik}
H.~Weyl, ``Quantenmechanik und gruppentheorie,'' \emph{Zeitschrift f{\"u}r Physik}, vol.~46, no. 1-2, pp. 1--46, 1927. doi: \href{https://doi.org/10.1007/BF02055756}{10.1007/BF02055756}\,.

\bibitem{helsen2022general}
J.~Helsen, I.~Roth, E.~Onorati, A.~H. Werner, and J.~Eisert, ``General framework for randomized benchmarking,'' \emph{PRX Quantum}, vol.~3, no.~2, p. 020357, 2022. doi: \href{https://doi.org/10.1103/PRXQuantum.3.020357}{10.1103/PRXQuantum.3.020357}\,.

\bibitem{isaacs2006character}
I.~M. Isaacs, \emph{Character theory of finite groups}.\hskip 1em plus 0.5em minus 0.4em\relax American Mathematical Soc., 2006, vol. 359. doi: \href{https://doi.org/10.1090/chel/359}{10.1090/chel/359}\,.

\bibitem{terras1999fourier}
A.~Terras, \emph{Fourier analysis on finite groups and applications}.\hskip 1em plus 0.5em minus 0.4em\relax Cambridge University Press, 1999. doi: \href{https://doi.org/10.1017/CBO9780511626265}{10.1017/CBO9780511626265}\,.

\bibitem{diaconis1988group}
P.~Diaconis, ``Group representations in probability and statistics,'' \emph{Lecture notes-monograph series}, vol.~11, pp. i--192, 1988. doi: \href{https://doi.org/10.1214/lnms/1215467407}{10.1214/lnms/1215467407}\,.

\bibitem{wilde2011classical}
M.~M. Wilde, \emph{Quantum information theory}.\hskip 1em plus 0.5em minus 0.4em\relax Cambridge University Press, 2013. doi: \href{https://doi.org/10.1017/9781316809976}{10.1017/9781316809976}\,.

\bibitem{scarani2009security}
V.~Scarani, H.~Bechmann-Pasquinucci, N.~J. Cerf, M.~Du{\v{s}}ek, N.~L{\"u}tkenhaus, and M.~Peev, ``The security of practical quantum key distribution,'' \emph{Reviews of modern physics}, vol.~81, no.~3, p. 1301, 2009. doi: \href{https://doi.org/10.1103/RevModPhys.81.1301}{10.1103/RevModPhys.81.1301}\,.

\bibitem{bernardes2011rate}
N.~K. Bernardes, L.~Praxmeyer, and P.~van Loock, ``Rate analysis for a hybrid quantum repeater,'' \emph{Physical Review A}, vol.~83, no.~1, p. 012323, 2011. doi: \href{https://doi.org/10.1103/PhysRevA.83.012323}{10.1103/PhysRevA.83.012323}\,.

\end{thebibliography}

\onecolumngrid
\appendix

\section{Other cut-off policies}\label{sec:policies}

For completeness, we mention here several other cut-off policies, but which will not be considered in this paper. As previously mentioned, decoherence is a prime obstacle to quantum communication. A cut-off policy is a condition for discarding the present entanglement.  Discarding the generated entanglement will reduce the rate at which entanglement is delivered, but will mitigate the time spent on distributing end-to-end entanglement which is of too low quality. A cut-off policy thus allows for a trade-off between the rate and the quality of the delivered end-to-end entanglement. Cut-off policies and their trade-offs have been studied in for example~\cite{rozpkedek2018parameter, rozpkedek2019near, collins2007multiplexed, avis2022requirements, li2020efficient, kamin2023exact, shchukin2019waiting, reiss2023deep}, where they were shown to be indispensable for near-term devices.

\begin{figure}[h!]
    \centerfloat
    \vspace{-30mm}
        \begin{tikzpicture}[scale=1.25, line cap=round, entanglement/.style={shade, shading=ring, line width = 0.7mm, Purple,  double=RedViolet!40!Periwinkle!90,double distance=1.2pt}, repeater/.style={diamond, NavyBlue}, noentanglement/.style={shade, shading=ring, line width = 0.7mm, Purple!20,  double=RedViolet!10!Periwinkle!50,double distance=1.2pt, repeater/.style={diamond, NavyBlue}}]

\node[scale=1.5] at (1,3) {\large\textbf{Local}};

\draw[line width=1pt] (-0.0,-0.25) -- (-0.0,3.4);
\node[diamond, scale=6] at (2,5) {};

\draw[line width=1pt] (0.0,-0.25) -- (6.5,-0.25);
\node[rotate=90] at (-0.4,1.5) {\large\textbf{Time}};
  \draw [-{Latex[length=2.5mm]}, line width = 1.5pt] (-0.15,1)   -- (-0.15,2);

  \node[rotate=90, scale=1.5] at (3.4,1.05) {$T_c$};
  \draw [decorate, decoration={brace}] (3.8, 0.045) -- (3.8, 2.115);

\draw[line width=1pt, dashed, green] (4,2.855) -- (4,0.065);







\draw[line width=5pt, entanglement] (1, 2) -- (2,2) node[above] {};
\draw[line width=5pt, entanglement] (2, 1) -- (3,1) node[right] {};
\draw[line width=5pt, noentanglement] (3, 3) -- (4,3) node[right] {};
\draw[line width=5pt, entanglement] (4, 0) -- (5,0) node[right] {};
\draw[line width=5pt, entanglement] (5, 1) -- (6,1) node[right] {};

\node[] at (1.5,2.25) {$t_1 = 3$};
\node[] at (2.5,1.25) {$t_2 = 2$};
\node[] at (3.5,3.25) {$t_3 = 4$};
\node[] at (4.5,0.25) {$t_4 = 1$};
\node[] at (5.5,1.25) {$t_5 = 2$};

\node[diamond, scale=6] at (5.2,1.25) {};

\end{tikzpicture}
    \caption{Graphical description of the local cut-off policy. Each node has agreed on a parameter $T_c$. For the local cut-off, a reset is performed when any node has to store entanglement for more than $T_c$ rounds. We consider the local cut-off in follow-up work.}
    \label{fig:local_cutoff}
\end{figure}
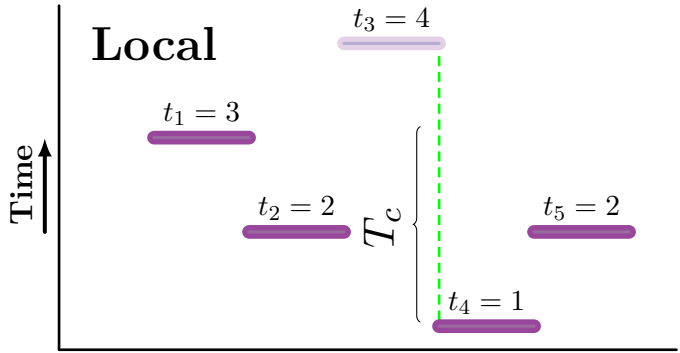

The \emph{local cut-off policy} limits the time a single memory is allowed to store entanglement for. As soon as a memory not belonging to Alice or Bob holds entanglement, a timer is started. As before, this timer increases by $1$ every round, and is reset to $0$ after discarding. A reset occurs when any timer exceeds $T_c$, where we use the same notation for the global and local cut-off. In Fig.~\ref{fig:local_cutoff} we show an example of the local cut-off. Here, the left memory of the fourth link had to store entanglement for more than $T_c = 2$ rounds.

The policy considered in~\cite{praxmeyer2013reposition} is similar to the global cut-off policy in this paper. However, instead of starting a timer at the beginning of all attempts, a timer is started at the time the first link succeeds. Another policy would be based on predicting the delivery time (see~\cite{inesta2023optimal}) and quality of the delivered state, conditioned on the current state of the repeater chain~\cite{inesta2023optimal}. Note that \emph{all} entanglement is reset after a cut-off condition has been reached in the policy of~\cite{praxmeyer2013reposition} and both the global and local cut-off.

We have numerically observed through Monte Carlo simulations that the local cut-off performs slightly better than the global cut-off, but not by a significant amount.

\section{Example of the Fibonacci generating function}\label{sec:fib_gen}
We motivate and explain here generating functions through a simple and more familiar example, before applying generating functions to swap ASAP repeater chains. In this example we will estimate the asymptotic behavior of the Fibonacci numbers $F_n$ by first finding the associated generating function $G(x)$ and studying its complex-analytic behavior.
\kg{To wit, the Fibonacci numbers $F_n$ are defined as $F_0=0$, $F_1=1$, and $F_n = F_{n-1} + F_{n-2},~n\geq 2$. We now show a well-known derivation of the Fibonacci generating function $G(x)=\sum_{n=0}^{\infty}F_nx^n$. Through a judiciously chosen starting expression we find}
\begin{align}
&\left(1-x-x^2\right)G(x)\nonumber\\
= &\left(1-x-x^2\right)\sum_{n=0}^\infty F_n x^n\nonumber\\
= &\sum_{n=0}^\infty F_n x^n - \sum_{n=0}^\infty F_n x^{n+1} - \sum_{n=0}^\infty F_n x^{n+2}\nonumber\\
= &~\sum_{n=0}^\infty F_n x^n - \sum_{n=1}^\infty F_{n-1} x^{n} - \sum_{n=2}^\infty F_{n-2} x^{n}\nonumber\\
= &~F_0 + (F_1-F_0)x\nonumber\\
= &~x\quad \implies G(x) = \frac{x}{1-x-x^2}\ ,
\end{align}
\kg{where we canceled all like terms going from the fourth to the fifth line.}


\kg{Note that the infinite sum $\sum F_n x^n$ does not converge for all $x\in \mathbb{C}$. Contrast this with $G(x) = \frac{x}{1-x-x^2}$, which is defined for all $x\in\mathbb{C}$, save for the two roots $\rho_1, \rho_2$ of $1-x-x^2$. The function $G(x)$ is an example of an analytic continuation of $\sum F_nx^n$, meaning that the original domain can be uniquely extended in a well-defined manner~\cite{lang2013complex}. Such machinery allows us to conveniently ignore issues of convergence. More importantly, under mild conditions a generating function $G(x)$ will be a \emph{meromorphic function}. This means that, save for a set of isolated points $\lbrace{\rho_i\rbrace}$, $G(x)$ is complex-differentiable on $\mathbb{C}$.}

\kg{The $\lbrace \rho_i \rbrace$ are called \emph{singularities} of a meromorphic function, and reveal key information about the underlying sequence $a_n$. For example, a standard result in analytic combinatorics (see theorem IV.10 from~\cite{flajolet2009analytic}) tells us that the ratio between $a_n$ and $\left(\frac{1}{\rho}\right)^{n+1}$ converges to a constant $c$, where $\rho$ is the singularity closest to the origin\footnote{For technical reasons the singularities are implicitly required to be \emph{simple poles}, see~\cite{lang2013complex}. This can be easily verified to be the case for the cases considered here. }. In other words, $a_n$ is asymptotically equal to $c\cdot \left(\frac{1}{\rho}\right)^{n+1}$. Applying this to the Fibonacci generating function, we find that $\rho$ is given by the smallest solution to $1-x-x^2$, i.e.~$\rho = -\phi' = \frac{1}{\phi} = \frac{\sqrt{5}-1}{2}= 0.618\ldots$ with $\phi$ the golden ratio and $\phi'$ its conjugate. In other words, the limit of the ratio between $F_n$ and $\left(\frac{1}{\rho}\right)^{n+1} = \phi^{n+1}$ converges to some constant. This immediately reclaims the classic result by Kepler that $\lim_{n\to\infty}\frac{F_{n+1}}{F_n} = \phi$~\cite{kepler1966strena, schneer1960kepler}.}

\kg{The location of the closest singularity $\rho$ dictates the exponential growth/decay, but not the constant $c$. Fortunately, the constant $c$ depends on the closest singularity as well, see theorem IV.10 of~\cite{flajolet2009analytic}. Specifically, $c$ is given by minus the \emph{residue} of the function $G$ at $\rho$. As noted in the main text, this can be derived more explicitly by expanding $f(z)$ in Theorem IV.10 of~\cite{flajolet2009analytic} around $z=\rho$ when $\rho$ is a simple pole. An explicit demonstration can be found in Example IV.7.~of~\cite{flajolet2009analytic}.
For the cases of interest to this manuscript (i.e.~where all singularities are simple poles), we have that $\textrm{Res}_{\rho_i}\left(G\right) = \lim_{x\rightarrow \rho_i}\left(x-\rho_i\right)G(x)$~\cite{lang2013complex}. A straightforward calculation then shows that the residue of $G$ at $\rho$ is given by $-\frac{1}{\sqrt{5}\phi}$, from which we find that} 

\begin{align}
    F_n &\sim \left[-\textrm{Res}_{\rho}\left(G\right)\right] \cdot \left(\frac{1}{\rho}\right)^{n+1}\label{eq:fib_example1}\\
    &= \frac{\phi^n}{\sqrt{5}} \label{eq:fib_example2} \ .
\end{align} 
\kg{Indeed, $F_{10}=55$, while the approximation in Eq.~\eqref{eq:fib_example2} yields the excellent agreement of $55.0036\ldots$. Theorem IV.10 of~\cite{flajolet2009analytic} tells us that the approximation can be further refined by including the other singularity $\rho'$. In fact, doing so would recover a closed-form expression of the Fibonacci numbers, known as Binet's formula~\cite{flajolet2009analytic}.}

\kg{To further highlight the insight generating functions provide, we conclude with one more fact that can be gleaned from $G$. Formally rewriting we have}

\begin{align}
G(x) =&~\frac{x}{1-x-x^2}\\
=&~x\cdot \frac{1}{1-(x+x^2)}\\
=&~x\cdot \left((x+x^2)^0 + (x+x^2)^1 + (x+x^2)^2 + \ldots \right)\\
\equiv&~\sum_{n=0}^\infty F_nx^n\ .
\end{align}
\kg{Expanding the penultimate line, we find that $F_n$ is the number of ways all the different terms $\left(x+x^2\right)^i$ contribute a $x^{n-1}$ term. In other words, this provides a purely algebraic proof of the classic fact that $F_{n}$ is the number of ordered sums of $1$ and $2$ adding up to $n-1$~\cite{stanley2011enumerative}, e.g.~$F_{5}=5$ since}
\begin{align}
4=&~1+1+1+1\\
=&~1+1+2= 1+2+1\\
=&~2+1+1=2+2 \ ,
\end{align}
\kg{and $F_{6} = 8$ since}
\begin{align}
5=&~1+1+1+1+1=1+1+1+2\\
=&~1+1+2+1=1+2+1+1\\
=&~2+1+1+1=1+2+2\\
=&~2+1+2=2+2+1\ .
\end{align}

\section{Generalization to $X$-symmetric qudit Pauli noise}\label{sec:general_noise}
Our analysis in the main text is based on the depolarizing channel as noise model for memory decoherence (see Section~\ref{sec:model}).
There were two key features of this noise model that made it amenable for analysis:
\begin{enumerate}
\item
{\textit{The key parameter multiplies when composing two noise channels.} 
The depolarizing noise model from Section~\ref{sec:model} is parameterized by two values: the probability of preserving the state perfectly (i.e.~$\lambda$) and the probability of transforming the state into the maximally mixed state. Due to normalization, this last probability is just $1-\lambda$. Applying a noise map with a given $\lambda_\textrm{noise}$ corresponds to the transformations $\lambda \rightarrow \lambda \cdot \lambda_\textrm{noise}$ and $1-\lambda \rightarrow 1-\lambda \cdot \lambda_\textrm{noise}$. The fact that the first map corresponds to multiplication was a key part of our analysis.
}
\item{
\textit{The key parameter multiplies when performing the entanglement swap.}
Next, we used the fact that performing a swap between two states described by noise parameters $\lambda$ and $\lambda'$, respectively, leads to a new state with noise parameter $\lambda\cdot \lambda'$.
}
\end{enumerate}

In this Appendix, we will consider the more general scenario where the nodes hold two-\emph{qudit} entangled states.
We will show for a large class of qudit channels that we call $X$-symmetric channels, how to find a parameterization into parameters $\vec{\lambda}$ such that the two above properties still hold.
To be precise, an $X$-symmetric channel $\mathcal{N}$ is a $d$-dimensional qudit Pauli channel (definition below in Section~\ref{sec:qudit-pauli-channels}) for which $p_{X^a Z^b} = p_{X^{-a} Z^b}$ for all $a, b \in \{0, 1, \dots, d-1\}$.
We will give a procedure how to find a parameterization $\mathcal{N} = \mathcal{N}_{\vec{\lambda}}$ into at most $d^2$ parameters $\vec{\lambda} = (\lambda_1, \lambda_2, \dots, \lambda_{d^n})$, such that both composition (item \ref{item:composition} below) and entanglement swapping (item \ref{item:swap}) are represented by pointwise multiplication of the entries of $\vec{\lambda}$:
\begin{enumerate}
\item{
\label{item:composition}
for any two $\vec{\lambda}, \vec{\lambda}'$, we have $\mathcal{N}_{\vec{\lambda}} \circ \mathcal{N}_{\vec{\lambda}'} = \mathcal{N}_{\vec{\lambda} \cdot \vec{\lambda}'}$.
}
\item{
\label{item:swap}
let $\psi_d = \ket{\psi_d}\bra{\psi_d}$ be a maximally entangled state on two $d$-dimensional qudits.  Then for any two $\vec{\lambda}, \vec{\lambda}'$, an entanglement swap on states 
$(\mathds{I} \otimes \mathcal{N}_{\vec{\lambda}}) \left[\psi\right]$
and
$(\mathds{I} \otimes \mathcal{N}_{\vec{\lambda}'}) \left[\psi_d\right]$
yields the state
$(\mathds{I} \otimes \mathcal{N}_{\vec{\lambda} \cdot \vec{\lambda}'}) \left[\psi_d\right]$.
}
\end{enumerate}
where $\vec{\lambda} \cdot \vec{\lambda}' = (\lambda_1\lambda_1', \ldots, \lambda_{d^2} \lambda_{d^2}')$ represents entrywise multiplication and $\mathds{I}$ is the $d$-dimensional qudit identity channel.

Consequently, the analysis in the main text is straightforwardly extendible to qudit links as long as the noise model is an $X$-symmetric channel.
The $X$-symmetric channels form a subset of qudit Pauli channels, and include (biased) depolarizing noise as well as \emph{any} qubit Pauli channel (i.e. Pauli channels for $d=2$).

In what follows we first recall the definition of qudit Pauli channels.
We then study composition of qudit Pauli channels, and show that qudit Pauli channels admit a parameterization so that channel composition is represented by multiplication of the parameters.
Finally, we investigate how maximally entangled states that have undergone qudit Pauli channels are transformed under swap operations.
From the latter, we infer that in general, the entanglement swap is not represented by parameter multiplication, but for $X$-symmetric channels, it is.

\subsection{Qudit Pauli channels}
\label{sec:qudit-pauli-channels}
 A qudit Pauli channel is any channel $\mathcal{N}$ corresponding to a probabilistic mixture of operators from the set $\mathcal{U} = \lbrace {X^aZ^b\rbrace}_{a, b=0}^{d-1}$~\cite{shahbeigi2021log}, i.e.~

\begin{align}
\mathcal{N}\left(\cdot \right) = \sum_{U\in \mathcal{U}}p_U U\left(\cdot \right)\ .
\end{align}
 Here $0\leq a \leq d-1$, $0\leq b \leq d-1$~\cite{shahbeigi2021log}, and $X$ and $Z$ are generalized Pauli operators defined by their action on the standard basis of
$\mathbb{C}^d$,

\begin{align}
X\ket{j} &= \ket{j+1\hspace{-2mm}\mod d}\ ,\\
Z\ket{j} &= \omega^j\ket{j}\ ,
\end{align}
where $\omega = e^{\frac{2\pi i}{d}}$~\cite{weyl1927quantenmechanik}.

Note that $X^aZ^b$ and $X^kZ^l$ commute up to a phase. Since $ABA^\dagger = \left(e^{i\theta}A\right)B\left(e^{i\theta}A\right)^\dagger$ for any angle $\theta \in [0, 2\pi)$, we have that $\left(X^aZ^b\right)\left(X^kZ^l\right)\rho\left(\left(X^aZ^b\right)\left(X^kZ^l\right)\right)^\dagger = \left(X^kZ^l\right)\left(X^aZ^b\right)\rho\left(\left(X^kZ^l\right)\left(X^aZ^b\right)\right)^\dagger$; qudit Pauli channels commute. Now equip the set $\mathcal{U}$ with a group structure, where the group operation is given by matrix multiplication but where phases are ignored. Formally, this is the group $\bigslant{\langle X, Z\rangle}{\langle \omega I\rangle}$, of which $\lbrace{X^{a}Z^b\rbrace}_{a, b=0}^{d-1}$ forms a transversal of the underlying equivalence relation. This group is isomorphic to $C_d \times C_d$, the direct product of the cyclic group of order $d$ with itself.

\subsection{Mapping composition of qudit Pauli channels to multiplication}
We first investigate the composition of two qudit Pauli channels, where in particular we are interested in reducing the composition to multiplication of real numbers, i.e.~the analogues of the $\lambda$ parameter for qubit depolarizing noise. The composition of two channels $\mathcal{N}_1, \mathcal{N}_2$ with distributions $p_U \equiv f(U)$ and $p_U \equiv g(U)$, respectively, can be written as 

\begin{align}
\left(\mathcal{N}_2\circ \mathcal{N}_1\right)\left[\cdot \right] &= \sum_{U\in \mathcal{U}}g(U)U\left[\sum_{V\in \mathcal{U}}f(V)V\left[\cdot \right]\right] \\
&= \sum_{W\in\mathcal{U}}\left(\sum_{UV=W}g(U)f(V)\right)W\left[\cdot \right]\ .
\label{eq:fourier-result}
\end{align}

We recognize eq.~\eqref{eq:fourier-result} as a convolution over finite Abelian groups\footnote{Fourier transforms on (non-Abelian) groups have been applied to quantum information before, e.g. in randomized benchmarking~\cite{helsen2022general}.}. 
The Fourier transform of a function $h: \mathcal{G} \rightarrow \mathbb{C}$, where $\mathcal{G}$ is a finite Abelian group, is defined as the function $\hat{h}: \hat{\mathcal{G}} \rightarrow \mathbb{C}$ such that

$$\hat{h}\left(\chi\right) = \sum_{U \in \mathcal{G}} h(U) \chi(U)^* \ .$$

The inverse Fourier transform is given by 

\begin{align}
h(U) = \frac{1}{\left|\mathcal{G}\right|}\sum_{\chi \in \hat{\mathcal{G}}}\hat{h}(\chi) \chi(U)\ .
\end{align}

Here, the $\chi$ are also called the (unitary) characters of $\mathcal{G}$, i.e. the distinct homomorphisms from $\mathcal{G}$ to the group of unit complex numbers $\mathbb{T} = \lbrace z\in \mathbb{C}\mid \left|z\right|=1\rbrace$. That is, the set of all functions $\chi: \mathcal{G}\rightarrow \mathbb{T}$ such that $\chi\left(U\right)\chi\left(V\right) = \chi\left(UV\right)$.
We denote by $\hat{\mathcal{G}}$ the set of all characters of $\mathcal{G}$.

The procedure to find the desired parameterization now is as follows.
First, we recall that the coefficients of the channel are specified through the noise coefficient map $f$, defined as $f(U) = p_U$, where $U$ is an element of the phaseless qudit Pauli group $\mathcal{U}$.
The first step is to compute the Fourier transform of $f$.
The second step is to determine the characters $\chi$ of $\mathcal{U}$.
These are the same as the characters of $C_d \times C_d$, to which the phaseless Pauli group $\mathcal{U}$ is isomorphic; in turn, we can derive the characters of $C_d \times C_d$ from the characters of $C_d$~\cite{isaacs2006character}.
Finally, we evaluate the Fourier transform $\hat{f}$ at the characters $\chi$.
We will choose the resulting $\hat{f}(\chi)$ as our new parameters $\lambda_1, \lambda_2, \ldots$, as convolution of two functions becomes pointwise multiplication of their Fourier transforms~\cite{terras1999fourier, diaconis1988group}.

As special case, we consider $d=2$, i.e. we consider an arbitrary qubit Pauli channel. In that case, $\mathcal{U}$ is isomorphic to $C_2 \times C_2$ (also known as the Klein four-group), which has characters

\begin{alignat*}{5}
    &\chi^1(I) &&= 1, \quad    \chi^1(X) &&= 1,   \quad  &&\chi^1(Y) = 1,  \quad  &&\chi^1(Z) = 1\ ,\\
    &\chi^2(I) &&= 1, \quad    \chi^2(X) &&= 1,   \quad  &&\chi^2(Y) = -1, \quad  &&\chi^2(Z) = -1\ ,\\
    &\chi^3(I) &&= 1, \quad    \chi^3(X) &&= -1,  \quad   &&\chi^3(Y) = 1,  \quad &&\chi^3(Z) = -1\ ,\\
    &\chi^4(I) &&= 1, \quad    \chi^4(X) &&= -1,   \quad  &&\chi^4(Y) = -1, \quad  &&\chi^4(Z) = 1\ .
\end{alignat*}

From normalization of the probability, we see that $\lambda_1$ is always equal to 1. The remaining parameters are

\begin{align}
\lambda_2 = p_I + p_X - p_Y - p_Z\ ,\\
\lambda_3 = p_I - p_X + p_Y - p_Z\ ,\\
\lambda_4 = p_I - p_X - p_Y + p_Z\ \ 
\end{align}

which is indeed the parameterization given in the main text in Section~\ref{sec:model}.

Let us specialize to the case of depolarizing noise, where $p_X = p_Y = p_Z = \frac{1-p_I}{3}$. We then see that $\lambda^2 = \lambda^3 = \lambda^4 = p_I - \frac{1-p_I}{3} = \frac{4p_I-1}{3} = \frac{4F-1}{3}$, which we see is the original $\lambda$ treated in the main text.

\subsection{Swapping of states}
Let $\ket{\psi_d}\equiv \frac{1}{\sqrt{d}}\sum_{i=0}^{d-1}\ket{ii}$ be a qudit maximally entangled state and set $\ket{\psi_d}\bra{\psi_d} \equiv \psi_d$. Let $\mathcal{N}$ be a qudit Pauli channel acting on one-half of a qudit maximally entangled state. We will first show that the side the channel acts on is immaterial (i.e.~$\mathds{I}\otimes \mathcal{N} \left[\psi_d\right]= \mathcal{N}\otimes \mathds{I} \left[\psi_d\right]$) if $\mathcal{N}$ is $X$-symmetric (which we define shortly). Thus even if $X$-symmetric noise has acted on both sides of $\ket{\psi_d}$, we are free to pretend there is one qudit Pauli channel $\mathcal{N}$ acting on an arbitrary side, yielding a state $ \mathds{I}\otimes \mathcal{N}\left[\psi_d\right] = \mathcal{N}\otimes \mathds{I}\left[\psi_d\right] \equiv  \rho_\mathcal{N}$. Secondly, we will show that swapping two states $\rho_{\mathcal{N}_1}, \rho_{\mathcal{N}_2}$ yields a state $\rho_{\left(\mathcal{N}_1\circ \mathcal{N}_2\right)}$.

\subsubsection{Transpose trick for $X$ symmetric channels}
The well-known transpose trick~\cite{wilde2011classical} states that if $A$ is any square matrix (of the appropriate size), the following holds  

\begin{align}
\mathds{I} \otimes A \ket{\psi_d} = A^\transp \otimes \mathds{I} \ket{\psi_d}\ .    
\end{align}
Here $A^\transp$ is the transpose of $A$ with respect to the basis $\lbrace{\ket{i}\rbrace}_{i=1}^{d-1}$. Applying the transpose trick to $\mathds{I}\otimes \mathcal{N}\left[\psi_d\right]$ yields

\begin{align}
    \mathds{I}\otimes \mathcal{N}\left[\psi_d\right] &=\sum_{a, b=0}^{d-1} p_{X^aZ^b} \left(\mathds{I}\otimes X^aZ^b\right)\left[\psi\right]\\
    & = \sum_{a, b=0}^{d-1} p_{X^aZ^b} \left(\left(X^aZ^b\right)^\transp\otimes \mathds{I}\right)\left[\psi\right]\\
    & = \sum_{a, b=0}^{d-1} p_{X^aZ^b} \left(X^{-a}Z^b \otimes \mathds{I}\right)\left[\psi\right]\ ,
\end{align}

where we used that $\left(X^aZ^b\right)^\transp = Z^{b}X^{-a}$, which is equal to $X^{-a}Z^b$ up to a phase.

The above implies that the side a qudit Pauli channel is applied to is immaterial if and only if $p_{X^aZ^b} = p_{X^{-a}Z^b}$. This means that the condition holds when the probability of applying $X^aZ^b$ is equal to the probability of applying $X^{-a}Z^b$. This property holds for several relevant classes of channels, such as arbitrary qubit Pauli channels, arbitrary qudit dephasing noise (since the probability of applying $X^aZ^b$ is zero when $a \neq 0$) and qudit depolarizing noise.

\begin{figure}[h]
\centerfloat
\begin{tikzpicture}[line cap=round]
\node[scale=0.8, fill=white] at (0,0.15){$a)$};
\draw [opacity=0, -{Stealth[length=1mm, width=1.6mm]}] (-0.82,-1.25) -- (-.6,-1.25);
\node[scale=0.8, fill=white] at (-0.05,-0.5){$\ket{\psi_d}$};
\node[scale=0.8, fill=white] at (-0.05,-2){$\ket{\psi_d}$};

\draw[line width=1pt] (0.25,-0.5 ) -- (1-0.5,0);
\draw[line width=1pt] (0.25,-0.5 ) -- (0.5,-1);
\draw[line width=1pt] (0.5,0.0) -- (3,0.0);
\draw[line width=1pt] (0.5,-1) -- (1.75,-1);
\node[draw, scale=1.0, fill=white] at (1.5-0.5,-1){$\mathcal{N}_1$};

\draw[line width=1pt] (0.25,-0.5-1.5) -- (1-0.5,0-1.5);
\draw[line width=1pt] (0.25,-0.5-1.5) -- (0.5,-1-1.5);
\draw[line width=1pt] (0.5,0.0-1.5) -- (1.75,0.0-1.5);
\draw[line width=1pt] (0.5,-1-1.5) -- (3,-1-1.5);
\node[draw, scale=1.0, fill=white] at (1.5-0.5,-1.5){$\mathcal{N}_2$};

\draw[dotted]    (2.5-0.35,-1.25) to[out=0,in=90] (2.5, -2.2);
\node[draw, scale=1.0, fill=white] at (2.5,-2.5){$U_k^{\dagger}$};
\node[scale=0.8, fill=white] at (2.5-0.4,-1){$k$};

\filldraw[draw=black, fill=LimeGreen!20] (2-0.4, -2+0.25) -- (2-0.4, -0.5-0.25) -- (2.5-0.4, -1.25) -- cycle;
\draw [-{Stealth[length=1mm, width=1.6mm]}] (3,-1.25) -- (3.3,-1.25);
\end{tikzpicture}
\hspace{1mm}
\begin{tikzpicture}[line cap=round]
\node[scale=0.8, fill=white] at (0,0.15){$b)$};
\node[scale=0.8, fill=white] at (-0.05,-0.5){$\ket{\psi_d}$};
\node[scale=0.8, fill=white] at (-0.05,-2){$\ket{\psi_d}$};

\draw[line width=1pt] (0.25,-0.5 ) -- (1-0.5,0);
\draw[line width=1pt] (0.25,-0.5 ) -- (0.5,-1);
\draw[line width=1pt] (0.5,0.0) -- (3,0.0);
\draw[line width=1pt] (0.5,-1) -- (1.75,-1);
\node[draw, scale=1.0, fill=white] at (1.5-0.5,0){$\mathcal{N}_1$};

\draw[line width=1pt] (0.25,-0.5-1.5) -- (1-0.5,0-1.5);
\draw[line width=1pt] (0.25,-0.5-1.5) -- (0.5,-1-1.5);
\draw[line width=1pt] (0.5,0.0-1.5) -- (1.75,0.0-1.5);
\draw[line width=1pt] (0.5,-1-1.5) -- (3,-1-1.5);
\node[draw, scale=1.0, fill=white] at (1.5-0.5,-2.5){$\mathcal{N}_2$};

\draw[dotted]    (2.5-0.35,-1.25) to[out=0,in=90] (2.5, -2.2);
\node[draw, scale=1.0, fill=white] at (2.5,-2.5){$U_k^{\dagger}$};
\node[scale=0.8, fill=white] at (2.5-0.4,-1){$k$};

\filldraw[draw=black, fill=LimeGreen!20] (2-0.4, -2+0.25) -- (2-0.4, -0.5-0.25) -- (2.5-0.4, -1.25) -- cycle;
\draw [-{Stealth[length=1mm, width=1.6mm]}] (3,-1.25) -- (3.3,-1.25);
\end{tikzpicture}
\hspace{1mm}
\begin{tikzpicture}[line cap=round]
\node[scale=0.8, fill=white] at (0,0.15){$c)$};
\node[scale=0.8, fill=white] at (-0.05,-0.5){$\ket{\psi_d}$};
\node[scale=0.8, fill=white] at (-0.05,-2){$\ket{\psi_d}$};
\draw[line width=1pt] (0.25,-0.5 ) -- (1-0.5,0);
\draw[line width=1pt] (0.25,-0.5 ) -- (0.5,-1);
\draw[line width=1pt] (0.5,0.0) -- (3,0.0);
\draw[line width=1pt] (0.5,-1) -- (0.8,-1);
\node[draw, scale=1.0, fill=white] at (1.7,0){$\mathcal{N}_1$};

\draw[line width=1pt] (0.25,-0.5-1.5) -- (1-0.5,0-1.5);
\draw[line width=1pt] (0.25,-0.5-1.5) -- (0.5,-1-1.5);
\draw[line width=1pt] (0.5,0.0-1.5) -- (0.8,0.0-1.5);
\draw[line width=1pt] (0.5,-1-1.5) -- (3,-1-1.5);
\node[draw, scale=1.0, fill=white] at (1.7,-2.5){$\mathcal{N}_2$};

\draw[dotted]    (2.5-0.35-1,-1.25) to[out=0,in=90] (2.5, -2.2);
\node[draw, scale=1.0, fill=white] at (2.5,-2.5){$U_k^{\dagger}$};
\node[scale=0.8, fill=white] at (2.5-0.4-1,-1){$k$};

\filldraw[draw=black, fill=LimeGreen!20] (2-0.4-1, -2+0.25) -- (2-0.4-1, -0.5-0.25) -- (2.5-1-0.4, -1.25) -- cycle;

\end{tikzpicture}
\vspace{5mm}

\begin{tikzpicture}[line cap=round]
\node[scale=0.8, fill=white] at (0,0.15){$d)$};
\draw [opacity=1, -{Stealth[length=1mm, width=1.6mm]}] (-1,-1.25) -- (-.7,-1.25);
\node[scale=1, fill=white] at (-0.1,-1.25){$\ket{\psi_d}$};
\draw[line width=1pt] (0.25,-1.25 ) -- (1-0.5,0);
\draw[line width=1pt] (0.5,0.0) -- (3,0.0);
\node[draw, scale=1.0, fill=white] at (1.7,0){$\mathcal{N}_1$};

\draw[line width=1pt] (0.25,-1.25) -- (0.5,-1-1.5);
\draw[line width=1pt] (0.5,-1-1.5) -- (3,-1-1.5);
\node[draw, scale=1.0, fill=white] at (1.7,-2.5){$\mathcal{N}_2$};


\node[draw, scale=1.0, fill=white] at (0.9,-2.5){$U_k$};
\node[draw, scale=1.0, fill=white] at (2.5,-2.5){$U_k^{\dagger}$};

\draw [-{Stealth[length=1mm, width=1.6mm]}] (3,-1.25) -- (3.3,-1.25);
\end{tikzpicture}
\begin{tikzpicture}[line cap=round]
\node[scale=0.8, fill=white] at (0,0.15){$e)$};
\node[scale=1, fill=white] at (-0.1,-1.25){$\ket{\psi_d}$};
\draw[line width=1pt] (0.25,-1.25 ) -- (1-0.5,0);
\draw[line width=1pt] (0.5,0.0) -- (3,0.0);
\node[draw, scale=1.0, fill=white] at (1.7,0){$\mathcal{N}_1$};

\draw[line width=1pt] (0.25,-1.25) -- (0.5,-1-1.5);
\draw[line width=1pt] (0.5,-1-1.5) -- (3,-1-1.5);
\node[draw, scale=1.0, fill=white] at (1.7,-2.5){$\mathcal{N}_2$};



\draw [-{Stealth[length=1mm, width=1.6mm]}] (3,-1.25) -- (3.3,-1.25);
\end{tikzpicture}
\hspace{1mm}
\begin{tikzpicture}[line cap=round]
\node[scale=0.8, fill=white] at (0,0.15){$f)$};

\node[scale=1, fill=white] at (-0.1,-1.25){$\ket{\psi_d}$};
\draw[line width=1pt] (0.25,-1.25 ) -- (1-0.5,0);
\draw[line width=1pt] (0.5,0.0) -- (3,0.0);

\draw[line width=1pt] (0.25,-1.25) -- (0.5,-1-1.5);
\draw[line width=1pt] (0.5,-1-1.5) -- (3,-1-1.5);
\node[draw, scale=1.0, fill=white] at (1.3333,-2.5){$\mathcal{N}_1$};
\node[draw, scale=1.0, fill=white] at (2.16666667,-2.5){$\mathcal{N}_2$};



\draw [opacity=0.0, -{Stealth[length=1mm, width=1.6mm]}] (3,-1.25) -- (3.3,-1.25);
\end{tikzpicture}
\caption{Graphical description of the proof of the desired properties of $X$-symmetric channels. From a) to b) we use the transpose trick to show that the side on which the noise is applied is immaterial. From b) to c) we use that the application of the noise and the Bell state measurement commute. From c) to d) we performed the projection into one of the (generalized) Bell states. From d) to e) we use the commutativity of qudit Pauli operators (up to a phase). From e) to f) we use the transpose trick again to move the noise all to one side.}
\label{fig:lambdaproof}
\end{figure}
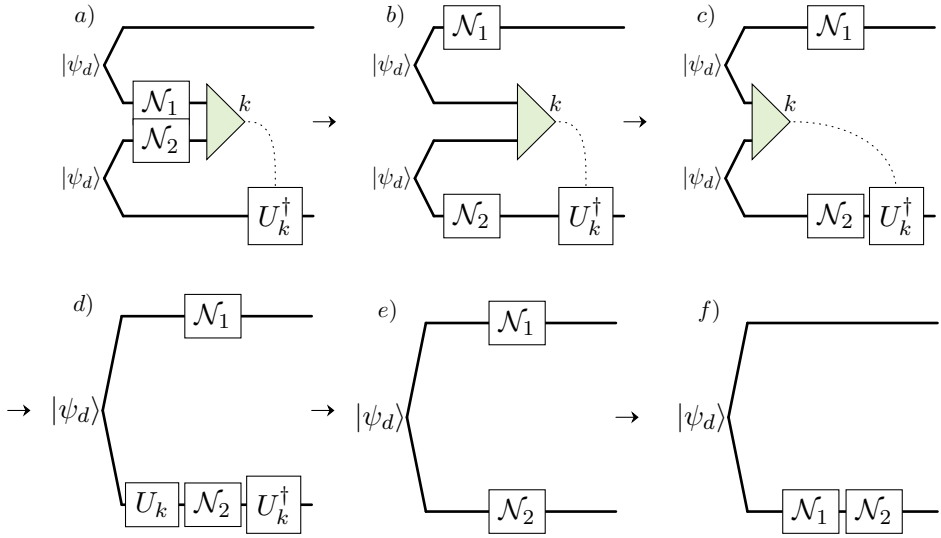

\subsubsection{Swapping of $X$-symmetric noisy states}
In the previous section we have shown that we have the freedom to move $X$-symmetric channels from one side of the maximally entangled state to the other. With this tool in hand, let us perform a swap on two such noisy states, see Fig.~\ref{fig:lambdaproof}. In a) we show the complete procedure of swapping, including the measurement and classical correction. We have assumed without loss of generality that the channels $\mathcal{N}_1$ and $\mathcal{N}_2$ act on the qudits on which the Bell state measurement (green triangle) is performed. After the Bell state measurement, a classical outcome $k$ is recorded, and is used to specify which correction $U_k^{\dagger}$ needs to be performed on one of the sides after measuring. Note that the corrections $U_k^{\dagger}$ are generalized Pauli operators.

In step b) we use the result from the previous section that we are free to move $X$-symmetric channels to the other side. In step c), we commute the Bell state measurement and the channels, which is allowed since the operations act on different qubits. In step d) we use that measuring outcome $k$ projects the state unto $\mathds{I}\otimes U_k\ket{\psi_d}$. From d) to e) we use that the qudit Pauli operators commute (up to an immaterial phase), and finally we use that $X$-symmetric channels can be moved freely between the two sides.

We close with one final remark. The average values for the $\lambda_i$ parameters can be interpreted as the $\lambda_i$ parameters of the average channel acting on the entangled state shared between Alice and Bob. This is because convex combinations of qudit Pauli channels map exactly to convex combinations of the $\vec{\lambda}$, since the Fourier transform is linear.

\section{Recursion calculations}\label{sec:app_recursion}
Here we provide the details for the recursive calculation for the analytics of the average noise parameter.

\subsection{Recursion calculation without a cut-off policy}
As discussed in Section~\ref{sec:recursion}, calculating the average $\Lambda_n$ parameter requires us first to express $H_n^t$ recursively, from which $H_n\equiv \mathds{E}\left[\Lambda_n\right]$ can be found. The main idea is to express the map from $H_n^t$ to $H_{n+1}^t$ in terms of a linear transformation on the real vector space spanned by $q^{at}\lambda^{bt},~a, b \in \mathbb{Z}$. That is, all finite linear combinations of the form $\sum_{m}c_{a, b} q^{at}\lambda^{bt} $, where $c_{a,b}\in \mathbb{R}$.

As a warm-up, let us consider the first recursion step. Here and in what follows, we will drop the $\frac{1-q}{q}$ prefactor, i.e.~we set

\begin{align}
 Z_n &= \sum_{\overline{t}}\left(q^{\sum_{i=1}^nt_i}\right)\cdot\left( \lambda^{\sum_{i=1}^{n-1}\left|t_i-t_{i+1}\right|}\right)\\
Z_n^{t} &= \sum_{\mathclap{\substack{\overline{t}~\textrm{s.t.}\\t_n =t}}}\left(q^{\sum_{i=1}^nt_i}\right)\cdot\left( \lambda^{\sum_{i=1}^{n-1}\left|t_i-t_{i+1}\right|}\right)\ ,
\end{align}
such that $H_n = \left(\frac{1-q}{q}\right)^nZ_n$. Note that $Z_1^t$ is given by $q^t$, since there is only the associated probability of succeeding at time $t$ and no decoherence.

Let $M$ correspond to the map in the recursion, i.e.~the map $Z_{i}^t\xrightarrow{M}Z_{i+1}^t$.
We then have for the first step that

\begin{align}
    Z_1^t\xrightarrow{M}Z_2^{t} =& \sum_{t'=1}^{\infty}q^t\lambda^{\left|t-t'\right|}Z_{1}^{t'}\\
    =&\left(\,\sum_{t'=1}^{t-1}q^t\lambda^{t-t'}Z_{1}^{t'}\right) + \left(\,\sum_{t'=t}^{\infty}q^t\lambda^{t'-t}Z_{1}^{t'}\right)\\
    =&\left(\,\sum_{t'=1}^{t-1}q^t\lambda^{t-t'}q^{t'}\right) + \left(\,\sum_{t'=t}^{\infty}q^t\lambda^{t'-t}q^{t'}\right)\\
     =&~q^t\lambda^t\left(\,\sum_{t'=1}^{t-1}\left(\frac{q}{\lambda}\right)^{t'}\right) + \left(\frac{q}{\lambda}\right)^t\left(\,\sum_{t'=t}^{\infty}\left(q\lambda \right)^{t'}\right)\\
          =&~q^t\lambda^t\left(\frac{\frac{q}{\lambda}-\left(\frac{q}{\lambda}\right)^t}{1-\frac{q}{\lambda}}\right) + \left(\frac{q}{\lambda}\right)^t   \left(\frac{q\lambda-\left(q\lambda\right)^t}{1-q\lambda}\right)\\
    = &\left(-\frac{q\left(\frac{1}{\lambda}-\lambda\right)}{\left(1-\lambda^{-1}q\right)\left(1-\lambda q\right)}\right)\cdot  q^{2t}\lambda^{0t} +\frac{q}{\lambda-q}\cdot q^t\lambda^t\ .
\end{align}

Note that the limit as $q$ goes to $\lambda$ and the limit of $q$ goes $\frac{1}{\lambda}$ exists.
Observe that we end up with a linear combination of terms of the form $q^{at}\lambda^{bt}$. This is not specific to starting with $Z_1^t =  q^t$, as we see in the following,

\begin{align}
q^{at}\lambda^{bt} \xrightarrow{M}& \sum_{t'=1}^{\infty}q^t\lambda^{\left|t-t'\right|}\cdot \left(q^{at'}\lambda^{bt'}\right)\label{eq:tmap}\\
    =&\left(\,\sum_{t'=1}^{t-1}q^t\lambda^{t-t'}\cdot \left(q^{at'}\lambda^{bt'}\right)\right)+
 \left(\,\sum_{t'=t}^{\infty}q^t\lambda^{t'-t}\cdot \left(q^{at'}\lambda^{bt'}\right)\right)\\
 =&~C^{a, b} q^{\left(a+1\right)t}\lambda^{bt} + D^{a, b} q^t\lambda^t\ ,
\end{align}

where

\begin{align}
C^{a,b} &= -\frac{q^a\lambda^b \left(\frac{1}{\lambda}-\lambda\right)}{\left(1-\lambda^{b-1}q^a\right)\left(1-\lambda^{b+1}q^a\right)}\ \\
D^{a,b} &= \frac{q^a}{\lambda^{1-b}-q^a}\ .
\end{align}
The $C^{a, b}$ and $D^{a, b}$ depend only on the values of $a, b\in \mathbb{Z}$ and $q$ and $\lambda$. 

We can think of the $q^{at}\lambda^{bt} \equiv \ket{a, b}$ for $a, b \in \mathbb{Z}$ as the basis vectors of a vector space $V$. The map $M$ defined for an arbitrary $\ket{a, b}$ is, in fact, a linear map from $V$ to $V$, and so works for all expressions that are sums of terms of the form $q^{at}\lambda^{bt}$.

This allows us to calculate $Z_n^t$ fast. Setting $q^{at}\lambda^{bt}\equiv \ket{a, b}$, the map $M$ takes $\ket{a, b}$ to $C^{a,b}\ket{a+1, b} + D^{a,b}\ket{1, 1}$.
As such, we find that

\begin{gather}
    Z_1^t = \ket{1, 0}\\
    \xrightarrow{M}Z_2^t = C^{1, 0}\ket{2, 0}+D^{1, 0}\ket{1, 1} \\
    \xrightarrow{M}Z_3^t = C^{1, 0}\left(C^{2, 0}\ket{3, 0}+ D^{2, 0}\ket{1, 1}\right)+D^{1, 0}\left(C^{1, 1}\ket{2,1}+D^{1, 1}\ket{1, 1}\right)\\
	=
    C^{1, 0}C^{2, 0}\ket{3, 0}
    +D^{1, 0}C^{1, 1}\ket{2, 1}+\left(C^{1, 0}D^{2, 0}+D^{1, 0}D^{1, 1}\right)\ket{1, 1}\label{eq:3linksexp}\ .
\end{gather}

This already gives us a method to fairly straightforwardly compute $Z_n^t$, especially when done on a computer.

Finally, we need to extract $Z_n^t$ from $Z_n$. Fortunately, since

\begin{align}
\sum_{t=1}^{\infty}q^{at}\lambda^{bt} = \frac{q^{a}\lambda^{b}}{1-q^{a}\lambda^{b}}\ \label{eq:lform1},
\end{align}
this can be seen as a linear form defined on the basis vectors by $\ket{a, b} \mapsto \frac{q^{a}\lambda^{b}}{1-q^a\lambda^b}$. Using the above idea and equation \eqref{eq:3linksexp}, we find that the expectation value for $\Lambda_3$ for a homogeneous repeater chain of three segments is given by

\begin{gather}
\mathds{E}\left[\Lambda_3\right] = \left(\frac{1-q}{q}\right)^3\left(C^{1, 0}C^{2, 0}\frac{q^{3}}{1-q^{3}}
    +D^{1, 0}C^{1, 1}\frac{q^{2}\lambda}{1-q^{2}\lambda}+\left(C^{1, 0}D^{2, 0}+D^{1, 0}D^{1, 1}\right)\frac{q\lambda}{1-q\lambda}\right)\label{eq:3linksexp_total} \ .
    \end{gather}

\subsection{Recursion calculation with a global cut-off policy}\label{sec:recursion_cutoff}

We take the same approach as in the previous section; the only difference now is that $t'$ ranges from $1$ to $T_c$. Define

\begin{align}
 \bar{Z}_n^{t} &= \sum_{\mathclap{\substack{\overline{t}~\textrm{s.t.}\\t_n =t\\t_i \leq T_c}}}\left(q^{\sum_{i}^nt_i}\right)\cdot\left( \lambda^{\sum_{i}^{n-1}\left|t_i-t_{i+1}\right|}\right)\ ,
\end{align}

such that $\bar{Z}_n = \sum_{t=1}^{T_c} \bar{Z}_n^{t}$ and $\bar{H}_n = \left(\frac{1-q}{q\left(1-q^{T_c}\right)}\right)^n\bar{Z}_n$.

This gives the following recursion relation,

\begin{align}
q^{at}\lambda^{bt} \xrightarrow{\bar{M}}& \sum_{t'=1}^{T_c}q^t\lambda^{\left|t-t'\right|}\cdot \left(q^{at'}\lambda^{bt'}\right)\\
    =&\left(\sum_{t'=1}^{t-1}q^t\lambda^{t-t'}\cdot \left(q^{at'}\lambda^{bt'}\right)\right)+
 \left(\sum_{t'=t}^{T_c}q^t\lambda^{t'-t}\cdot \left(q^{at'}\lambda^{bt'}\right)\right)\label{eq:cutoff_step}\\
 =&~C^{a, b} q^{\left(a+1\right)t}\lambda^{bt} + D^{a, b} q^t\lambda^t + E^{a, b}q^t \lambda^{-t}\\
 =&~C^{a, b} \ket{a+1, b} + D^{a, b} \ket{1, 1} + E^{a, b}\ket{1, -1}\ ,
\end{align}
where $C^{a, b}$ and $D^{a,b}$ are as before and 

\begin{gather}
E^{a, b} = -\frac{q^{a(1+T_c)}\lambda^{\left(b+1\right)\left(T_c+1\right)}}{1-\lambda^{b+1}q^a} \ .
\end{gather}
As in the previous section, by repeating this map $\bar{Z}_n^t$ can be calculated. Finally, since $\sum_{t'=1}^{T_c}q^{at}\lambda^{bt} = \frac{q^a\lambda^b - q^{a\left(T_c+1\right)}\lambda^{b\left(T_c+1\right)}}{1-q^a\lambda^b}$ we find that the linear form that takes $\bar{Z}_n^t$ to $\bar{Z}_n$ is defined by $\ket{a, b} \mapsto \frac{q^a\lambda^b - q^{a\left(T_c+1\right)}\lambda^{b\left(T_c+1\right)}}{1-q^a\lambda^b}$.

\section{Expected delivery time and secret-key rate calculation}\label{sec:exp_delivery_time}
The secret-key rate is defined as the ratio between the secret-key fraction and the average delivery time. In the following we discuss how to calculate the secret-key fraction, how to determine the average delivery time when using a global cut-off policy, and finally an upper and lower bound to the average delivery time.

\subsection{Secret-key fraction} \label{sec:secret-key-rate-calculation}
The secret-key rate is defined as the ratio between the secret-key fraction and the average delivery time. The secret-key fraction is the ratio between the number of secret bits that can be extracted and the number of measured states.
The secret-key fraction depends on the used protocol. Here we consider fully asymmetric BB84~\cite{bennett2020quantum, rozpkedek2018parameter, rozpkedek2019near, scarani2009security}. The `BB84' refers to the original quantum key distribution protocol~\cite{bennett2020quantum}, where both parties measure in two bases ($X$ and $Z$). The `fully asymmetric' refers to the fact that only in a vanishingly small subset of cases either of the two parties measures in $X$. Furthermore, we will also assume that we run the protocol for an asymptotic number of rounds, such that finite-size effects can be ignored.

For fully asymmetric BB84, the secret-key fraction can be expressed in terms of $\Lambda$ as follows~\cite{scarani2009security},

\begin{align}
\textrm{SKF}\left(\Lambda\right) = \max\left(0,\,\,1-2\cdot H\left(\frac{1-\Lambda}{2}\right)\right)\ ,\\
\textrm{where } H(p) = -p\log_2(p)-(1-p)\log_2(1-p) \ .
\end{align}

\subsection{Expected delivery time}
Calculating the secret-key rate above, in Section~\ref{sec:secret-key-rate-calculation}, includes the expected delivery time.
Without a cut-off, this is given by

\begin{align}
    \mathds{E}\left[\max\left(T_1, T_2, \ldots, T_n\right)\right]= \sum_{k=1}^{n}\binom{n}{k}\frac{(-1)^{k+1}}{1-\left(1-p\right)^k}\ ,
\end{align}
see~\cite{shchukin2019waiting, bernardes2011rate}.

For the cut-off case, we first define an `attempt' as the consecutive collection of rounds in which entanglement generation was attempted before a success or a reset.
In the presence of a global cut-off, the expected delivery time is the sum of two terms: the time that the protocol has spent in attempts that were eventually restarted because the cut-off was met, and the time until successful entanglement generation in the final attempt, which occurred before the cut-off time.

For the first term, we note that the expected number of resets is $\frac{1}{P_{\textrm{succ}}}-1$, where $P_{\textrm{succ}} = \left(1-(1-p)^{T_c}\right)^n$ is the probability that the cut-off is not reached in a single attempt and $T_c$ the cut-off time.
Since every time a reset is enforced $T_c$ rounds have been performed, the first term equals $\left(\frac{1}{P_{\textrm{succ}}}-1\right)T_c$, 

The second term is the expectation value of the delivery time, \textit{conditioned} on the fact that entanglement generation succeeds before $T_c$:

\begin{align}
&
\sum_{\mathclap{t_\textrm{max}=1}}^{T_c} t_{\textrm{max}} \cdot
	\left(
	\frac{f(t_{\max})}{
\left(1-(1-p)^{T_c}\right)^n
	}
	\right)
\end{align}
where
\[
	f(t_{\max}) = \left(1-\left(1-p\right)^{t_\textrm{max}}\right)^n-\left(1-\left(1-p\right)^{t_\textrm{max}-1}\right)^n
\]
is the probability that an attempt succeeds exactly at timestep $t_{\max}$, and the denominator $\left(1-(1-p)^{T_c}\right)^n$ is the probability that entanglement generation succeeds before $T_c$.

Combining the two terms, we find that the expected delivery time in the presence of a cut-off $T_c$ is given by

\begin{align}
\left(\frac{1}{\left(1-(1-p)^{T_c}\right)^n}-1\right)T_c + \sum_{\mathclap{t_\textrm{max}=1}}^{T_c} t_{\textrm{max}} \cdot
	\left(
	\frac{f(t_{\max})}{
\left(1-(1-p)^{T_c}\right)^n
	}
	\right)\ .
	\label{eq:delivery-time-with-cutoff}
\end{align}

\kg{From here we straightforwardly calculate the generation time when all segments always wait until the cut-off timer has been reached before starting entanglement generation,}

\begin{align}
\frac{T_c}{\left(1-(1-p)^{T_c}\right)^n}\ .
\end{align}

\section{Derivation of the generating function}\label{sec:appendix_gen_function}
We will show here the derivation of the function discussed in Section~\ref{sec:gen_function}, i.e.~a closed-form expression for

\begin{align}
G \equiv &\sum_{n=1}^{\infty}\mathds{E}\left[\Lambda_n\right]x^n \ .
\end{align}

It will turn out to be more convenient to consider the sum
\begin{align}
\overline{G} = &\sum_{n=1}^{\infty}Z_nx^n\ ,\\
\textrm{where } Z_n = &\sum_{\overline{t}}\left(q^{\sum_{i=1}^nt_i}\right)\cdot\left( \lambda^{\sum_{i=1}^{n-1}\left|t_i-t_{i+1}\right|}\right)\ \label{eq:def_G}.
\end{align}

Here, $\overline{t}$ runs over all length-$n$ sequences of strictly positive integers, i.e. all possible combinations of times at which a link is generated over the $n$ repeater segments individually.
Since $\mathds{E}\left[\Lambda_n\right] = \left(\frac{1-q}{q}\right)^n Z_n$, we have that

\begin{align}
    G(x) = \overline{G}\left(x \left(\frac{1-q}{q}\right) \right)\ \label{eq:replaceq} \ .
\end{align}

That is, by replacing $x$ with $x\left(\frac{1-q}{q}\right)$ in $\overline{G}$ we retrieve $G$, the function of interest.

Before continuing, let us make a few technical remarks. In analytic combinatorics, \emph{combinatorial classes} are studied through the analytical properties of their associated generating function. A combinatorial class can be thought of as a countable collection of objects such that the number of objects with a given `size' $n$ is finite. However, the number of allowed $\overline{t}$ of a fixed length $n$ is infinite when no cut-off is considered. Fortunately, the analytical and combinatorical tools that we employ remain valid. Furthermore, the function $G$ is technically a \emph{multivariate} generating function~\cite{pemantle2013analytic}, since the sum has not only a variable $x$ (corresponding to the number of entries $n$ in $\overline{t}$), but also two auxiliary variables $q$ and $\lambda$. Finally, a key idea of analytic combinatorics is to treat generating functions as formal power series, ignoring issues of convergence. This can be justified rigorously; see, for example,~\cite{flajolet2009analytic}.

In what follows, we first find an expression of $G$ in terms of so-called $q$-hypergeometric series, after which we find a similar expression for the case of a global cut-off, corresponding to restricting the sum in Eq.~\eqref{eq:def_G} to those $\overline{t}$ satisfying $\max(\overline{t}) \leq T_c$.

\subsection{No cut-off}
We now give a high-level overview of the proof. First, we will use the recursive relation found in Section~\ref{sec:app_recursion} that relates $Z_n^t$ to $Z_{n+1}^t$. This relation can be thought of as a linear map on the vector space spanned by basis vectors of the form $\ket{a, b}\equiv q^{at}\lambda^{bt}$ to itself. Using the well-known correspondence between linear maps and unnormalized Markov chains, the calculation of $Z_n^t$ can be formulated in terms of walks of length $n-1$ on an unnormalized Markov chain. The possible walks on the Markov chain can be split into two disjoint types. The first type is easy to characterize; the second type is more involved. In particular, the second type is the composition of three separate subwalks, the second of which can repeat an arbitrarily large number of times before moving to the third subwalk. The composition of subwalks is most naturally understood in terms of generating functions. That is, the product of the generating functions corresponding to the three aforementioned subwalks keeps track of the relevant data for the second type, as we shall see.

Let us start the proof by recalling the recursive relation from Section~\ref{sec:app_recursion}. As before, setting $q^{at}\lambda^{bt} \equiv \ket{a, b}$, we have $Z_1^t = \ket{1, 0}$. Each subsequent $Z_n^t$ can be found by repeatedly applying the linear map $M$, which is defined on the basis vectors by $\ket{a, b} \xrightarrow{M} C^{a,b}\ket{a+1, b} + D^{a,b}\ket{1, 1}$, with

\begin{align}
C^{a,b} &= -\frac{q^a\lambda^b \left(\frac{1}{\lambda}-\lambda\right)}{\left(1-\lambda^{b-1}q^a\right)\left(1-\lambda^{b+1}q^a\right)}\  ,\\
D^{a,b} &=  \frac{q^a}{\lambda^{1-b}-q^a}\ .
\end{align}

With the above approach, we found in Sec.~\ref{sec:app_recursion}, Eq.~\eqref{eq:3linksexp} for example that

\begin{gather}
    Z_1^t = \ket{1, 0}\\
    \xrightarrow{M}Z_2^t = C^{1, 0}\ket{2, 0}+D^{1, 0}\ket{1, 1} \\
    \xrightarrow{M}Z_3^t = C^{1, 0}\left(C^{2, 0}\ket{3, 0}+ D^{2, 0}\ket{1, 1}\right)+D^{1, 0}\left(C^{1, 1}\ket{2,1}+D^{1, 1}\ket{1, 1}\right)\\
	=
    C^{1, 0}C^{2, 0}\ket{3, 0}
    +D^{1, 0}C^{1, 1}\ket{2, 1}+\left(C^{1, 0}D^{2, 0}+D^{1, 0}D^{1, 1}\right)\ket{1, 1}\ \label{eq:example_eq}.
\end{gather}

\begin{figure}[h]
\centerfloat
\begin{tikzpicture}[node distance=1.8cm]
\node [circle , draw ] (10) at (-2 , 0) {$\ket{1, 0}$};
\node [circle , draw ] (20) [ above of= 10] {$\ket{2, 0}$};
\node [circle , draw ] (30) [ above of= 20] {$\ket{3, 0}$};
\node [] (40) [ above of= 30, rotate=90] {$\cdots$};

\node [circle , draw ] (11)  at (2, 0) {$\ket{1, 1}$};
\node [circle , draw] (21) [ above of= 11] {$\ket{2, 1}$};
\node [circle , draw ] (31) [ above of= 21] {$\ket{3, 1}$};
\node [] (41) [ above of= 31, rotate=90] {$\cdots$};

\path[->] (10) edge node [left] {$C^{1, 0}$} (20) ;
\path[->] (20) edge node [left] {$C^{2, 0}$} (30) ;
\path[->] (30) edge node [left] {$C^{3, 0}$} (40) ;

\path[->] (10) edge node [below] {$D^{1, 0}$} (11) ;
\path[->] (20) edge node [above=0.1cm] {$D^{2, 0}$} (11) ;
\path[->] (30) edge node [above=0.35cm] {$D^{3, 0}$} (11) ;

\path[->] (11) edge node [left] {$C^{1, 1}$} (21) ;
\path[->] (21) edge node [left] {$C^{2, 1}$} (31) ;
\path[->] (31) edge node [left] {$C^{3, 1}$} (41) ;


\path (11) edge [loop below] node[below]{$D^{1, 1}$} (11);
\path[->] (21) edge [ bend left ] node [right] {$D^{2,1}$} (11) ;
\path[->] (31) edge [ bend left=80] node [right] {$D^{3,1}$} (11) ;


\end{tikzpicture}
\caption{An unnormalized Markov chain with transition `probabilities' given by the $C^{a, b}$ and $D^{a, b}$. The starting point is $\ket{1, 0}$. There are two types of finite walks, those that stay in the left branch (corresponding to $b=0$), and those that transition to the right branch at a certain point. The second type can be decomposed into three parts---a walk staying in the left branch, then any number of ascents followed by a transition back to $\ket{1, 1}$, and finally one more ascent.}
\label{fig:markovchain1}
\end{figure}
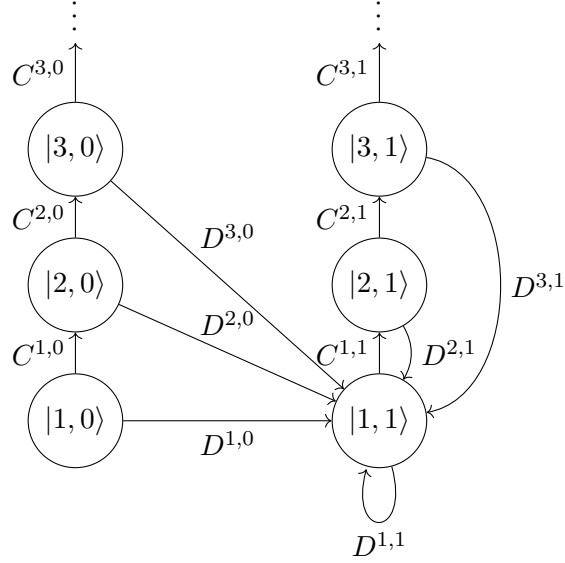

An approach to determine $Z_n^t$ or $Z_n$ would correspond to understanding each of the allowed terms $\ket{a', b'}$ and their associated \emph{weights}, e.g.~the weight of $\ket{1, 1}$ after two applications of $M$ is given by $C^{1, 0}D^{2, 0}+D^{1, 0}D^{1, 1}$. We can think of the repeated application of $M$ as summing each of the possible walks that one can take in the Markov chain below, see Fig.~\ref{fig:markovchain1}. For each step, an additional multiplicative factor of $C^{a,b}$ or $D^{a,b}$ is incurred. After $n-1$ steps (i.e.~$n-1$ applications of $M$), there will be a number of walks that end up at a given $\ket{a', b'}$. The sum of the total weight of each walk that ends up at a given $\ket{a', b'}$ after $n-1$ steps is exactly the coefficient (i.e.~the weight) corresponding to $\ket{a', b'}$, when expanding $Z_n^t$ in the basis $\lbrace{\ket{a, b}\rbrace}_{a,b}$. 

Thus, doing this for each possible $\ket{a', b'}$ allows us to retrieve the full $Z_n^t$. At first sight, this seems like a hopeless task---it would require not only understanding the possible walks from $\ket{1, 0}$ to $\ket{a', b'}$ (for each $a',~b'$) in $n-1$ steps, but also keeping track of the associated weights. Fortunately, such problems can be dealt with using a generating function approach, as we will show in the following. For an extensive overview on generating functions see~\cite{flajolet2009analytic}.

Note that the above Markov chain has two distinct `branches' --- one with $b=0$ and one with $b=1$. First, we deal with the case of $b=0$. The only walk that ends in the $b=0$ branch is the one that stays in it for each of the $n-1$ steps, corresponding to multiplying by $C^{a, b}$ for each of the $n-1$ steps. This is because after taking a step corresponding to the $D^{a,b}$ coefficient, the $b$ parameter will remain equal to $1$.

The case for $b=1$ is a bit more complicated. Any possible such walk decomposes into three distinct \emph{subwalks}. The first subwalk stays in the $b=0$ branch for a number of steps before moving to the $b=1$ branch. The second subwalk consists of an arbitrary number of `loops' in the $b=1$ branch. A `loop' here corresponds to starting from $\ket{1, 1}$, `ascending' to some $\ket{a',1}$, and dropping down to $\ket{1, 1}$ again. Finally, there will be a final ascent to some $\ket{a', 1}$. The final ascent is considered separately, since the weight associated to $\ket{a', 1}$ will acquire a multiplicative factor of $\frac{1-q^{a'}\lambda}{q^{a'}\lambda}$, due to the linear form taking $Z_n^t$ to $Z_n$, see Eqs.~\eqref{eq:3linksexp} and~\eqref{eq:3linksexp_total}.

The above characterization of the walks allows us to reduce the calculation of $\overline{G}$ to the calculation of the generating functions of the above walks. In other words, we have that
\begin{align}
\overline{G} =&~\overline{G}_{0\rightarrow \textrm{end}} + \overline{G}_{b=1}\\
 =&~\overline{G}_{0\rightarrow \textrm{end}} + \overline{G}_{0\rightarrow 1} \cdot \overline{G}_{1\rightarrow 1}^{\textrm{seq}} \cdot \overline{G}_{1\rightarrow \textrm{end}}\label{eq:sep_gen_functions}\ .
\end{align}
Here $\overline{G}_{0\rightarrow 1}$ is the generating function corresponding to staying in the $b=0$ branch, while $\overline{G}_{b=1}$ is the generating function for walks ending up in the $b=1$ branch. Furthermore, $\overline{G}_{0\rightarrow 1}$ is the generating function for walks starting in the $b=0$ branch ending with a transition to the $b=1$ branch, $\overline{G}_{1\rightarrow 1}^{\textrm{seq}}$ is the generating function of all possible walks starting and ending at $\ket{1, 1}$, and $\overline{G}_{1\rightarrow \textrm{end}}$ the generating function of all ascents starting at $\ket{1, 1}$ (including the term corresponding to the linear form). 
In the first line we used that the generating function of two disjoint sets is the sum of the generating functions. In the second line we used the simple but powerful fact that the generating function of the Cartesian product of two classes equals the product of the generating functions of the two classes~\cite{flajolet2009analytic}. The last statement can be seen as follows. Let there be two walks of two different types, where $a_i$ and $b_j$ are the weights associated to walks of length $i$ and $j$ for the two types, respectively. Summing the product of the weights over every possible path of total length $n$ yields $\sum_{k=0}^{n}a_kb_{n-k}$. The claim then follows since 
\begin{align}
\left(\sum_{n=0}^\infty a_n x^n\right)\left(\sum_{n=0}^\infty b_n x^n\right) =& \sum_{n=0}^\infty c_n x^n\\
\textrm{where } c_n = \sum_{k=0}^n a_kb_{n-k}\ .
\end{align}
Now, it remains to determine the generating functions $\overline{G}_{0\rightarrow \textrm{end}},~\overline{G}_{0\rightarrow 1},~ \overline{G}_{1\rightarrow 1}^{\textrm{seq}} $ and $\overline{G}_{1\rightarrow \textrm{end}}$. First, we consider the $\overline{G}_1$ case. The transition from $\ket{1, 0}$ to $\ket{2, 0}$ corresponds to a weight of $C^{1,0}$. Similarly, the transition from $\ket{2, 0}$ to $\ket{3, 0}$ corresponds to a weight of $C^{2, 0}$. This yields a total contribution of $C^{1, 0}C^{2, 0}$ --- exactly the term in front of the $\ket{3, 0}$ term in Eq.~\eqref{eq:example_eq}. Generalizing, a single ascent from $\ket{1, 0}$ to $\ket{m+1, 0}$ acquires a weight of

\begin{align}
x^{m+1}\prod_{a=1}^{m}C^{a, 0} &= x^{m+1}\prod_{a=1}^{m}\left(-\frac{q^a\left(\frac{1}{\lambda}-\lambda\right)}{\left(1-\frac{1}{\lambda}q^a\right)\left(1-\lambda q^a\right)}\right)\\
&= x^{n+1}\prod_{a=0}^{n-1}\left(-\frac{q^aq \left(\frac{1}{\lambda}-\lambda\right)}{\left(1-\frac{q}{\lambda}q^a\right)\left(1-\left(\lambda q\right) q^a\right)}\right)\label{eq:gen_func_simp1}\\
&= x \left(-1\right)^nq^{\binom{n}{2}}\frac{\left[\left(\frac{1}{\lambda}-\lambda\right)qx\right]^n}{\left(\frac{q}{\lambda};q\right)_n\left(\lambda q;q\right)_n}\label{eq:gen_func1}\ .
\end{align}

where $\left(a;q\right)_m \equiv \prod_{i=0}^{m-1}(1-aq^{i})$ is the $q$-Pochhammer symbol~\cite{gasper2004basic}. The $x$ term keeps track of the corresponding number of segments in the repeater chain.

Furthermore, due to the linear form taking $Z_n^t$ to $Z_n$ (see Eq.~\eqref{eq:lform1}), each weight corresponding to $\ket{m+1, 0}$ also acquires a multiplicative factor of $\frac{q^{m+1}}{1-q^{m+1}}$, leading to a total weight of

\begin{align}
&~x \left(-1\right)^mq^{\binom{m}{2}}\frac{\left[\left(\frac{1}{\lambda}-\lambda\right)qx\right]^m}{\left(\frac{q}{\lambda};q\right)_m\left(\lambda q;q\right)_m}\cdot \frac{q^{m+1}}{1-q^{m+1}}\\
 = &~qx \left(-1\right)^mq^{\binom{m}{2}}\frac{\left[\left(\frac{1}{\lambda}-\lambda\right)q^2x\right]^m}{\left(\frac{q}{\lambda};q\right)_m\left(\lambda q;q\right)_m}\cdot \frac{1}{1-q^{m+1}}\\
  = &~\frac{qx}{1-q} \left(-1\right)^mq^{\binom{m}{2}}\frac{\left(q;q\right)_m\cdot \left[\left(\frac{1}{\lambda}-\lambda\right)q^2x\right]^m}{\left(\frac{q}{\lambda};q\right)_m\left(\lambda q;q\right)_m  \left(q^2;q\right)_m}\\
    = &~\frac{qx}{1-q} \left(-1\right)^mq^{\binom{m}{2}}\frac{\left(q;q\right)_m\left(q;q\right)_m\cdot \left[\left(\frac{1}{\lambda}-\lambda\right)q^2x\right]^m}{\left(q;q\right)_m\left(\frac{q}{\lambda};q\right)_m\left(\lambda q;q\right)_m  \left(q^2;q\right)_m}\ \label{eq:gen_func2}.
\end{align}
In the first step we took the expression from Eq.~\eqref{eq:gen_func1} and multiplied by the term acquired from the linear map, i.e.~$\frac{q^{m+1}}{1-q^{m+1}}$. In the second step we pulled the numerator of $\frac{q^{m+1}}{1-q^{m+1}}$ into the first part. In the third line we use that 
\begin{gather}
\frac{1}{1-aq^{m+1}} = \frac{\left(aq;q\right)_m}{\left(1-aq\right)\left(aq^2;q\right)_m} \label{eq:q_identity}\ .
\end{gather}
In the final step we multiply by $\frac{\left(q;q\right)_m}{\left(q;q\right)_m}$, which will turn out to be convenient in the next step.

The expression in Eq.~\eqref{eq:gen_func2} corresponds to the weight of $\ket{m+1,0}$. The total contribution is then given over all $m$, i.e.~

\begin{align}
\overline{G}_{0\rightarrow\textrm{end}} = &~\frac{qx}{1-q} \sum_{m=0}^{\infty}\left(-1\right)^mq^{\binom{m}{2}}\frac{\left(q;q\right)_m\left(q;q\right)_m\cdot \left[\left(\frac{1}{\lambda}-\lambda\right)q^2x\right]^m}{\left(q;q\right)_m\left(\frac{q}{\lambda};q\right)_m\left(\lambda q;q\right)_m  \left(q^2;q\right)_m}\\
 =& \frac{qx}{1-q} \pFq{3}{3}{q,q,0}{\frac{q}{\lambda}, \lambda q, q^2}{q,x q^2\left(\frac{1}{\lambda}-\lambda\right)} \label{eq:gen_expr1}\ , 
\end{align}
where 

\begin{align}
\pFq{j}{k}{c_1,c_2,\ldots, c_j}{d_1,d_2,\ldots, d_k}{q,x} \equiv \sum_{m=0}^{\infty} \frac{\left(c_1, c_2, \ldots , c_j;q\right)_m}{\left(d_1, d_2, \ldots , d_k, q;q\right)_m} \left(\left(-1\right)q^{\binom{m}{2}}\right)^{1+k-j}x^m\label{eq:qhyper}
\end{align}

is a so-called $q$-\emph{hypergeometric series}~\cite{gasper2004basic}. In the above we used the common shorthand $\left(c_1;q\right)_m \left(c_2;q\right)_m\cdots \left(c_k; q\right)_m\equiv \left(c_1, c_2, \ldots, c_k;q\right)_m$. Note the additional $q$ in the $q$-Pochhammer symbol in the denominator of Eq.~\ref{eq:qhyper}, motivating the introduction of the term $\frac{\left(q;q\right)_m}{\left(q;q\right)_m}$ in Eq.~\eqref{eq:gen_func2}.

We now move on to the $b=1$ case. As mentioned above, any possible walk that ends in the $b=1$ branch consists of three possible \emph{subwalks}, see Eq.~\eqref{eq:sep_gen_functions}. Let us first focus on the first subwalk corresponding to staying in the $b=0$ branch before moving to $\ket{1, 1}$, i.e.~the generating function $\overline{G}_{0\rightarrow 1}$. The associated weights are similar to before, but with an additional factor of $D^{m+1, 0}$. That is,  

\begin{align}
    x^{m+1}D^{m+1, 0}\prod_{a=1}^{m}C^{a, 0} &=x^{m+1}\frac{q^{m+1}}{\lambda-q^{m+1}}\prod_{a=1}^{m}\left(-\frac{q^a\left(\frac{1}{\lambda}-\lambda\right)}{\left(1-\frac{1}{\lambda}q^a\right)\left(1-\lambda q^a\right)}\right)\\
    &=x^{m+1}\left(\frac{q}{\lambda}\right)\frac{q^{m}}{1-\frac{1}{\lambda}q^{m+1}}\prod_{a=0}^{m-1}\left(-\frac{q^aq \left(\frac{1}{\lambda}-\lambda\right)}{\left(1-\frac{q}{\lambda}q^a\right)\left(1-\left(\lambda q\right) q^a\right)}\right)\\
     &=x^{m+1}\left(\frac{q}{\lambda}\right)\frac{q^{m}}{1-\frac{1}{\lambda}q^{m+1}}\left(-1\right)^mq^{\binom{m}{2}}\frac{\left[\left(\frac{1}{\lambda}-\lambda\right)qx\right]^m}{\left(\frac{q}{\lambda};q\right)_m\left(\lambda q;q\right)_m}\\
      &=x\left(\frac{q}{\lambda}\right)\frac{1}{1-\frac{1}{\lambda}q^{m+1}}\left(-1\right)^mq^{\binom{m}{2}}\frac{\left[\left(\frac{1}{\lambda}-\lambda\right)q^2x\right]^m}{\left(\frac{q}{\lambda};q\right)_m\left(\lambda q;q\right)_m}\\
            &=x\left(\frac{q}{\lambda}\right)\frac{1}{1-\left(\frac{q}{\lambda}\right)}\left(-1\right)^mq^{\binom{m}{2}}\frac{\left(q;q\right)_m\left(\frac{q}{\lambda};q\right)_m\left[\left(\frac{1}{\lambda}-\lambda\right)q^2x\right]^m}{\left(q;q\right)_m\left(\frac{q^2}{\lambda};q\right)_m\left(\frac{q}{\lambda};q\right)_m\left(\lambda q;q\right)_m}\label{eq:replace_id}\\
                        &=x\frac{\left(\frac{q}{\lambda}\right)}{1-\frac{q}{\lambda}}\left(-1\right)^mq^{\binom{m}{2}}\frac{\left(q;q\right)_m\left[\left(\frac{1}{\lambda}-\lambda\right)q^2x\right]^m}{\left(q;q\right)_m\left(\frac{q^2}{\lambda};q\right)_m\left(\lambda q;q\right)_m} \ .
\end{align}

where we used the identity from Eq.~\eqref{eq:q_identity} and added once more a factor of $\frac{\left(q;q\right)_m}{\left(q;q\right)_m}$ in Eq.~\eqref{eq:replace_id}. Summing once more over every value of $m$ we find

\begin{align}
 \overline{G}_{0\rightarrow 1}=&~x\frac{\left(\frac{q}{\lambda}\right)}{1-\frac{q}{\lambda}}\sum_{m=0}^{\infty}\left(-1\right)^mq^{\binom{m}{2}}\frac{\left(q;q\right)_m\left[\left(\frac{1}{\lambda}-\lambda\right)q^2x\right]^m}{\left(q;q\right)_m\left(\frac{q^2}{\lambda};q\right)_m\left(\lambda q;q\right)_m}\\
=&~x\frac{\left(\frac{q}{\lambda}\right)}{1-\frac{q}{\lambda}}  \pFq{2}{2}{q,0}{\frac{q^2}{\lambda},\lambda q}{q, x q^2\left(\frac{1}{\lambda}-\lambda\right)} \label{eq:gen_expr2}\ . 
\end{align}

We now deal with the $ \overline{G}_{1\rightarrow 1}^{\textrm{seq}}$ term. This is the generating function that corresponds to all walks from $\ket{1, 1}$ back to $\ket{1, 1}$. Note that such a walk can always be decomposed into several `loops', each consisting of a number of ascents before dropping back down to $\ket{1, 1}$. Let $ \overline{G}_{1\rightarrow 1}$ be the generating function associated to such a single loop. We then have that 

\begin{align}
\overline{G}_{1\rightarrow 1}^{\textrm{seq}} &= 1 + \left(\overline{G}_{1\rightarrow 1}\right) + \left(\overline{G}_{1\rightarrow 1} \cdot \overline{G}_{1\rightarrow 1}\right) + \left(\overline{G}_{1\rightarrow 1} \cdot \overline{G}_{1\rightarrow 1} \cdot \overline{G}_{1\rightarrow 1}\right) + \ldots\\
&=\frac{1}{1-\overline{G}_{1\rightarrow 1}} \ \label{eq:gen_quasi_inv} ,
\end{align}

at least in terms of formal power series. \kg{That is, we can safely ignore details of convergence, see~\cite{flajolet2009analytic}.} This follows from the fact that the generating function $\left(\overline{G}_{1\rightarrow 1}\right)^k$ corresponds to taking exactly $k$ loops, and that walks with different number of loops $k$ are disjoint. As noted in the main

A loop that ascends to $\ket{m+1, 1}$ has weight

\begin{align}
x^{m+1} D^{m+1, 1} \prod_{a=1}^{m} C^{a, 1}  = &~x^{m+1} \frac{q^{m+1}}{1-q^{m+1}} \prod_{a=1}^{m}\left( -\frac{q^a\lambda \left(\frac{1}{\lambda}-\lambda\right)}{\left(1-q^a\right)\left(1-\lambda^{2}q^a\right)}\right)\\
= &~x^{m+1} \frac{q^{m+1}}{1-q^{m+1}} \prod_{a=0}^{m-1}\left( -\frac{q^aq \left(1-\lambda^2\right)}{\left(1-q q^a\right)\left(1-q\lambda^{2}q^a\right)}\right)\\
= &~xq \frac{1}{1-qq^{m}} \prod_{a=0}^{m-1}\left( -\frac{q^axq^2 \left(1-\lambda^2\right)}{\left(1-q q^a\right)\left(1-q\lambda^{2}q^a\right)}\right)\\
= &~x\frac{1}{1-qq^m} \left(-1\right)^mq^{\binom{m}{2}}\frac{\left[xq^2\left(1-\lambda^2\right)\right]^m}{\left(q;q\right)_m \left(q\lambda^2;q\right)_m}\\
= &~\frac{x}{1-q}\left(-1\right)^mq^{\binom{m}{2}}\frac{\left[xq^2\left(1-\lambda^2\right)\right]^m}{\left(q^2;q\right)_m \left(q\lambda^2;q\right)_m}\label{eq:replace_with_identity}\\
= &~\frac{x}{1-q}\left(-1\right)^mq^{\binom{m}{2}}\frac{\left(q;q\right)_m\left[xq^2\left(1-\lambda^2\right)\right]^m}{\left(q;q\right)_m\left(q^2;q\right)_m \left(q\lambda^2;q\right)_m}\ ,
\end{align}

Where we used that $\left(1-qq^m\right)\left(q;q\right)_m = \left(1-q\right)\left(q^2;q\right)_m$ in Eq.~\eqref{eq:replace_with_identity}. Summing over all values of $m$ then gives

\begin{align}
\overline{G}_{1\rightarrow 1}=&~ \frac{xq}{\left(1-q\right)} \sum_{m=0}^{\infty}(-1)^mq^{\binom{m}{2}}\frac{\left(q;q\right)_m\left[xq^2\left(1-\lambda^2\right)\right]^m}{\left(q\lambda^2;q\right)_m\left(q;q\right)_m\left(q^2;q\right)_m}\\
= &~\frac{xq}{1-q} \pFq{2}{2}{q,0}{q^2,q\lambda^2}{q, x q^2\left(1-\lambda^2\right)}\ \label{eq:gen_expr3}.
\end{align}

We now move to the final piece---the calculation of $\overline{G}_{1\rightarrow \textrm{end}}$. This generating function corresponds to a number of ascents $m$, before applying the linear form. As such, we find

\begin{align}
x^{m+1}\frac{\lambda q^{m+1}}{1-\lambda q^{m+1}} \prod_{a=1}^{m} C^{a, 1} = &~x^{m+1}\frac{\lambda q^{m+1}}{1-\lambda q^{m+1}} \prod_{a=0}^{m-1} \left(-\frac{q^aq\left(1-\lambda^2\right)}{\left(1-qq^a\right)\left(1-q\lambda^2q^a\right)}\right)\\
= &~x^{m+1}\frac{\lambda q^{m+1}}{1-\lambda q^{m+1}} \left(-1\right)^mq^{\binom{m}{2}}\frac{\left[q\left(1-\lambda^2\right)\right]^m}{\left(q;q\right)_m\left(q\lambda^2;q\right)_m}\\
= &~x\frac{\lambda q}{1-\lambda q} \left(-1\right)^mq^{\binom{m}{2}}\frac{\left(\lambda q;q\right)_m\left[xq^2\left(1-\lambda^2\right)\right]^m}{\left(\lambda q^2;q\right)_m\left(q;q\right)_m\left(q\lambda^2;q\right)_m}\ .
\end{align}

Summing over $m$ yields

\begin{align}
\overline{G}_{1\rightarrow \textrm{end}}=&~x\frac{\lambda q}{1-\lambda q} \sum_{m=0}^{\infty}\left(-1\right)^mq^{\binom{m}{2}}\frac{\left(\lambda q;q\right)_m\left[xq^2\left(1-\lambda^2\right)\right]^m}{\left(\lambda q^2;q\right)_m\left(q;q\right)_m\left(q\lambda^2;q\right)_m}\\
=&~x\frac{\lambda q}{1-\lambda q}\pFq{2}{2}{\lambda q,0}{\lambda q^2,q\lambda^2}{q, x q^2\left(1-\lambda^2\right)}\ \label{eq:gen_expr4}.
\end{align}

We have now all ingredients; using Eqs.~\eqref{eq:def_G},~\eqref{eq:replaceq},~\eqref{eq:sep_gen_functions},~\eqref{eq:gen_expr1},~\eqref{eq:gen_expr2},~\eqref{eq:gen_quasi_inv},~\eqref{eq:gen_expr3},~\eqref{eq:gen_expr4} and $q\equiv 1- p$ we find that

\begin{align}
G = &\sum_{n=1}^{\infty}\mathds{E}\left[\Lambda_n\right]x^n\label{eq:final_gen_func}\\
=  &~x\cdot \Phi_1 +x^2\left(\frac{\left(1-q\right)^2\lambda}{\left(\lambda-q\right)\left(1-\lambda q\right)}\right) \frac{\Phi_2\cdot \Phi_4}{1-x\cdot \Phi_3}\ \label{eq:final_gen_func1},\\
\textrm{with } \Phi_1 =&~ \pFq{3}{3}{q,q,0}{q^2,\lambda q, \frac{q}{\lambda}}{q,x pq\left(\frac{1}{\lambda}-\lambda\right)}\ \label{eq:final_gen_func2}, \\
\Phi_2 = &~  \pFq{2}{2}{q,0}{\frac{q^2}{\lambda},\lambda q}{q, x pq\left(\frac{1}{\lambda}-\lambda\right)}\ , \\
\Phi_3  = &~\pFq{2}{2}{q,0}{q^2,q\lambda^2}{q, x pq\left(1-\lambda^2\right)}\ \label{eq:final_gen_func3},\\
\Phi_4 =&~\pFq{2}{2}{\lambda q,0}{\lambda q^2,q\lambda^2}{q, x pq\left(1-\lambda^2\right)} \label{eq:final_gen_func4}\ .
\end{align}

In the above we have changed $x$ to $x\left(\frac{1-q}{q}\right)$ (see Eq.~\eqref{eq:replaceq}), and collected the prefactors in front of the $q$-hypergeometric series.

\subsection{Cut-off case}
The goal of this section is to find the generating function for the case of a global cut-off. The construction is the same as the case without a cut-off. That is, the recurrence relation can be mapped unto a Markov chain, and the generating function can be recovered by splitting the possible walks into different subwalks (and finding the generating functions of those subwalks). In this section we will only discuss the splitting into subwalks---the determination of the individual generating functions is similar to the previous section. As such, we do not report these calculations here and instead refer to our Mathematica code~\cite{mathematicafiles}.

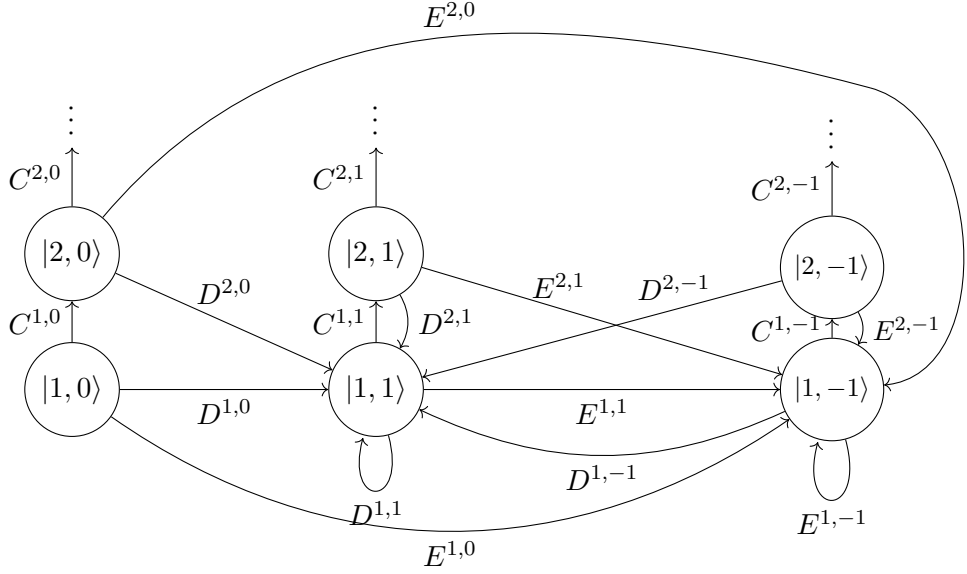
\begin{figure}[h]
\centerfloat
\begin{tikzpicture}[node distance=1.8cm]
\tikzset{invisible/.style={minimum width=0mm,inner sep=0mm,outer sep=0mm}}
\node [circle , draw ] (10) at (-2 , 0) {$\ket{1, 0}$};
\node [circle , draw ] (20) [ above of= 10] {$\ket{2, 0}$};
\node [] (30) [ above of= 20, rotate=90] {$\cdots$};

\node [circle , draw ] (1L)  at (2, 0) {$\ket{1, 1}$};
\node [circle , draw] (2L) [ above of= 1L] {$\ket{2, 1}$};
\node [] (3L) [ above of= 2L, rotate=90] {$\cdots$};

\node [circle , draw, scale=0.9] (1R)  at (8, 0) {$\ket{1,-1}$};
\node [circle , draw, scale=0.9] (2R) [ above of= 1R] {$\ket{2, -1}$};
\node [] (3R) [ above of= 2R, rotate=90] {$\cdots$};

\path[->] (10) edge node [left] {$C^{1, 0}$} (20) ;
\path[->] (20) edge node [left] {$C^{2, 0}$} (30) ;

\path[->] (10) edge node [below] {$D^{1, 0}$} (1L) ;
\path[->] (20) edge node [above=0.1cm] {$D^{2, 0}$} (1L) ;

\node[invisible] (bC) at (8.5, 4) {};
\path[->] (10) edge [ bend right=35]  node [below] {$E^{1, 0}$} (1R) ;
\path (20) edge [out=50, in=165]  node [above] {$E^{2, 0}$} (bC) ;
\path[->] (bC) edge [ out=-15,in=5] (1R) ;

\path[->] (1L) edge node [left] {$C^{1, 1}$} (2L) ;
\path[->] (2L) edge node [left] {$C^{2, 1}$} (3L) ;


\path (1L) edge [loop below] node[below]{$D^{1, 1}$} (1L);
\path[->] (2L) edge [ bend left ] node [right] {$D^{2,1}$} (1L) ;


\path[->] (1R) edge node [left] {$C^{1, -1}$} (2R) ;
\path[->] (2R) edge node [left] {$C^{2, -1}$} (3R) ;


\path (1R) edge [loop below] node[below]{$E^{1, -1}$} (1R);
\path[->] (2R) edge [ bend left ] node [right, scale=0.95] {$E^{2,-1}$} (1R) ;


\path[->]  (1R) edge [ bend left=25 ] node[below] {$D^{1, -1}$} (1L);
\path[->]  (1L) edge [ bend left=0 ] node[below ] {$E^{1, 1}$} (1R);

\path[->] (2R) edge node [above, pos=0.3] {$D^{2, -1}$} (1L) ;
\path[->] (2L) edge node [above, pos=0.38] {$E^{2, 1}$} (1R) ;

\end{tikzpicture}
\caption{An unnormalized Markov chain with transition `probabilities' given by $C^{a, b}$, $D^{a, b}$ and $E^{a, b}$. The starting point is $\ket{1, 0}$. There are now several different types of walks, explained in the main text.}
\label{fig:markovchain2}
\end{figure}

In the case of a cut-off, the linear map taking $Z_{n}^t$ to $Z_{n}^t$ is defined by $\ket{a, b} \xrightarrow{M} C^{a,b}\ket{a+1, b} + D^{a,b}\ket{1, 1} + E^{a,b}\ket{1, -1}$, with $C^{a, b}$, $D^{a, b}$ as before, and

\begin{align}
E^{a,b} = -\frac{q^{a\left(t_c+1\right)} \lambda^{1+t_c+b\left(1+t_c\right)} }{1-q^a\lambda^{b+1}} \ ,
\end{align}
see Section~\ref{sec:recursion_cutoff}. This map can also be seen in terms of an unnormalized Markov chain, where there are now three branches, $b=0, 1, -1$, see Fig.~\ref{fig:markovchain2}. This case is more complicated, since a walk might now alternate between the $b=-1$ and $b=1$ branch. In what follows, we list all possible subwalks. We change notation slightly; for a generating function of the form $\G{x}{y}$,  $x$ refers to the start of the subwalk, while $y$ refers to the end of the subwalk. Furthermore, instead of referring to the branches by their $b$ parameter, we will use the convention that the $b=0, 1, -1$ branches correspond to $c=0, 1, 2$, respectively. Finally, we will drop the overline notation, with the understanding that the distinction between the generating functions with and without the replacement $x\mapsto x \frac{1-q}{q\left(1-q^{T_c}\right)}$ of is clear.

As before, a walk might stay in the $c=0$ branch, corresponding to the $\G{0}{\tend}$ generating function. On the other hand, a walk might stay in the $c=0$ branch for some time, before moving to the $c=1$ or $c=2$ branch, corresponding to $\G{0}{1}$ and $\G{0}{2}$, respectively. Let us focus on the walks that first go to the $c=1$ branch and also finish there. As before, the walk can stay for an arbitrary number of attempts in the $c=1$ branch before falling back to $\ket{1, 1}$. A single such `loop' is described by a generating function $\G{1}{1}$, such that an arbitrary sequence of such `loops' is given by $\frac{1}{1-\G{1}{1}}$.

Since the walk needs to end in the $c=1$ branch, it needs to transition back and forth between the $c=1$ and $c=2$ branches several times (including potentially zero times). Let us consider the case where it transitions only once to the $c=2$ branch. The transition itself is captured by some generating function $\G{1}{2}$. As with the $c=1$ branch, the walk can stay in the $c=2$ branch for a number of `loops', corresponding to $\frac{1}{1-\G{2}{2}}$. Afterwards, the walk transitions back to the $c=1$ branch with associated generating function $\G{2}{1}$. Up to this point, the generating function of a walk that transitions once back and forth between the $c=1$ and $c=2$ branch is given by

\begin{gather}
\G{0}{1}\cdot \left(\frac{\G{1}{2}}{1-\G{1}{1}}\cdot \frac{\G{2}{1}}{1-\G{2}{2}}\right)\ .
\end{gather}

The transition from $c=1$ to $c=2$ and back can now happen an arbitrary number of times, leading to the following,
\begin{gather}
\G{0}{1} \cdot \frac{1}{1-\left(\frac{\G{1}{2}}{1-\G{1}{1}}\cdot \frac{\G{2}{1}}{1-\G{2}{2}}\right)} \ .
\end{gather}
Finally, the walk will have an arbitrary number of loops in the $c=1$ branch, before having one final ascent,
\begin{gather}
\G{0}{1}\cdot \frac{1}{1-\left(\frac{\G{1}{2}}{1-\G{1}{1}}\cdot \frac{\G{2}{1}}{1-\G{2}{2}}\right)}\cdot  \frac{1}{1-\G{1}{1}}\cdot \G{1}{\tend} \ .
\end{gather}
In a similar fashion, the generating functions for the walk starting and finishing at $c=2$ is given by
\begin{gather}
\G{0}{2}\cdot \frac{1}{1-\left(\frac{\G{2}{1}}{1-\G{2}{2}}\cdot \frac{\G{1}{2}}{1-\G{1}{1}}\right)}\cdot  \frac{1}{1-\G{2}{2}}\cdot \G{2}{\tend} \ .
\end{gather}
Let us now consider the walk that starts at $c=1$ and ends at $c=2$. Note that we can think of this as a path that is very similar to a path that starts and ends at $c=1$. That is, instead of finishing in the $c=1$ branch after oscillating between the $c=1$ and $c=2$ branch, there is a transition between to the $c=2$ branch. Finally, there will be a number of `loops' and one last ascent in the $c=2$ branch. This yields
\begin{gather}
\G{0}{1}\cdot \frac{1}{1-\left(\frac{\G{1}{2}}{1-\G{1}{1}}\cdot \frac{\G{2}{1}}{1-\G{2}{2}}\right)}\cdot  \frac{1}{1-\G{1}{1}}\cdot   \G{1}{2} \cdot \frac{1}{1-\G{2}{2}}\cdot \G{2}{\tend}\ .
\end{gather}
Similarly, for the walks that start and end at $c=2$ and $c=1$, respectively, we find that
\begin{gather}
\G{0}{2}\cdot \frac{1}{1-\left(\frac{\G{2}{1}}{1-\G{2}{2}}\cdot \frac{\G{1}{2}}{1-\G{1}{1}}\right)}\cdot  \frac{1}{1-\G{2}{2}}\cdot   \G{2}{1} \cdot \frac{1}{1-\G{1}{1}}\cdot \G{1}{\tend}\ .
\end{gather}
Adding the generating function of all five types of path and collecting like terms give a generating function of the form
\begin{align}
&\G{0}{\tend}+ \frac{\frac{\G{0}{1}}{1-\G{1}{1}}\cdot\left(\G{1}{\tend}+\frac{\G{1}{2}\cdot \G{2}{\tend}}{1-\G{2}{2}}\right)+\frac{\G{1}{2}}{1-\G{2}{2}}\cdot\left(\G{2}{\tend}+\frac{\G{2}{1}\cdot \G{1}{\tend}}{1-\G{1}{1}}\right)}{1-\left(\frac{\G{1}{2}}{1-\G{1}{1}}\cdot \frac{\G{2}{1}}{1-\G{2}{2}}\right)}\\
=&~\G{0}{\tend}+ \frac{  \G{0}{1}\left(\left(1-\G{2}{2}\right)\G{1}{\tend} + \G{1}{2}\cdot \G{2}{\tend}\right)    + \G{0}{2}\left(\left(1-\G{1}{1}\right)\G{2}{\tend} +     \G{2}{1}\cdot \G{2}{\tend}        \right)      }{1-\left(\G{1}{1}     + \G{2}{2} +   \G{1}{2}\cdot  \G{2}{1}  -  \G{1}{1}\cdot \G{2}{2}\right)   }\ .
\end{align}

As mentioned previously, the individual generating functions $\G{x}{y}$ themselves are tedious but straightforward to calculate, and we leave them to the Mathematica file~\cite{mathematicafiles}. After making the replacement $x\mapsto x \frac{1-q}{q\left(1-q^{T_c}\right)} \equiv R \cdot x $ we find that the generating function is given by

\begin{align}
G(x) = &\sum_{n=1}^{\infty}\mathds{E}\left[\Lambda_n\left(T_c\right)\right]x^n\label{eq:final_gen_func_cutoff}\\
= &~\Phit{0}{\tend}+ Rx\cdot \frac{  \Phit{0}{1}\left(\left(1-\Phit{2}{2}\right)\Phit{1}{\tend} + \Phit{1}{2}\cdot \Phit{2}{\tend}\right)    + \Phit{0}{2}\left(\left(1-\Phit{1}{1}\right)\Phit{2}{\tend} +     \Phit{2}{1}\cdot \Phit{2}{\tend}        \right)      }{1-\left(\Phit{1}{1}     + \Phit{2}{2} +   \Phit{1}{2}\cdot  \Phit{2}{1}  -  \Phit{1}{1}\cdot \Phit{2}{2}\right)   }\ ,\\
\textrm{with } \Phit{0}{\tend} =&~\sum_{t=1}^{T_c}q^t R x \cdot \pFq{2}{2}{q,0}{\frac{q}{\lambda},\lambda q}{q, q^{t+1} R x \left(\frac{1}{\lambda}-\lambda\right)}\ , \\
\Phit{0}{1} =&~ -\frac{Rx}{1-\frac{\lambda}{q}}\pFq{2}{2}{q, 0}{\frac{q^2}{\lambda}, q\lambda}{q,   q^2 R x \left(\frac{1}{\lambda}-\lambda\right)}\ ,\\
\Phit{0}{2} =&~ -\frac{Rx\left(q\lambda\right)^{T_c+1}}{1-\lambda q} \pFq{2}{2}{q, 0}{\frac{q}{\lambda}, q^2\lambda}{q, q^{T_c+2}Rx\left(\frac{1}{\lambda}-\lambda\right)}\ ,\\
\Phit{1}{1} =&~ \frac{qRx}{1-q} \pFq{2}{2}{q, 0}{q^2, q\lambda^2}{q, q^2Rx\left(1-\lambda^2\right)}\ , \\
\Phit{2}{2} =& - \frac{q^{T_c+1}Rx}{1-q} \pFq{2}{2}{q, 0}{q^2, \frac{q}{\lambda^2}}{q, q^{T_c+2}Rx\left(\frac{1}{\lambda^2}-1\right)}\ , \\
\Phit{1}{2} =& -\frac{Rx\left(q\lambda^2\right)^{T_c+1}}{1-q\lambda^2} \pFq{2}{2}{q, 0}{q, q^2\lambda^2}{q, q^{T_c+2}Rx\left(1-\lambda^2\right)}\ , \\
\Phit{2}{1} =& -\frac{Rx}{1-\frac{\lambda^2}{q}} \pFq{2}{2}{q, 0}{q, \frac{q^2}{\lambda^2}}{q, q^2Rx\left(\frac{1}{\lambda^2}-1\right)}\ ,\\
\Phit{1}{\tend} =\sum_{t=1}^{T_c}&\left(q\lambda\right)^t\pFq{1}{1}{0}{q\lambda^2}{q, q^{t+1}Rx\left(1-\lambda^2\right)}\ ,\\
\Phit{2}{\tend} =&\sum_{t=1}^{T_c}\left(\frac{q}{\lambda}\right)^{t}\pFq{1}{1}{0}{\frac{q}{\lambda^2}}{q, q^{t+1}Rx\left(\frac{1}{\lambda^2}-1\right)}\ .\\
\end{align}

\subsection{Approximations for the global cut-off}\label{sec:appendix_cutoff_approx}
Here we find a tight approximation similar to the case without a cut-off, see Section \ref{sec:gen_func_exploration}. The approximation is given by

\begin{gather}
\mathds{E}\left[\Lambda_n(T_c)\right] \sim  \left[-\textrm{Res}_\rho\left(G\right)\right]]\left(\frac{1}{\rho}\right)^{n+1}\nonumber\\
\equiv \left[\overline{A}(\lambda, q, T_c)\right]\cdot \overline{B}(\lambda, q, T_c)^n\ \label{eq:approximation_cutoff} \ ,
\end{gather}

As with the case without a cut-off, the behavior of the average noise parameter is governed by the simple pole of $G(x)$ closest to $x=0$. Denote by $\rho$ the location of this singularity. As before, $\rho$ is given by the solution to $1-\left(\Phit{1}{1}     + \Phit{2}{2} +   \Phit{1}{2}\cdot  \Phit{2}{1}  -  \Phit{1}{1}\cdot \Phit{2}{2}\right)  =0$. Denote this solution by $x=\rho$. The decay factor $\overline{B}$ is then given by

\begin{gather}
\overline{B}\left(\lambda, q, T_c\right) = \frac{1}{\rho}\ .
\end{gather}

The other factor $\overline{A}$ is a bit more complicated,

\begin{gather}
\overline{A}\left(\lambda, q, T_c\right) = R\cdot \frac{\left.\left(\Phit{0}{1}\left(\left(1-\Phit{2}{2}\right)\Phit{1}{\tend} + \Phit{1}{2}\cdot \Phit{2}{\tend}\right)    + \Phit{0}{2}\left(\left(1-\Phit{1}{1}\right)\Phit{2}{\tend} +     \Phit{2}{1}\cdot \Phit{2}{\tend}   \right)     \right)\right|_{x=\rho} }{\left.\left(\left(\Phit{1}{1}     + \Phit{2}{2} +   \Phit{1}{2}\cdot  \Phit{2}{1}  -  \Phit{1}{1}\cdot \Phit{2}{2}\right)'\right)\right|_{x=\rho}} \ .
\end{gather}

Note that only the $\overline{A}$ function involves a sum over $t$ while $\overline{B}$ does not.

\end{document}